%% file: Higgs_EFT_limits.tex
\documentclass[amsmath,amssymb,aps,prd,11pt,tightenlines,superscriptaddress,
nofootinbib,preprintnumbers,notitlepage]{revtex4-1}
\usepackage{multirow}
\usepackage{graphicx}
\usepackage{tabularx}
\usepackage{slashed}
\usepackage{url}
\usepackage{hyperref}
\usepackage[maxfloats=100]{morefloats}
\usepackage{color,xcolor}
\usepackage{booktabs}
\usepackage[normalem]{ulem}
\definecolor{dark-red}{rgb}{0.8,0.0,0.0}
\definecolor{dark-blue}{rgb}{0.0,0.0,0.8}
\definecolor{dark-green}{rgb}{0.,0.6,0.}

\makeatletter
\def\l@subsubsection#1#2{}
\makeatother

\input{declare}

\AtBeginDocument{
  \heavyrulewidth=.08em
  \lightrulewidth=.05em
  \cmidrulewidth=.03em
  \belowrulesep=.65ex
  \belowbottomsep=0pt
  \aboverulesep=.4ex
  \abovetopsep=0pt
  \cmidrulesep=\doublerulesep
  \cmidrulekern=.5em
  \defaultaddspace= .5em
  \setlength{\tabcolsep}{0.5em}
}

\begin{document}


\title{Pushing Higgs Effective Theory to its Limits} 

\author{Johann Brehmer}
\affiliation{Institut f\"ur Theoretische Physik, Universit\"at Heidelberg, Germany}

\author{Ayres Freitas}
\affiliation{PITT-PACC, Department of Physics \& Astronomy, University of Pittsburgh, USA}

\author{David L\'opez-Val}
\affiliation{Centre for Cosmology, Particle Physics \& Phenomenology CP3, Universit\'e catholique de Louvain, Belgium}
 
\author{Tilman Plehn}
\affiliation{Institut f\"ur Theoretische Physik, Universit\"at Heidelberg, Germany}

\date{\today}

\begin{abstract} 
At the LHC, an effective theory of the Higgs sector allows us to
analyze kinematic distributions in addition to inclusive rates,
although there is no clear hierarchy of scales. We systematically
analyze how well dimension-6 operators describe LHC observables in
comparison to the full theory, and in a range where the LHC will be
sensitive. The key question is how the breakdown of the dimension-6
description affects Higgs measurements during the upcoming LHC run for
weakly interacting models. We cover modified Higgs sectors with a
singlet and doublet extension, new top partners, and a vector triplet.
First, weakly interacting models only generate small relevant subsets
of dimension-6 operators.  Second, the dimension-6 description tends
to be justified at the LHC. Scanning over model parameters,
significant discrepancies can nevertheless arise; their main source is
the matching procedure in the absence of a well-defined hierarchy of
scales. This purely theoretical problem should not affect future LHC
analyses. 
\end{abstract}

\maketitle
\tableofcontents

\clearpage

\section{Introduction}
\label{sec:intro}

The Higgs boson~\cite{higgs} discovery announced on July 4th
2012~\cite{discovery} is a historical milestone in the physics of the
21st century.  The thorough scrutiny of the LHC Run I data has so far
confirmed that the narrow resonance observed at a mass around 125~GeV
is compatible with the minimal Standard Model (SM) agent of electroweak
symmetry breaking~\cite{lecture}. To date, this agreement is limited to
around $20\%$ precision in the Higgs
couplings~\cite{sfitter_higgs,sfitter_last,Lopez-Val:2013yba, coupling_fits},
which is not sensitive to the deviations that one would expect from
typical perturbatively extended Higgs
sectors.  This accuracy, based on a large set of on-shell and most
recently off-shell Higgs measurements~\cite{sfitter_last}, will soon
improve with data from Run~II.  Odds are high that the upcoming
runs will shed light on a possible UV completion of the Standard
Model~\cite{higgsreview,bsmreview}.

Based on everything we know, such an underlying theory should be
described by a gauge field theory. While the measurement of Higgs
couplings from inclusive rates has been extremely successful at Run~I,
it needs to be extended, for example to include kinematic
distributions. For this purpose, Higgs effective field theories
(EFT)~\cite{eftfoundations,eftorig,eftreviews} have become the
\emph{koin\'e} for discussing the phenomenology of extended Higgs
sectors.  In the effective field theory language, beyond the Standard
Model (BSM) effects are described in terms of a Lagrangian with local
operators of increasing mass dimension $d > 4$. Each of them includes
a suppression by inverse powers of a new physics scale, which should
be well separated from the experimentally accessible scale, in our
case the electroweak scale, $\luv \gg v$.

Despite its generality, the EFT approach is known to suffer from its
limited applicability when the hierarchy of scales is not guaranteed.
This has fueled intense investigation in the context of dark matter
searches~\cite{dm}.  While in that field EFT-based predictions are
usually robust for early-universe and late-time annihilation rates as
well as for dark matter-nucleon scattering, the required hierarchy of
scales can break down for dark matter signals at colliders. Because
hadron collider do not have a well-defined partonic energy,
strategies relying on boosted objects and large
recoils are the most critical. While it is not clear that a marginal separation of scales
invalidates the EFT approach, such observables clearly pose a
challenge.\medskip

There exists a first set of studies of the applicability of EFTs to
Higgs physics at the
LHC~\cite{Biekoetter:2014jwa,heft_limitations,heft_limitations2}.
These questions first arose in studies of tagging jet kinematics in
weak boson fusion, which are sensitive to the UV structure of the
theory~\cite{bad_one,spins,phi_jj,higgs_pole}. Similar issues appear
in Higgs-strahlung~\cite{Biekoetter:2014jwa} and in the production of
off-shell Higgs bosons in gluon
fusion~\cite{taming,Buschmann:2014sia}. A key problem is that Higgs
production at hadron colliders does not probe a single energy scale
over the full relevant phase space.

On the other hand, in Ref.~\cite{sfitter_last} is has been shown that a fit of
dimension-6 operators to the Higgs data at Run~I is a sensible and
practicable extension of the usual Higgs couplings fit.  Dimension-6
operators including derivatives complement the Higgs coupling
modifications and allow us to extract information from kinematic
distributions. Because the LHC constraints do not induce a hierarchy
of scales, the EFT approach is not formally well defined. However,
there appears to be no problem in describing the LHC Higgs data in
terms of a truncated dimension-6 Lagrangian. This description induces
theory uncertainties if we want to interpret the LHC results in terms
of an effective field theory~\cite{trott}. On the other hand, these
and other theory uncertainties can and should be separated from the
experimental uncertainties~\cite{Cranmer:2013hia,fichet_moreau}.

Related to the topic of the validity of the effective theory is the
question if, given the experimental performance, the analysis of a
UV-complete model offers an advantage compared to the effective
theory~\cite{Henning:2014wua}. The two approaches are only equivalent
if we account for the full correlations between the effective
operators in all analysis steps, and if the effective theory is
applicable over the full relevant phase space. Unless the experimental
collaborations provide their fully correlated results beyond a
Gaussian approximation~\cite{sfitter_last}, a direct analysis of full
models will be superior.\medskip

Given these arguments, the applicability of the dimension-6 description of the
Higgs sector has to be tested on a process-to-process as well as
model-to-model basis.  In this paper we present a comprehensive
comparison of full models and their truncated EFT description during
the LHC Run~II. We select extensions of the Higgs sector of the
Standard Model by \textit{i)} a scalar singlet, \textit{ii)} a scalar
doublet, \textit{iii)} a colored top-partner scalar, and \textit{iv)}
a massive vector triplet.  Each of these models is mapped onto an EFT,
which we obtain by integrating out the heavy fields and expanding the
operators to dimension 6. We then derive predictions for selected
Higgs observables in the full model and compare them to the EFT
results.  The key questions we aim to address are:
\begin{enumerate}
\item Given the LHC sensitivity, how large do relevant new
  physics effects have to be?
\item Does the corresponding new physics scale respect a
  self-consistency condition $\luv \gg v$?
\item Which observables are correctly described by the truncated EFT?
\item What are the reasons for the potential failure of this EFT?
\item Do they pose a problem for LHC analyses?
\end{enumerate}
For weakly interacting models, visible effects at the LHC lead us to
scenarios in which the heavy scale is not sufficiently separated from
the electroweak scale, and the EFT description is not obviously
justified. We will analyze what problems the lack of
a clear hierarchy of scales leads to in practice, and discuss how these
might affect global LHC-Higgs fits including kinematic distributions~\cite{sfitter_last}.

It will turn out that two limitations of the EFT description will
guide us through the different models. First, we need to ensure that
the new physics scale and with it all new particles are properly
decoupled, in particular when we go beyond total cross
sections. Second, when we define our effective field theory
in terms of a Higgs-Goldstone doublet, it is crucial that the
electroweak vacuum expectation value (VEV) does not have a destabilizing
effect on the hierarchy of scales.\medskip

The remainder of the paper is organized as follows: in
Section~\ref{sec:theory} we review our theoretical framework. We discuss
how new physics effects in the Higgs sector are accounted for in the
full model and EFT languages, and we identify the reasons why the two
methods can deviate from each other.  In Section~\ref{sec:analysis} we
show these ideas at work by explicitly confronting full model versus
EFT predictions for a variety of UV completions and Higgs observables.
We give our conclusions in Section~\ref{sec:summary}. We hope that the
Appendices~\ref{sec:ap-eft}\,--\,\ref{sec:ap-triplet} with 
exhaustive details on the different models and their EFT
parametrizations will be particularly useful to practitioners.

\section{Effective theory basics}
\label{sec:theory}

Extensions of the SM Higgs sector involve new degrees of freedom with
electroweak charges and\,/\,or color charges, coupled to or mixing
with the SM-like Higgs boson. Hidden sectors coupled to the Higgs
potential without any SM charge lead to non-standard Higgs
decays. Since the Higgs potential is closely linked to the electroweak
sector, any model that affects the SM gauge bosons will also affect
Higgs physics. This way, a wide range of new physics models can be
probed in Higgs signatures at the LHC, both in total rates and kinematic
distributions.  The simplest effect are shifted couplings of the
observed Higgs boson at 125~GeV~\cite{sfitter_higgs},
\begin{align}
g_{xxH} = g_{xxH}^\text{SM} \left( 1 + \delx_x \right) \,.
\label{eq:shift2}
\end{align}
In this notation $\Delta$ can reflect both, a truncated EFT or a full
new physics model. These coupling deviations have been used to test an
effective light Higgs model with either free or model-specific
couplings~\cite{sfitter_higgs,Lopez-Val:2013yba,coupling_fits,sfitter_last}.

\subsection{Higgs effective theory}
\label{sec:theory_eff}

Effective field theories provide a systematic method to link Higgs
measurements to a large class of high-scale UV completions. Their
ingredients are \textit{i)} the dynamic degrees of freedom and
\textit{ii)} the symmetries at low energies.  The Higgs EFT framework
keeps the SM fields and requires an invariance under the SM gauge
group $SU(3) \times SU(2) \times U(1)$,
\begin{align}
 \lageff &= \lag_\text{SM} + \sum_{d=5}^\infty \, \sum_{a_d} \,
           \dfrac{C_{a_d}^{(d)}}{\Lambda^{d-4}}\,\mathcal{O}_{a_d}^{(d)} \,.
 \label{eq:efflaggen}
\end{align}
We assume a linear realization of electroweak symmetry breaking.  This
implies that the Higgs scalar and the Goldstones of the Standard Model
form an $SU(2)$ doublet $\phi$ with the vacuum expectation value
$v=246$~GeV. This is justified by the level of agreement of the
Standard Model with all available data on the electroweak sector. A
non-linear formulation in terms of a general scalar field $h$ is
also possible~\cite{nonlin}.  The higher-dimensional terms denote a linear
combination of local operators with mass dimension $d$, weighted by
Wilson coefficients $C_a$ and suppressed by inverse powers of the
new physics scale $\Lambda$.

Higher-dimensional operators can be classified depending on whether
they include derivatives to compensate for the mass dimension in
$1/\Lambda^2$. This leads to momentum-dependent couplings, scattering
amplitudes growing with energy, and eventually the violation of
perturbative unitarity. It reflects the onset of new on-shell
contributions, which are integrated out in the EFT.

When we link full models to an EFT description it is useful to
categorize the higher-dimensional operators according to whether they
arise from the tree-level exchange of heavy mediators or through loop
effects mediated by the heavy
fields~\cite{Passarino:2012cb,Arzt:1994gp}. This categorization is
only meaningful for weakly interacting complete models.\medskip

For the linear realization there exists a set of 59 dimension-6
operators. Popular bases are the Warsaw~\cite{Grzadkowski:2010es},
HISZ~\cite{Hagiwara:1993ck}, and SILH bases~\cite{silh}. All three
maximize the use of bosonic operators to describe Higgs and
electroweak observables. They can be mapped onto each other using
equations of motion, integration by parts, field redefinitions, and
Fierz transformations~\cite{Alonso:2014rga}.
We use the SILH basis and retain only those operators relevant for
Higgs physics at the LHC~\cite{silh}.  The effective
Lagrangian truncated to dimension 6 reads
\begin{align}
  \label{eq:EFT}
  \lag_\text{EFT} =& \lag_\text{SM} 
+ \frac{\bar{c}_H}{2v^2} \, \partial^\mu (\pbp) \, \partial_\mu (\pbp)
+ \frac{\bar{c}_T}{2v^2} \, (\phi^\dagger \, \overleftrightarrow{D}^\mu \, \phi) \, (\phi^\dagger \, \overleftrightarrow{D}_\mu \, \phi)
-\frac{\bar{c}_6\lambda}{v^2} (\pbp)^3 \notag \\
&+ \frac{ig\bar{c}_W}{2m^2_W} \, (\phi^\dagger \, \sigma^k \overleftrightarrow{D}^\mu\phi) \,  D^\nu \, {W^k}_{\mu\nu}
                         + \frac{ig'\bar{c}_B}{2m_W^2} \, (\phi^\dagger \, \overleftrightarrow{D}^\mu \, \phi) \, \partial^\nu \, B_{\mu\nu} \notag \\
&+ \frac{ig \, \bar{c}_{HW}}{m_W^2} \, (D^\mu \, \phi^\dagger) \, \sigma^k \, (D^\nu \, \phi) \, W^k_{\mu\nu}
                       + \frac{ig'\bar{c}_{HB}}{m_W^2} (D^\mu\phi^\dagger) \, (D^\nu \, \phi) \, B_{\mu\nu}\notag \\
&+ \frac{g'^2 \bar{c}_\gamma}{m_W^2} \, (\pbp) \, B_{\mu\nu} \, B^{\mu\nu} + \frac{g_s^2 \bar{c}_g}{m_W^2} \, (\pbp) \, G^A_{\mu\nu} \, G^{\mu\nu\, A} \notag \\
&- \left[
  \frac{\bar{c}_u}{v^2} \, y_u \, (\pbp) (\phi^\dagger\cdot \, \overline{Q}_L) \, u_R
+ \frac{\bar{c}_d}{v^2} \, y_d \, (\pbp) (\phi \,  \overline{Q}_L) \, d_R  
+ \frac{\bar{c}_\ell}{v^2} \, y_\ell \, (\pbp) (\phi \,  \overline{L}_L) \, \ell_R
                                 + \text{h.c.} \, \right] \,.
\end{align}
Here, $g = e/\sw, g' = e/\cw$, and $g_s$ stand for the SM gauge
couplings and $\lambda$ denotes the usual Higgs quartic coupling. The
normalization of the dimension-6 Wilson coefficients $\overline{c}_i$
does not follow Eq.\,\eqref{eq:efflaggen}, but includes conventional
prefactors which reflect a bias concerning their origin. We present
further details on the EFT setup, the translation between the
different bases, and the connection to Higgs observables in
Appendix~\ref{sec:ap-eft}.

\subsection{Default vs $v$-improved matching}
\label{sec:theory_improved}

Matching the dimension-6 Lagrangian to a full model
is a three-step procedure.
Its starting point is the definition of a heavy mass scale $\Lambda$.
Second, we integrate out the degrees of freedom above $\Lambda$,
which leads to an infinite tower of higher-dimensional operators.
Finally, this effective action is truncated so that only the dimension-6 terms,
suppressed by $1 / \Lambda^2$, remain.
The matching is not unambiguous:
on the one hand, $\Lambda$ is usually not uniquely defined.
Further ambiguities arise in the third step because a dimension-6 truncation
does not tell us how $\ord(\Lambda^{-4})$ contributions to the Wilson coefficients
of the dimension-6 operators should be treated.
\medskip

For the linear dimension-6 Lagrangian in terms of the doublet field
$\phi$ the underlying assumption $\Lambda \gg v$ suggests to match the
linear EFT to the full theory in the unbroken electroweak phase. An obvious choice
for the matching scale is then the mass scale of new particles in the limit of
$v \to 0$. We expand the effective action and drop all terms of $\ord(\Lambda^{-4})$.
This way, the truncation removes the parts
of the Wilson coefficients of the dimension-6 operators that are suppressed
by additional factors of $1/ \Lambda$.
This procedure is our default matching scheme.

In the absence of a clear hierarchy of scales, multiple heavy mass
scales of the type $\Lambda \pm g v$ occur for instance through mixing
effects in mass matrices, even if just one dimensionful parameter
$\Lambda$ governs the new physics. 
This raises the question if we can improve the agreement between
full model and dimension-6 Lagrangian by
incorporating effects of the non-zero electroweak VEV in the
matching. 
In the first matching step, we can define $\Lambda$ as the physical mass of the
new particles in the broken phase, including contributions from $v$, rather
than the mass scale in the unbroken phase. In addition, the third step gives us the
choice to keep (part of) the $\ord(\Lambda^{-4})$ terms of the
Wilson coefficients. This is equivalent to expressing the
coefficients in terms of phenomenologically relevant quantities such
as mixing angles and physical masses, again defined in the broken
phase.
Both of these prescriptions effectively include effects from dimension-8 operators
into the dimension-6 Lagrangian by once replacing $\phi^\dagger \phi \to v^2 / 2$.
We will use the term \emph{$v$-improved} matching for these
alternative EFT definitions. 
\medskip

The truncation of the EFT Lagrangian is formally justified as long as
$v \ll \Lambda$ and we only probe energies $\ephys \ll \Lambda$.  In
this limit the dimension-8 operators as well as the
$\Lambda$-suppressed terms in the Wilson coefficients are negligible;
our two matching procedures then give identical results.
In the absence of a large enough scale separation, our bottom-up approach
allows us to treat them independently. This way we can use the $v$-improved
matching to enhance the validity of the dimension-6 Lagrangian.

The external energy scale depends on the specific process and observable, \eg
$\ephys \sim m_h$ for on-shell Higgs coupling measurements, $\ephys
\sim m_{4 \ell}$ for off-shell Higgs coupling measurements, $\ephys
\sim m_{hh}$ for Higgs pair production at threshold, or $\ephys \sim
p_{T,h}$ for boosted single or double Higgs production. In kinematic
distributions the high-energy tails can probe significantly larger
energy scales.
This implies that the energy range where the EFT description is
applicable is model-dependent and observable-dependent. Successively
adding higher-dimensional operators should improve the situation, as long
as the key scales $\ephys, \Lambda$ are sufficiently separated. Of
course, the EFT description fails spectacularly in the presence of new
resonances in the relevant energy range, and we have to adjust the
field content of the effective Lagrangian.

\subsection{Self consistency at the LHC}
\label{sec:theory_self}

Interpreting LHC physics in terms of an effective theory involves a
delicate balance between energy scales.  On the one hand, new physics
searches often rely on selection criteria which demand $\ephys > m_h$
to separate a high-energy signal from the QCD background.  On the
other hand, a model-specific scale $\luv$ limits the validity of the effective
theory, as discussed above.  

The extraction of Higgs properties during the LHC Run~I essentially
relies on on-shell single Higgs production and decay.  This allows us
to roughly estimate the new physics scales they are able to probe.
Assuming no loop suppression, a deviation from the total single Higgs
production and decay rate lies within the experimental reach of the
LHC if
\begin{align}
\left| \frac{\sigma \times \br}{\left( \sigma \times \br \right)_\text{SM}} - 1 \right|
= \frac{g^2 m_h^2}{\Lambda^2} \gtrsim 0.1
\qquad \Leftrightarrow \qquad 
\luv < \sqrt{10} \, g \, m_h \simeq 280~\gev \,,
\label{eq:lmax}
\end{align}
where we assume a weakly interacting theory with $g^2 \sim 1/2$.
Because of the limited precision of the available data, current Higgs
results cannot test very high energy scales, at least for
weakly coupled new physics~\cite{sfitter_last}.  
For this simple power-counting argument we
ignore that new physics might also change distributions and
especially affect the high-energy tails. In this case the EFT expansion 
develops in two different directions, $E/\Lambda$ and $gv/\Lambda$.

For loop-induced new
physics effects, the corresponding loop suppression factor pulls
$\luv$ to even lower values,
\begin{align}
\left| \frac{\sigma \times \br}{\left( \sigma \times \br \right)_\text{SM}} -1 \right |
= \frac{g^2 m_h^2}{16\pi^2 \Lambda^2} \gtrsim 0.1
\qquad \Leftrightarrow \qquad 
\luv < \frac{\sqrt{10} \, g\, m_h}{4\pi}
\simeq 20~\gev \,.
\label{eq:lmax_loop}
\end{align}
This implies that the cut-off of the effective theory is below the
electroweak scale. We can compensate for this by probing phase space
regions where $m_h$ is not the relevant scale in the numerator.  Only
for moderately strongly coupled dynamics with $g = 1\dots\sqrt{4\pi}$ 
one can
probe large enough energy scales for the EFT approach to be valid
given the precision of the LHC Higgs program,
\begin{align}
\left| \frac{\sigma \times \br}{\left( \sigma \times \br \right)_\text{SM}} -1 \right| 
= \frac{g^2 m_h^2}{\Lambda^2} \gtrsim 0.1
\qquad \Leftrightarrow \qquad 
\luv < \sqrt{10} \, g\,m_h
\simeq 400~\gev\dots1.4~\tev \, .
\label{eq:lmax_strong}
\end{align}
In fact, the EFT approach to Higgs observables has largely been
motivated by the desire to describe models with strongly interacting
electroweak symmetry breaking~\cite{silh}.

The increased statistics and Higgs production cross sections at 
Run II will enable us to add a wide range of distributions and
off-shell processes to the Higgs observables. They can probe higher
energy scales $\ephys \gg m_h$, which are more sensitive to
differences between the dimension-6 and full model predictions.  A well-known
example is weak boson fusion, where the details of the ultraviolet
completion can have a huge effect for example on the transverse
momenta of the tagging jets~\cite{bad_one,spins,phi_jj,higgs_pole}.

\section{Models vs effective theory}
\label{sec:analysis}

The aim of this paper is to compare a comprehensive set of LHC
predictions from specific new physics models with their corresponding
effective field theory predictions.  As discussed in
Section~\ref{sec:theory_self}, the applicability of the effective
Lagrangian given in Eq.\,\eqref{eq:EFT} is by no means guaranteed. We
test it based on detailed comparisons of matched EFTs with the
original, more or less UV-complete models, namely
\begin{itemize}
\item[A.] a scalar singlet extension with mixing effects and a 
  second scalar resonance;
\item[B.] two Higgs doublets, adding a variable Yukawa
  structure, a CP-odd, and a charged Higgs;
\item[C.] scalar top partners, contributing to Higgs couplings at
  one loop; and
\item[D.] a vector triplet with gauge boson mixing.
\end{itemize}
For each of these four models we introduce the setup and the main LHC
features, discuss the decoupling in the Higgs sector, define the dimension-6
setup, and finally give a detailed account of the full
and dimension-6 phenomenology at the LHC.\medskip

Our comparison covers the most relevant observables for LHC Higgs
physics.  We evaluate all amplitudes at tree level and take into
account interference terms between Higgs and gauge amplitudes.  Our
acceptance and background rejections cuts are minimal, to be able to
test the effective field theory approach over as much of the
phase space as possible.\medskip

In the case of Higgs production through gluon fusion, we analyze the
production process with a Higgs decay to four
leptons or to photons,
\begin{align}
  p p \to h \to 4 \ell \qqqquad \qquad
  p p \to h \to \gamma \gamma \,.
  \label{eq:gg_4l_process}
\end{align}
For the photons we do not apply any cuts, while for $\ell = e, \mu$ we require
\begin{align}
  m_{4 \ell} > 100 \ \gev \quad \text{and} \quad m_{\ell^+ \ell^-}^\text{same flavor} > 10 \ \gev \, 
\end{align}
to avoid too large contributions from the $Z$ peak and
bremsstrahlung.\medskip

For Higgs production in weak boson fusion (WBF), we evaluate the production process
\begin{align}
  u d 
\to h \, u d 
\to W^+ W^- \, ud 
\to (\ell^+ \nu) \, (\ell^- \bar{\nu}) \, ud \,.
\label{eq:wbf_proc}
\end{align}
We require the standard WBF cuts
\begin{align}
  p_{T,j} &> 20 \ \gev \,, \quad   &    \Delta \eta_{jj} &> 3.6 \,,   \quad &    m_{jj} &> 500 \ \gev \,, \notag \\
  p_{T,\ell} &> 10 \ \gev \,, \quad   &    \met &> 10 \ \gev \,.
\label{eq:wbf_cuts}
\end{align}
Unlike for gluon fusion, the kinematics of the final state can now
introduce new scales and a dependence on the UV structure of the
model. The process is particularly interesting in the
context of perturbative unitarity~\cite{general-unitarity}. While the latter
is satisfied in a UV-complete model by construction, 
deviations from the SM Higgs-gauge couplings in the EFT may lead to an increasing
rate at very large energies~\cite{Han:2009em,higgs_pole}, well outside
the EFT validity range $E / \Lambda \ll 1$.
To look for such signatures, we focus on the high-energy tail of the
transverse mass distribution,
\begin{align}
m_T^2 =  \left( E_{T,\ell \ell } + E_{T,\nu \nu } \right)^2 
       - \left( \mathbf{p}_{T,\ell \ell } + \mathbf{p}_T^{\text{miss}} \right)^2 
\qquad \text{with} \qquad 
  E_{T,\ell \ell } &= \sqrt{\mathbf{p}_{T,\ell \ell }^2 + m_{\ell \ell }^2} \,, \notag \\
  E_{T,\nu \nu } &= \sqrt{\met + m_{\ell \ell }^2} \,.
  \label{eq:mT}
\end{align}
\medskip

As the last single Higgs production process we evaluate Higgs-strahlung
\begin{align}
  q q \to V h
\end{align}
with $V = W^\pm, Z$. We do not simulate the Higgs and gauge boson
decays, assuming that we can always reconstruct for example the full
$Zh \to \ell^+ \ell^- \, b \bar{b}$ final state. No cuts are
applied.\medskip

Finally, Higgs pair production is well known to be problematic when it
comes to the effective theory description~\cite{hh-breakdown},
\begin{align}
  g g \to h h \,.
\end{align}
Again, neither Higgs decays nor kinematic cuts are expected to affect
our analysis, so we leave them out.

We test all these channels for the singlet and doublet Higgs sector
extensions. For the top partner and vector triplet models we focus on
the WBF and Higgs-strahlung modes, which are the most sensitive.  In
the dimension-6 simulations we always include the square of the dimension-6
operator contributions. While these terms are technically of the same
mass dimension as dimension-8 operators, which we neglect, we must keep them to
avoid negative values of the squared matrix element in extreme
phase-space regions. Notice that these situations do not
necessarily imply a breakdown of the EFT expansion. On the contrary,
they may appear in scenarios where new physics contributions
dominate over the SM part, while the EFT expansion is fully valid
(with $E/\Lambda \ll 1$). In such cases, the bulk 
effects stem from the squared dimension-6 terms instead of 
the interference with the SM, while the effects from
dimension-8 operators are smaller and can be safely neglected. 
\medskip

Tree-level processes we generate with
\toolfont{MadGraph5}~\cite{Alwall:2014hca}, using publicly available
model files~\cite{feynrules-site} and our own implementations though
\toolfont{FeynRules}~\cite{Alloul:2013bka}, which also provides the
corresponding UFO files~\cite{Degrande:2011ua}.  For the dimension-6
predictions we resort to an in-house version of the \toolfont{HEL}
model file~\cite{Alloul:2013naa}.  For all models we evaluate the
Higgs-gluon and Higgs-photon couplings with the full one-loop form
factors~\cite{ggh-analytical}, including top, bottom and $W$ loops as
well as new particles present in the respective models. For Higgs pair
production, we use a modified version of Ref.~\cite{higgspair-ucl}.

Other loop effects are analyzed using reweighting: we generate event
samples using appropriate general couplings. Next, we compute the
one-loop matrix element for each phase space point and reweight the
events with the ratio of the renormalized one-loop matrix element squared to the
tree-level model. For the one-loop matrix elements we utilize 
\toolfont{FeynArts} and \toolfont{FormCalc}~\cite{Hahn:2000kx} with our
own model files that include the necessary counterterms. The loop form factors are
handled with dimensional regularization in the 't~Hooft-Veltman
scheme, and written in terms of standard loop integrals. These are
further reduced via Passarino-Veltman decomposition and evaluated with
the help of \toolfont{LoopTools}~\cite{Hahn:1998yk}.

Generally we create event samples of at least 100\,000 events per
benchmark point and process for $pp$ collisions at $\sqrt{s} =
13$~\tev. We use the \toolfont{CTEQ6L} pdf~\cite{CTEQ6L} and the
default dynamical choices of the factorization and renormalization
scale implemented in \toolfont{MadGraph}. For the purpose of this
project we limit ourselves to parton level and do not apply a detector
simulation.  The mass of the SM-like Higgs is fixed to $m_h =
125$~\gev~\cite{higgsmass}. For the top mass we take $m_t =
173.2$~\gev~\cite{topmass}. The Higgs width in each model is based on calculations
with \toolfont{Hdecay}~\cite{hdecay}, which we conveniently rescale
and complement with additional decay channels if applicable.

\subsection{Singlet extension}
\label{sec:singlet}

The simplest extension of the minimal Higgs sector of the Standard
Model is by a real scalar singlet~\cite{singlet}. The extended scalar
potential has the form
\begin{alignat}{5}
V(\phi,S) = 
  \mu^2_1\,(\phi^\dagger\,\phi) 
+ \lambda_1\,|\phi^{\dagger}\phi|^2 
+ \mu^2_2\,S^2
+ \lambda_2\,S^4 
+ \lambda_3\,|\phi^{\dagger}\,\phi|S^2 \,,
\label{eq:singlet-potential}
\end{alignat}
where the new scalar $S$ can mix with the SM doublet $\phi$ 
provided the singlet develops a VEV, $\langle S \rangle = v_s/\sqrt{2}$. Details
on the parametrization, Higgs mass spectrum and coupling patterns are
given in Appendix~\ref{sec:ap-singlet}.

The additional scalar singlet affects Higgs physics in three ways:
\textit{i)} mixing with the Higgs via the mixing angle $\alpha$, which
leads to a universal rescaling of all Higgs couplings to fermions and
vectors; \textit{ii)} a modified Higgs self-coupling; and
\textit{iii)} a new, heavy resonance $H$ coupled to the Standard Model
through mixing.

The key parameter is the portal interaction between the doublet and
the singlet fields $\lambda_3(\pbp)\,S^2$, which is responsible for
the mixed mass eigenstates.  The mixing reduces the coupling of the
SM-like Higgs $h$ to all Standard Model particles universally,
\begin{align}
 \Delta_x = \cos\alpha -1 
\quad \text{for $x=W,Z,t,b,\tau,g,\gamma,\dots$\;.}
\label{eq:singlet-shift}
\end{align}
It also affects the self-coupling of the light Higgs, which takes on the form
\begin{equation}
  g_{hhh} = 6 \cos^3 \alpha\, \lambda_1 v - 3 \cos^2 \alpha \sin \alpha\, \lambda_3 v_s
  + 3 \cos \alpha \sin^2 \alpha\, \lambda_3 v - 6 \sin^3 \alpha\, \lambda_2 v_s \,.
\end{equation}
The parameter $\sin\alpha \simeq \alpha$ quantifies the departure from
the SM limit $\alpha \to 0$.  This limit can be attained in two ways:
first, a small mixing angle can be caused by a weak portal
interaction,
\begin{align}
\left| \tan(2\alpha) \right| = \left| \frac{\lambda_3\,v\,v_s}{\lambda_2 v_s^2 - \lambda_1 v^2} \right| \ll 1 \qquad \text{if} \qquad \lambda_3 \ll 1 \,.
\label{eq:limit1} 
\end{align}
The Higgs couplings to SM particles approach their SM values, but
there is no large mass scale associated with this limit. In the
extreme case of $\lambda_2,\lambda_3 \ll \lambda_1$ we find small
$\alpha \approx - \lambda_3/\lambda_1 \times v_s/(2v)$ even for $v_s
\lesssim v$.  This situation is to some extent the singlet model counterpart
of the \textit{alignment without decoupling} scenario in the
Two-Higgs-doublet model (2HDM)~\cite{Gunion:2002zf,Craig:2013hca} or the
MSSM~\cite{Carena:2013ooa,Delgado:2013zfa}. 
It relies nonetheless
on a weak portal coupling and a small scale separation, which cannot be properly 
described by an effective field theory. 

Second, the additional singlet can introduce a large mass scale
$v_s \gg v$, giving us
\begin{align}
\tan\alpha \approx \frac{\lambda_3}{2\lambda_2}\,\frac{v}{v_s} 
\ll 1 \qquad \text{if} \qquad v \ll v_s \,,
\label{eq:limit2}
\end{align}
where $\lambda_3/(2\lambda_2)$ is an effective coupling of up to order
one. In this limit the heavy Higgs mass, which we identify as the heavy mass scale,
is given by 
\begin{equation}
  m_H \approx \sqrt{2\lambda_2} \, v_s \equiv \Lambda \,.
\end{equation}

In terms of the heavy scale $\Lambda$ the Higgs couplings scale like
\begin{align}
\Delta_x 
= - \frac{\alpha^2}{2} + \mathcal{O}(\alpha^3)  
\approx - \frac{\lambda_3^2}{4 \lambda_2}\, \left( \frac{v}{\Lambda} \right)^2 \,.
\label{eq:singlet_decoup}
\end{align}
This is a dimension-6 effect.  If we require $|\Delta_x| \gtrsim 10\%$
to keep our discussion relevant for the LHC, this implies
\begin{align}
m_H \approx \Lambda <  \frac{\sqrt{5} \lambda_3}{\sqrt{2 \lambda_2}} \, v
    = 390~\text{GeV} \times \frac{\lambda_3}{\sqrt{\lambda_2}} \,. 
 \label{eq:singlet-delta3}
\end{align}
If we also assume that the ratio of quartic couplings is of the order
of a perturbative coupling, $\lambda_3/\sqrt{\lambda_2} \lesssim 0.5$,
the LHC reach in the Higgs coupling analysis translates into heavy
Higgs masses below 200~GeV. For strongly coupled scenarios,
$\lambda_3/\sqrt{\lambda_2} \lesssim 1 \dots \sqrt{4\pi}$,
the heavy mass reach increases to $m_H \lesssim 0.4 \dots 1.5$ TeV.
This suggests that a weakly coupled Higgs portal will
fail to produce a sizable separation of scales when looking at
realistic Higgs coupling analyses. The question becomes if and where
this lack of scale separation hampers our LHC analyses.\medskip

\begin{table}[t]
  \renewcommand{\arraystretch}{1.2}
  \centering
    \begin{tabular}{c c rccr c ccc c cc}
      \toprule
      \multirow{2}{*}{Benchmark} &\hspace*{1em}& \multicolumn{4}{c}{Singlet} &\hspace*{1em}& \multicolumn{3}{c}{EFT} &\hspace*{1em}& \multicolumn{2}{c}{EFT ($v$-improved)} \\
      \cmidrule{3-6} \cmidrule{8-10} \cmidrule{12-13}
                && $m_H$ & $\sin\alpha$ & $v_s/v$ & $\Delta_x^\text{singlet}$ 
                &&  $\Lambda$ & $\bar{c}_H$ & $\Delta_x^\text{EFT}$ 
                &&  $\bar{c}_H$ & $\Delta_x^\text{EFT}$\\
      \midrule
      S1 &&  500 & 0.2 & 10 & $-0.020$ && 491 & 0.036 & $-0.018$ && 0.040 &  $-0.020$ \\
      S2 &&  350 & 0.3 & 10 & $-0.046$ && 336 & 0.073 & $-0.037$ && 0.092 &  $-0.046$ \\
      S3 &&  200 & 0.4 & 10 & $-0.083$ && 190 & 0.061 & $-0.031$ && 0.167 &  $-0.083$ \\
      S4 && 1000 & 0.4 & 10 & $-0.083$ && 918 & 0.183 & $-0.092$ && 0.167 & $-0.092$ \\
      S5 && 500 & 0.6 & 10 & $-0.200$ && 407 & 0.461 & $-0.231$ && 0.400 &  $-0.200$ \\
      \bottomrule
    \end{tabular}
  \caption{Benchmarks for the singlet extension. We show the model
    parameters and the universal coupling modification for the
    complete model, as well as the matching scale $\Lambda$,
    the Wilson coefficient $\bar{c}_H$, and
    the universal coupling modification in the EFT truncated to
    dimension 6. We also give these results for an alternative,
    $v$-improved construction. $m_H$ and $\Lambda$ are in GeV.}
  \label{tab:singlet_benchmarks}
\end{table}

In the EFT approach the singlet model only generates $\ope{H}$ at
dimension 6, with the Wilson coefficient
\begin{align}
  \bar{c}_H = \dfrac{\lambda_3^2}{2\lambda_2} \, \left(\dfrac {v} {\Lambda}\right)^2 \,.
\end{align}
We give the details of the EFT description in
Appendix~\ref{sec:ap-singlet}.  As discussed in the previous section,
the construction of the EFT is not unique.
Instead of keeping only the leading term in the expansion in $1/\Lambda$, we
can match the dimension-6 operators to the full, untruncated singlet model.  
In the broken phase 
the Higgs couplings are fully expressed through the mixing angle $\alpha$, 
so the $v$-improved EFT truncated to dimension-6 operators gives the Wilson coefficient
\begin{align}
  \bar{c}_H = 2 ( 1-\cos \alpha)\,.
\end{align}
\medskip

We start our numerical analysis by defining five singlet benchmark points in
Tab.~\ref{tab:singlet_benchmarks}.  The first three
scenarios are in agreement
with current experimental and theoretical constraints.  This includes
direct mass bounds from heavy Higgs searches at colliders, Higgs
coupling measurements, electroweak precision observables, perturbative
unitarity and vacuum stability~\cite{singlet_bounds}. We note that for S4 and S5
the combination of large heavy Higgs masses together with large mixing angles
is incompatible with
perturbative unitarity and electroweak precision constraints.
We nevertheless keep such benchmarks for
illustration purposes. Table~\ref{tab:singlet_benchmarks} also includes
the universal shift of the light Higgs couplings, both for the
full singlet model and its dimension-6 EFT descriptions.\medskip

In Tab.~\ref{tab:singlet_rates} we give the ratio of the total Higgs
production cross sections in gluon fusion, WBF and
Higgs-strahlung. They confirm what we expect from the coupling
modification shown in Tab.~\ref{tab:singlet_benchmarks}:
qualitatively, the full singlet and the dimension-6 model predict similar
shifts in the total rates.  But there are differences in the coupling
modifications $\Delta_x^\text{singlet}$ and $\Delta_x^\text{EFT}$ of
up to $5\%$, translating into a rate deviation of up to $10 \%$. In
the $v$-improved EFT we find that the Higgs couplings and total rates
agree exactly with the full model predictions. The dimension-6
operators are entirely sufficient to capture the coupling shifts, but
a significant part of their coefficients are formally of
$\ord(v^4/\Lambda^4)$.\medskip

\begin{table}[t]
  \renewcommand{\arraystretch}{1.2}
  \centering
  \begin{tabular}{c c ccc c ccc}
    \toprule
    \multirow{2}{*}{Benchmark} &\hspace*{1em}& \multicolumn{3}{c}{$\sigma_\text{EFT} / \sigma_\text{singlet}$} 
    &\hspace*{1em}& \multicolumn{3}{c}{$\sigma_\text{$v$-improved EFT} / \sigma_\text{singlet}$}\\
    \cmidrule{3-5} \cmidrule{7-9}
              && ggF & WBF & $Vh$ && ggF & WBF & $Vh$\\
    \midrule
    S1 && 1.006 & 1.006 & 1.004 && 1.001 & 1.001 & 1.000 \\
    S2 && 1.019 & 1.021 & 1.019 && 1.000 & 1.001 & 1.000 \\
    S3 && 1.119 & 1.118 & 1.118 && 1.000 & 0.999 & 1.000 \\
    S4 && 0.982 & 0.982 & 0.982 && 0.999 & 0.999 & 1.000 \\
    S5 && 0.925 & 0.925 & 0.925 && 0.999 & 0.999 & 1.000 \\
    \bottomrule
  \end{tabular}
  \caption{Cross section ratios of the matched dimension-6 EFT
    approximation to the full singlet model at the LHC. We show the
    leading Higgs production channels for all singlet benchmark
    points. The statistical uncertainties on these ratios are below
    0.4\%.}
  \label{tab:singlet_rates}
\end{table}

\begin{figure}[t]
  \centering
  \includegraphics[width=0.49\textwidth]{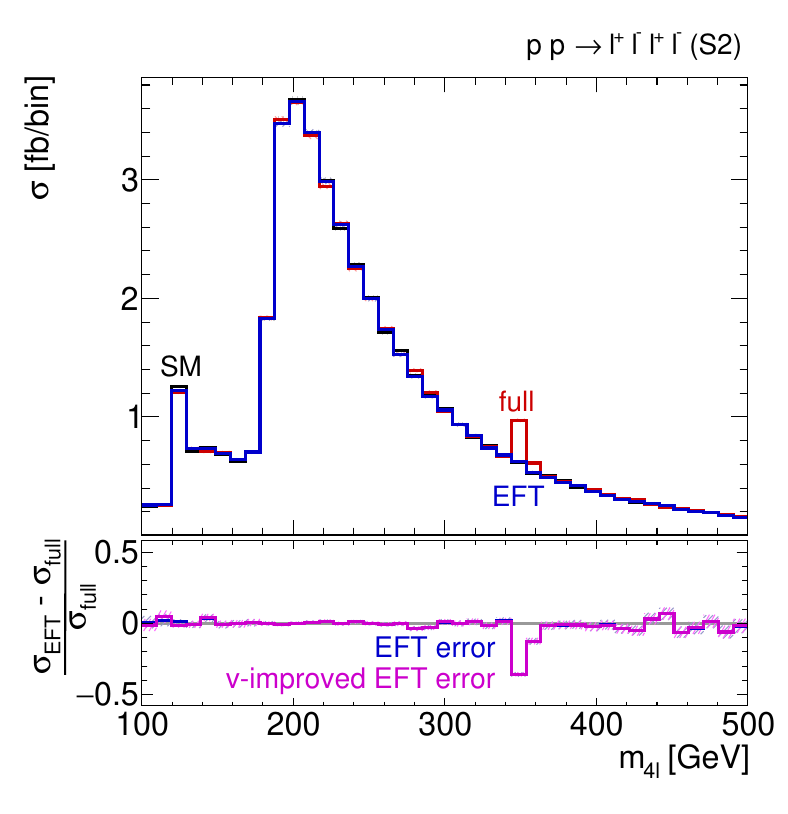}
  \includegraphics[width=0.49\textwidth]{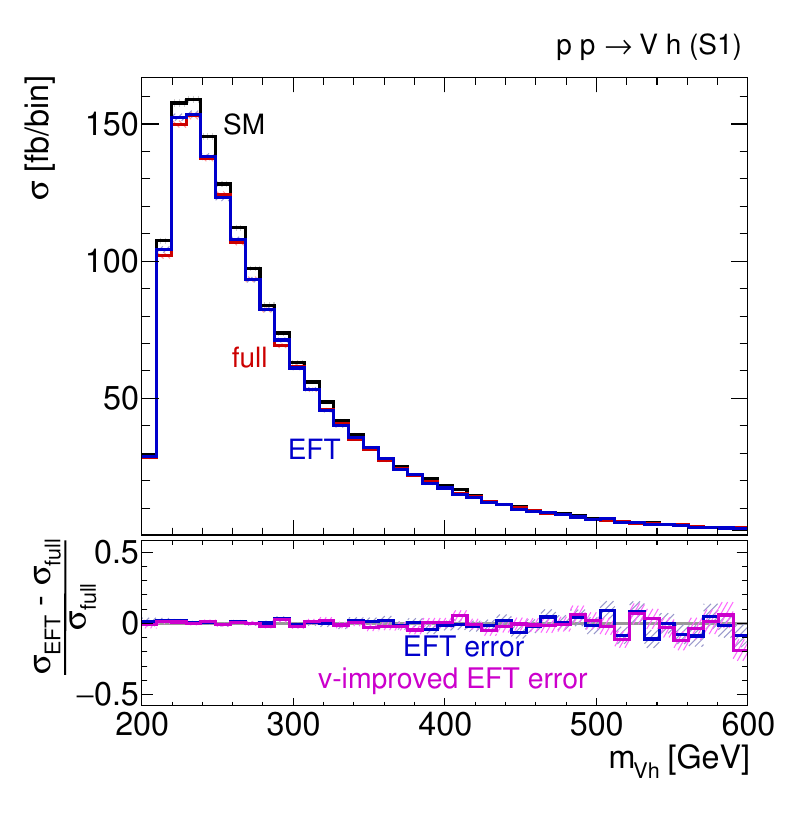} \\
  \includegraphics[width=0.49\textwidth]{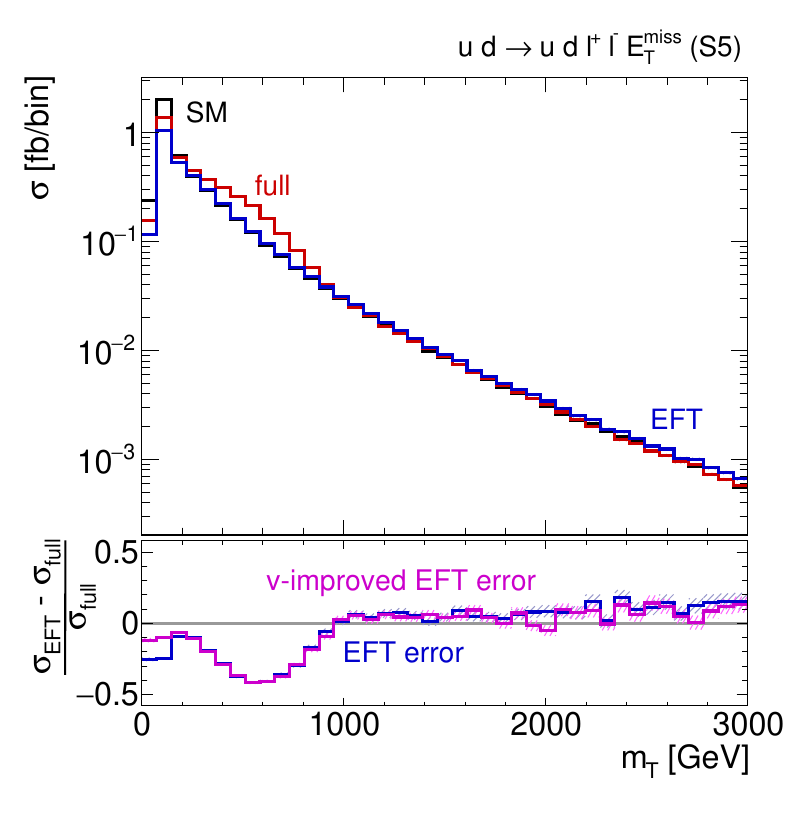}
  \includegraphics[width=0.49\textwidth]{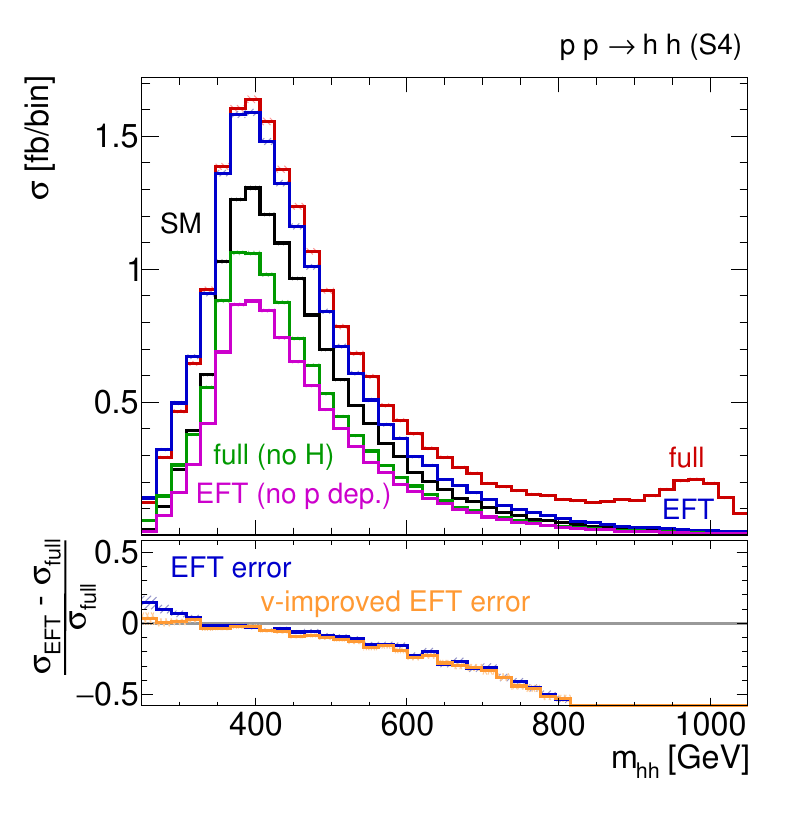}
  \caption{Kinematic distributions in the singlet model.  The
    different curves show the SM, full singlet model and
    singlet-matched dimension-6 predictions respectively, as indicated
    in each panel.  Top left: $m_{4\ell}$ distribution in the $gg \to
    h \to 4 \ell$ channel after loose acceptance cuts for S2 in the
    full and effective models. Top right: $m_{Vh}$ distribution in
    $Vh$ production for S1.  Bottom left: $m_T$ distribution in the
    WBF $h \to \ell^+ \ell^- \; \met$ channel for S5. Bottom right:
    $m_{hh}$ distribution in Higgs pair production for S4. For
    $m_{hh}$ we show several contributions in the full theory and the
    dimension-6 approach. In all plots, the error bars give the
    statistical uncertainties.}
  \label{fig:singlet}
\end{figure}

The most obvious source of discrepancy between the full model and the
EFT is the heavy resonance $H$. It can for example be
produced in gluon fusion and then observed as a peak in the 
$m_{4\ell}$ distribution. By construction, it will not be captured by
the dimension-6 model. We illustrate this in the upper left panel of
Fig.~\ref{fig:singlet}. For Higgs-strahlung production (Fig.~\ref{fig:singlet}, right panel),
where the novel $H$ resonance does not appear in an intermediate Born-level propagator
and hence has no impact,
we find instead excellent agreement between both descriptions over the entire
phase space.

The second Higgs has a second, more subtle effect.  In the full model,
both Higgs exchange diagrams are needed to unitarize $WW$
scattering. Correspondingly, the EFT description breaks perturbative
unitarity roughly at the scale~\cite{Han:2009em}
\begin{align}
  m_{WW}^2 
= \frac{16 \pi v^2}
       {\bar{c}_H \left( 1 - \dfrac{\bar{c}_H}{4 (1+\bar{c}_H)} \right)}
\approx \left( \frac {1.7 \ \tev} {\sin \alpha} \right)^2 \,. 
  \label{eq:UnitarityViolation}
\end{align}
In our benchmark point S5, this is around 2.8~TeV. The incomplete
cancellations between Higgs and gauge amplitudes means that the dimension-6
model tends to have a larger rate at energies already below this
scale. For this specific benchmark choice, this can be seen in the lower left panel of
Fig.~\ref{fig:singlet}, where we show the distribution of the
transverse mass defined in Eq.\,\eqref{eq:mT} in the process $ u d \to
W^+ W^- \, ud \to (\ell^+ \nu) \, (\ell^- \bar{\nu}) \, ud$, to which
WBF production of both $h$ and $H$ contributes.  We observe that the
dimension-6 model predicts a slightly higher rate at large $m_T$ than both the
full singlet model and the SM. Given the very mild signal, which results
from the fast decrease in the parton densities and the small mixing angle 
for realistic scenarios, such effect is likely of no relevance
for LHC physics.\medskip

A more interesting channel to study in the singlet model is Higgs pair
production. The Higgs self-coupling is the only Higgs coupling which
gains a momentum dependence in the matched EFT. In addition, there
exists an approximate cancellation between the two leading amplitudes
in the SM at threshold~\cite{hh_threshold}.  This induces a second
relevant scale and with it a sensitivity to small deviations in the
Higgs couplings.  In Fig.~\ref{fig:singlet} we give the $m_{hh}$
distribution in the full and dimension-6 models.
In addition, we show how the
distributions would look in the full model without a $H$ state, and in
the EFT without the momentum-dependent (derivative) terms given in
Eq.\,\eqref{eq:singlet-self}.  Already at threshold and far away from
the $H$ resonance, the interference of the SM-like terms with the $H$
diagrams makes up a significant part of the amplitude.  In the EFT,
the derivative terms are similarly relevant already at low
energies. Close to threshold, the dimension-6 approximates the full model
well. This agreement becomes worse
towards the $H$ pole~\cite{hh-breakdown}. The question of how the
Wilson coefficients are expanded in $v^2/\Lambda^2$ does not play a
role here.\medskip

If we limit ourselves to Higgs properties relevant for single Higgs
production at the LHC, the modifications from a singlet extension are
very simple: all Standard Model couplings acquire a common scaling
factor, and no relevant new Lorentz structures appear at tree-level.
The dimension-6 setup reproduces this effect correctly:
the reduced couplings to all SM fields alone
do not require a large hierarchy of scales.  An EFT construction in
which the dimension-6 coefficients are not truncated at
$\ord(v^2/\Lambda^2)$ gives perfect agreement with the full theory,
while expanding the coefficients to leading order in $v^2/\Lambda^2$
may lead to sizeable deviations from the full model.  Higgs pair
production is different. There is a large contribution from off-shell
$H$, while in the EFT the $h$ self-coupling involves a
derivative. These different structures lead to discrepancies between
full and effective description that increase with momentum
transfer. Finally, the effective theory by definition does not include
the second resonance, so it fails whenever a heavy Higgs appears
on-shell in the full theory.

\subsection{Two-Higgs-doublet model}
\label{sec:2hdm}

The two-Higgs-doublet model (2HDM)~\cite{2hdm_review} adds a second
weak doublet with weak hypercharge $Y = +1$ to the SM Higgs sector.
The combined potential reads
\begin{alignat}{5}
 V(\phi_1,\phi_2) 
=& \, m^2_{11}\,\phi_1^\dagger\phi_1
 + m^2_{22}\,\phi_2^\dagger\phi_2
 + \frac{\lambda_1}{2} \, (\phi_1^\dagger\phi_1)^2
 + \frac{\lambda_2}{2} \, (\phi_2^\dagger\phi_2)^2
 + \lambda_3 \, (\phi_1^\dagger\phi_1)\,(\phi_2^\dagger\phi_2) 
 + \lambda_4 \, |\phi_1^\dagger\,\phi_2|^2 \notag \\
& + \left[ - m^2_{12}\,\phi_1^\dagger\phi_2 
        + \frac{\lambda_5}{2} \, (\phi_1^\dagger\phi_2)^2 
        + \text{h.c.} 
   \right] \,.
\label{eq:2hdmpotential}
\end{alignat}
The physical degrees of freedom are two neutral CP-even scalars
$h^0,H^0$, one neutral CP-odd scalar $A^0$, and a set of charged
scalars $H^\pm$. The relevant model parameters are the mixing angle
between the CP-even scalars $\alpha$, the ratio of the VEVs $\tan
\beta = v_2/v_1$, and the mixed mass term $m_{12}$. The latter
induces a soft breaking of the discrete $\mathbb{Z}_2$ symmetry
$\phi_i \to (-1)^{i}\,\phi_i$ (for $i=1,2$).  The two-doublet structure allows for a
rich variety of Higgs couplings to fermions.  We refer the reader to
Appendix~\ref{sec:ap-2hdm} for a detailed account of the model setup,
Higgs spectrum, coupling patterns, and matched effective description.\medskip

Just as the singlet extension, the 2HDM predicts two types of LHC
signatures: \textit{i)} scalar and VEV mixing lead to modified light
Higgs couplings. Unlike for the singlet extension, these coupling
modifications are not universal and reflect the more flexible flavor
structure as well as the multiple scales of the model. \textit{ii)}
There exist three heavy resonances $H^0, A^0, H^\pm$, which should
have near-degenerate masses to avoid custodial symmetry breaking.
\medskip

The light Higgs coupling to weak bosons $V=W,Z$ always scales like
\begin{align}
\Delta_V = \sin (\beta - \alpha) - 1 
         = - \frac{ \cos^2(\beta - \alpha)}{2} 
           + \ord (\cos^4(\beta - \alpha)) \,.
\label{eq:2hdm_shift}
\end{align}
We can insert the leading contribution of a mass-degenerate heavy
Higgs sector and find 
\begin{align}
\Delta_V 
\approx \frac{\sin^2 (2\beta)}{8} \, \left( \dfrac{v}{m_{A^0}} \right)^4 \,.
\label{eq:2hdm_decoup}
\end{align}
While in the singlet model the light Higgs coupling to gauge
bosons is shifted at $\ord(v^2/\Lambda^2)$,
Eq.\,\eqref{eq:singlet_decoup}, the same coupling is now affected at
$\ord(v^4/m_{A^0}^4)$, corresponding to a dimension-8 effect.

Two aspects turn the decoupling in the general 2HDM into a challenge:
first, delayed decoupling effects appear after electroweak symmetry
breaking~\cite{Haber:2000kq}. 
For example, in type-II models we find~\cite{Lopez-Val:2013yba}
\begin{align}
\Delta_b 
= - \tan \beta \, \sqrt{|2 \Delta_V|} + \Delta_V +  \mathcal{O}(\Delta_V^{3/2}) 
\approx - \tan \beta \; \frac{\sin (2\beta)}{2} \, \left( \dfrac{v}{m_{A^0}} \right)^2 \,.
\label{eq:2hdm_delayed}
\end{align}
This correction to the bottom Yukawa coupling corresponds to a
dimension-6 effect, and already moderate values of $\tan \beta$
significantly delay the decoupling of the heavy 2HDM states in the
Yukawa sector.

Second, unlike in the MSSM the Higgs self-couplings $\lambda_1 \dots
\lambda_5$ and $m_{12}$ are not bounded from above. In combinations
like $\lambda_j v^2$ they contribute to the interactions of the
SM-like Higgs state, effectively inducing a new energy scale through
terms of the kind $\sqrt{|2\Delta_V|} \sqrt{\lambda_j} v$ or
proportional to $m_{12}$. They are significantly less suppressed than
we would expect for the usual suppression
$\sqrt{|2\Delta_V|}$\,---\,in particular if an additional factor $\tan
\beta$ appears in this coupling deviation.

This additional, effectively lower mass scale driven by $v$ leads to
problems with any EFT derived from and matched to the full theory
assuming only one new physics scale. While this should not be viewed
as a problem of the EFT approach in general, it will require a $v$-improved
matching procedure.
\medskip

We first match the effective theory to the 2HDM
in the unbroken phase. For this we define the
new physics scale in terms of the mass terms in the potential of
Eq.\,\eqref{eq:2hdmpotential} and ratio of VEVs~\cite{heft_limitations2} as
\begin{align}
\Lambda^2 
= \mheavy^2
\equiv m^2_{11}\sin^2\beta
  + m^2_{22}\cos^2\beta + m^2_{12} \sin (2\beta) \,.
\end{align}

The 2HDM generates a number of
dimension-6 operators at tree level, for which the Wilson
coefficients depend on the flavor structure. While the up-type Yukawa
coupling is always modified the same way, the down-type and lepton
couplings are different for type-I and type-II. We find
\begin{align}
  \bar{c}_u = \bar{c}_d^\text{I\phantom{I}} = \bar{c}_\ell^\text{I\phantom{I}} &=
\phantom{-} \dfrac{\sin (2\beta) \cot\beta}{2}
\left[\frac{\lambda_1}{2} -\frac{\lambda_2}{2}
     +\left(\frac{\lambda_1}{2} +\frac{\lambda_2}{2} -\lambda_3-\lambda_4-\lambda_5\right)\cos (2\beta)
\right]
\left(\dfrac {v}{\Lambda} \right)^2 \,, \notag \\
\bar{c}_d^\text{II} = \bar{c}_\ell^\text{II} &=
- \dfrac{\sin (2\beta) \tan \beta}{2} 
\left[\frac {\lambda_1} 2-\frac {\lambda_2} 2+\left(\frac {\lambda_1} 2
      +\frac {\lambda_2} 2-\lambda_3-\lambda_4-\lambda_5\right)\cos (2\beta)
\right] 
\left(\dfrac{v}{\Lambda} \right)^2 \,, \label{eq:2hdmmatching1}
\end{align}
where the superscripts I and II denote the type of the flavor structure.

Upon electroweak symmetry breaking, the physical heavy Higgs masses $m_{H^0}$,
$m_{A^0}$, and $m_{H^{\pm}}$ acquire VEV-induced contributions $\sim \lambda_i v^2$
in addition to contributions from the heavy scale $M$.
As in the singlet model, we therefore also consider a $v$-improved matching
where the matching scale is $\Lambda = m_{A^0}$ and the Wilson
coefficients are expressed in terms of mass eigenstates. 
In this setup, Eq.\,\eqref{eq:2hdmmatching1} remains unchanged, except that
$\Lambda$ is identified with $m_{A^0}$. 

The two matching
schemes exhibit significant differences in the 2HDM; for instance, the
pseudoscalar mass is given by $m^2_{A^0} = m_{12}^2/(\sin\beta\cos\beta)
-\lambda_5\,v^2$. This means that it does not coincide with $\mheavy$,
unless we enforce a single mass scale $m_{11} \approx m_{22} \approx
m_{12}$ and $\tan \beta \approx 1$.\medskip

\begin{table}[t]
  \renewcommand{\arraystretch}{1.2}
  \centering
  \begin{tabular}{c c rrrrrrr }
    \toprule
    \multirow{2}{*}{Benchmark} &\hspace*{1em}& \multicolumn{7}{c}{2HDM} \\
    \cmidrule{3-9} 
              && Type & $\tan\beta$ & $\alpha/\pi$ & $m_{12} $ & $m_{H^0} $ &  $m_{A^0} $  & $m_{H^\pm}$ \\
    \midrule
    D1 && I  & 1.5 & $-0.086$  & 45  & 230 & 300 & 350  \\
    D2 && II & 15  & $-0.023$ & 116 & 449 & 450 & 457 \\
    D3 && II & 10  & 0.032  & 157 & 500 & 500 & 500 \\
    D4 && I  & 20  & 0       & 45  & 200 & 500 & 500 \\
    \bottomrule
  \end{tabular}
 \caption{Benchmarks for the 2HDM extension. We show the model
   parameters and the heavy Higgs masses. All
   masses are in GeV.}
 \label{tab:2hdm_benchmarks}
\end{table}

\begin{table}[b!]
  \renewcommand{\arraystretch}{1.2}
  \centering
  \begin{tabular}{c c rrr c rrrr}
    \toprule
    \multirow{2}{*}{Benchmark} &\hspace*{1em}& \multicolumn{3}{c}{EFT}  &\hspace*{1em}& \multicolumn{4}{c}{EFT ($v$-improved)} \\
    \cmidrule{3-5} \cmidrule{7-10}
                               && $|\Lambda|$~[GeV] & $\bar{c}_u$ & $\bar{c}_{d,\ell}$
                               && $\Lambda$~[GeV] & $\bar{c}_u$ & $\bar{c}_{d,\ell}$ & $\bar{c}_\gamma$ \\
    \midrule
    D1 && 100 & $-0.744$ & $-0.744$ && 300 & 0.082 & 0.082 & $1.61 \cdot 10^{-4}$ \\
    D2 && 448 & 0.000 & 0.065 && 450 & 0.000 & 0.065 & $4.16 \cdot 10^{-6}$ \\
    D3 && 99 & 0.465 & $-46.5$ && 500 & 0.018 & $-1.835$ & $1.05 \cdot 10^{-4}$ \\
    D4 && 142 & 0.003 & 0.003 && 500 & 0.000 & 0.000 & $1.48 \cdot 10^{-4}$ \\
    \bottomrule
  \end{tabular}
 \caption{Matching scales and Wilson coefficients for the effective theory
   matched to the 2HDM. We give these results both for the EFT matching
   in the unbroken phase as well as for the $v$-improved matching with $\Lambda = m_{A^0}$.}
 \label{tab:2hdm_EFT}
\end{table}

The 2HDM benchmark points in Tab.~\ref{tab:2hdm_benchmarks} are in
agreement with all current constraints, implemented with the help of
\toolfont{2HDMC}~\cite{2hdmc},
\toolfont{HiggsBounds}~\cite{higgsbounds},
\toolfont{SuperIso}~\cite{superiso}, and
\toolfont{HiggsSignals}~\cite{higgssignals}. To better illustrate
certain model features, in some scenarios we tolerate deviations
between $1\,\sigma$ and $2\,\sigma$ in the Higgs couplings measurements.
The key physics properties of the different 2HDM scenarios can be
summarized as:
\begin{itemize}
\item[D1] moderate decoupling: with Higgs couplings shifts of up to
  $2\sigma$ in terms of the LHC constraints.  This generates
  $\Delta_{\tau,b,t} \approx \mathcal{O}(15\%)$ as well as a large
  $h^0 H^+ H^-$ coupling. Additional Higgs masses around
  $250\dots350$~GeV can leave visible imprints.
\item[D2] supersymmetric: reproducing the characteristic mass
  splittings and Higgs self-couplings of the MSSM with light
  stops~\cite{Carena:2013qia}.
\item[D3] sign-flipped bottom Yukawa: this is possible in type-II
  models at large $\tan\beta$, as shown in
  Eq.\,\eqref{eq:2hdm_delayed}~\cite{Ferreira:2014naa}. This can be
  viewed as a manifestation of a delayed
  decoupling~\cite{Haber:2000kq}.
\item[D4] fermiophobic heavy Higgs: possible only in type-I models
  for $\sin\alpha =0$. The heavy Higgs $H^0$ is relatively light, but
  essentially impossible to observe at the LHC~\cite{Hespel:2014sla, fermiophobic}.
\end{itemize}
In Tab.~\ref{tab:2hdm_EFT} we show the heavy scales $\Lambda$ and the
Wilson coefficients for both the EFT matched in the unbroken phase and
the $v$-improved EFT construction. In contrast to the singlet model, a
significant $v$-dependence of the heavy masses occurs even for
parameter points in agreement with all relevant experimental and
theoretical constraints. Only in one of our four benchmark scenarios
does the heavy scale $\mheavy$ approximate the physical mass
$m_{A^0}$. The matching in the unbroken phase is particular pathological
in benchmark D1, where $\mheavy^2$ is negative and the signs of the
Wilson coefficients are switched compared to the $v$-improved matching.

\begin{table}[t]
  \renewcommand{\arraystretch}{1.2}
  \setlength{\tabcolsep}{0.3em}
  \centering 
  \footnotesize
  \begin{tabular}{c c rr c rrr c rrr}
    \toprule
    \multirow{2}{*}{Benchmark} &\hspace*{0.5em}& \multicolumn{2}{c}{$\Delta_V$} &\hspace*{0.5em}& \multicolumn{3}{c}{$\Delta_t$}
    &\hspace*{1em}& \multicolumn{3}{c}{$\Delta_b=\Delta_\tau$}  \\ 
    \cmidrule{3-4} \cmidrule{6-8} \cmidrule{10-12}
    && 2HDM & EFT (both) && 2HDM & EFT & EFT  ($v$-improved) && 2HDM & EFT & EFT  ($v$-improved) \\ 
    \midrule
    D1 &&$-0.05$ & 0.00 &&  0.16  & $-0.74$ & 0.08 && 0.16  & $-0.74$ & 0.08  \\ 
    D2 && 0.00 & 0.00 &&  0.00  & 0.00 & 0.00 && 0.07  & 0.07 & 0.07  \\ 
    D3 &&$-0.02$ & 0.00 &&  0.00  & 0.46 & 0.02 && $-2.02$  & $-46.5$ & $-1.84$  \\ 
    D4 && 0.00 & 0.00 &&  0.00  & 0.00 & 0.00 && 0.00  & 0.00 & 0.00  \\ 
    \bottomrule
  \end{tabular}
  \caption{Normalized tree-level couplings of the light Higgs in our 2HDM benchmarks. }
  \label{tab:2HDM_couplings_tree}
\setlength{\tabcolsep}{0.5em}
\end{table} 
 
Table~\ref{tab:2HDM_couplings_tree} confirms that matching in the
unbroken phase does not reproduce the modified Higgs couplings, while
the $v$-improved matching essentially captures the coupling shifts
without a strong requirement on the hierarchy of scales. For our
purpose we conclude that the expansion in powers of $v/\mheavy$ is not
well controlled, and we have to rely on $v$-improved matching for the
2HDM.\medskip

\begin{table}[b!]
  \renewcommand{\arraystretch}{1.2}
  \centering
    \begin{tabular}{c c rrr}
      \toprule
      \multirow{2}{*}{Benchmark} &\hspace*{1em}& \multicolumn{3}{c}{$\sigma_\text{$v$-improved EFT} / \sigma_\text{2HDM}$} \\
      \cmidrule{3-5}
                && ggF & WBF & $Vh$ \\
      \midrule
      D1 && 0.872 & 1.109 & 1.108 \\
      D2 && 1.001 & 1.000 & 1.000 \\
      D3 && 1.022 & 1.042 & 1.042 \\
      D4 && 1.001 & 1.001 & 1.003\\
      \bottomrule
    \end{tabular}
  \caption{Cross section ratios of the matched dimension-6 EFT
    approximation to the full 2HDM at the LHC. We show the leading
    Higgs production channels for all 2HDM benchmark points.  The
    statistical uncertainties on these ratios are below 0.4\%.}
  \label{tab:2HDM_rates}
\end{table}

However, even in the $v$-improved EFT, the dimension-6 truncation can
present an important source of deviations. According to
Tab.~\ref{tab:2HDM_couplings_tree} the operators $\ope{u}$, $\ope{u}$,
and $\ope{\ell}$ modify the Higgs couplings similarly to the mixing,
at least in the limit of small mixing angles. This is clearly visible
\eg in the MSSM-like scenario D2 as well as the fermiophobic scenario
of benchmark D4, which are very well described by the dimension-6
Lagrangian, in spite of the lacking scale separation.

In Tab.~\ref{tab:2HDM_rates} we show LHC rate predictions by the
dimension-6 approach and the full 2HDM.  Depending on the benchmark,
the dimension-6 truncation leads to up to $10 \%$ departures.  A
particularly interesting scenario is described by benchmark D3. In the
full model, the bottom Yukawa is exactly sign-flipped, a signature
hardly visible at the LHC.
Generating such a signature from higher-dimensional operators
requires their contributions to be twice as large as the SM Yukawa coupling
due to the enhancement of $v/\Lambda$ by a large coupling.
The EFT with default matching is certainly not valid anymore,
and even the $v$-improved prescription fails to capture this coupling shift
fully, leading to a significantly different coupling pattern.
\medskip

\begin{figure}[tp]
  \centering
  \includegraphics[width=0.49\textwidth]{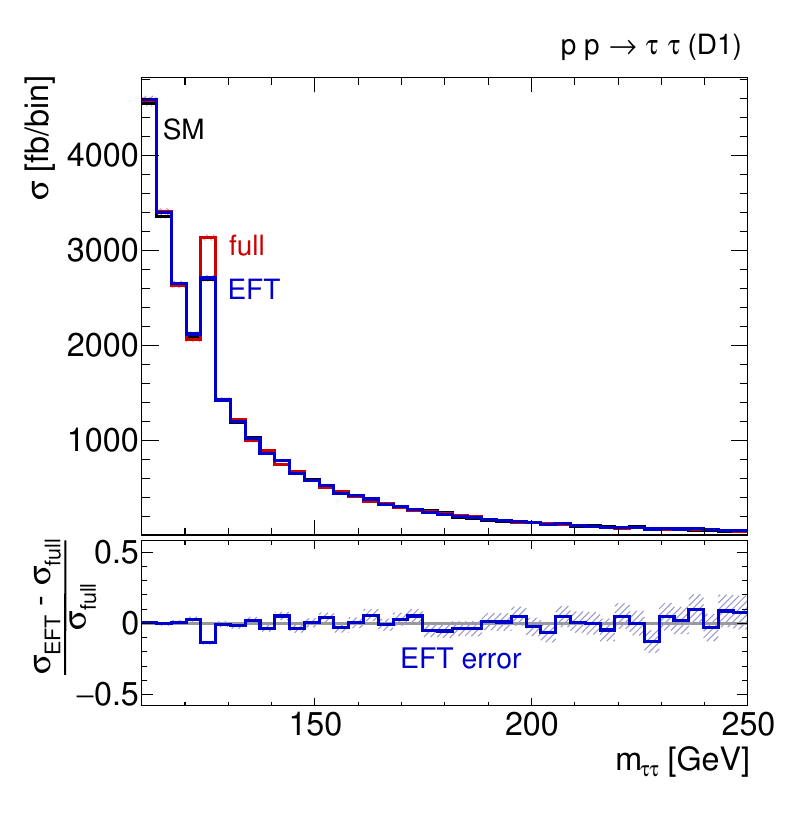}
  \includegraphics[width=0.49\textwidth]{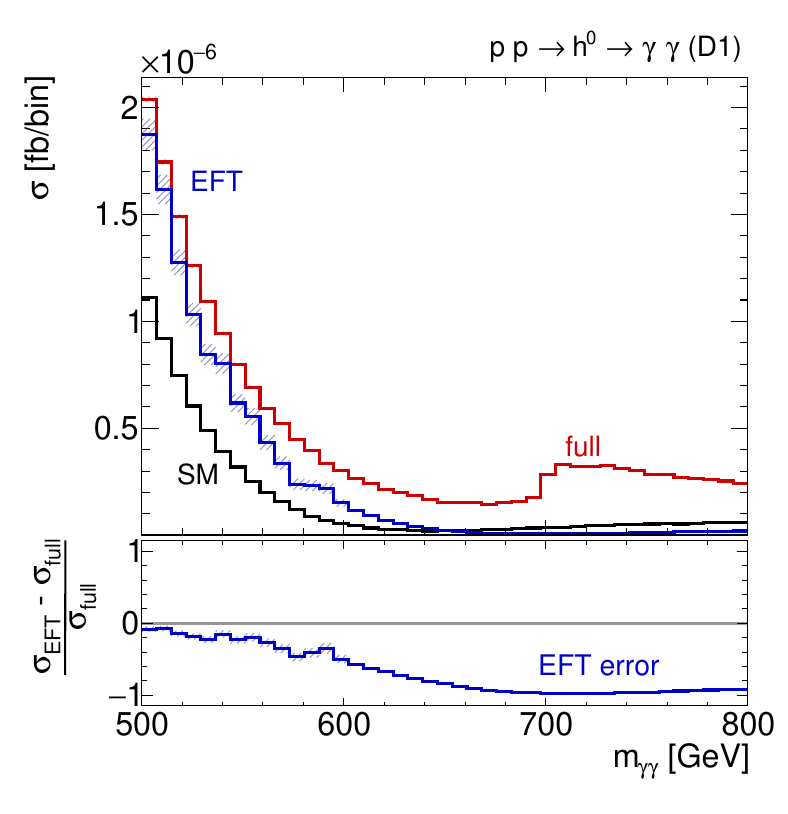}
  \vspace{-.5em}
  \caption{Left: $m_{\tau \tau}$ distribution in the ggF $h^0 \to
    \tau^+ \tau^-$ channel.  Right: off-shell behavior of the process
    $p p (gg) \to h^0 \to \gamma \gamma$ in 2HDM benchmark D1, only
    taking into account the Higgs diagrams. At $m_{\gamma \gamma}
    \gtrsim 2 m_{H^\pm} = 700\ \gev$, the charged Higgs threshold is
    visible.}
  \label{fig:2HDM_resolve_HAA}
\end{figure}

In the left panel of Fig.~\ref{fig:2HDM_resolve_HAA} we illustrate the
coupling deviations in gluon fusion Higgs production with a decay
$h\to \tau^+ \tau^-$. The full 2HDM and the EFT give substantially
different predictions for the size of the Higgs signal, but do not
affect the remaining distribution.\medskip

In addition, the charged Higgs contributes to the Higgs-photon
coupling, an effect which is mapped onto the operator
$\ope{\gamma}$. Within the $v$-improved EFT, one finds
\begin{align}
  c_\gamma &= - \frac {g^2 \, \left(\tan \beta + \cot \beta \right) } {12\,288 \, \pi^2} \,
             \Bigg[\left(\lambda_1 + \lambda_2 - 2 \lambda_3 + 6 \lambda_4 + 6 \lambda_5 - 8 \frac {m_{h^0}^2} {v^2} \right) \sin (2 \beta)  \notag \\
            &\quad   + 2 ( \lambda_1 - \lambda_2) \sin (4 \beta) 
             + (\lambda_1 + \lambda_2- 2 \lambda_3 - 2 \lambda_4 - 2 \lambda_5 ) \sin (6 \beta) \Bigg] \,
             \left(\frac {v}{m_{A^0}}\right)^2 \,.
              \label{eq:2hdmmatching2}
\end{align}
There appear no non-decoupling term of $\ord(\Lambda^0)$, because the
charged Higgs loop decouples in the limit $m_{A^0} \to \infty$ with
finite $\lambda_i$. If instead we keep $m_{12}$ fixed and let one of
the couplings $\lambda_i$ grow with $m_{A^0}$, the charged Higgs does
not decouple.
Interestingly, Eqs.\,\eqref{eq:2hdmmatching1}\,--\,\eqref{eq:2hdmmatching2}
show that in this model it is possible to realize 
\emph{alignment without decoupling} 
scenarios~\cite{Gunion:2002zf,Craig:2013hca,Carena:2013ooa,Delgado:2013zfa}, where the
limit of SM-like couplings is achieved via very small prefactors of $(v/m_{A^0})^2$,
while the additional
Higgs states can remain moderately light\,---\,and hence potentially within LHC reach. 

For all our benchmarks we find good agreement between the full
2HDM and the $v$-improved dimension-6 approach for on-shell Higgs decays to photons.
In Tab.~\ref{tab:2HDM_couplings_loop} the rescaling
of the Higgs-photon couplings shows slight discrepancies which can
nearly entirely be traced back to the different couplings of the Higgs
to the top and bottom in the loop due to the inaccurate truncation and are not related to the
$H^\pm$ contribution.

This changes for off-shell Higgs production. At $m_{\gamma \gamma}
\gtrsim 2 m_{H^\pm}$, the $H^\pm$ in the loop can resolve the charged
Higgs, enhancing the size of its contribution significantly. This
effect is not captured by the effective operator and leads to a
different behavior of the amplitude $g g \to h^0 \to \gamma \gamma$
between the full and effective model, as shown in the right panel of
Fig.~\ref{fig:2HDM_resolve_HAA}. However, the tiny rate and the
large combinatorial background mean that this discrepancy will be
irrelevant for LHC phenomenology. Similar threshold effects have been
computed for the top-induced Higgs-gluon coupling and appear to be
similarly irrelevant in practice~\cite{Buschmann:2014twa}.\medskip

\begin{table}[t]
  \renewcommand{\arraystretch}{1.2}
  \centering 
  \begin{tabular}{c c cc c rrrr}
      \toprule
      \multirow{2}{*}{Benchmark} &\hspace*{1em}& \multicolumn{2}{c}{$\Delta_g$}  &\hspace*{1em}& \multicolumn{4}{c}{$\Delta_\gamma$} \\ 
      \cmidrule{3-4} \cmidrule{6-9}
      && 2HDM & EFT  ($v$-improved) && \multicolumn{2}{c}{2HDM}  &  \multicolumn{2}{c}{EFT ($v$-improved)} \\ 
      \midrule
      D1 && $0.16 + 0.00 \,i$ & $0.08 + 0.00 \,i$ &&$-0.16$ & ($-0.05$) & $-0.10$ & ($-0.07$)  \\ 
      D2 && $0.00 + 0.00 \,i$ & $0.00 + 0.00 \,i$ && 0.00 & ( 0.00) & 0.00 & ( 0.00)  \\ 
      D3 && $0.07 - 0.09 \,i$ & $0.02 + 0.00 \,i$ && $-0.08$ & ($-0.05$) & $-0.05$ & ($-0.05$)  \\ 
      D4 && $0.00 + 0.00 \,i$ & $0.00 + 0.00 \,i$ && $-0.05$ & ($-0.05$) & $-0.05$ & ($-0.05$)  \\ 
      \bottomrule
  \end{tabular}
  \caption{Normalized couplings of the light Higgs to gluons
    and photons in our 2HDM benchmarks.  The bottom loop leads
    to small imaginary parts of $\Delta_g$ and $\Delta_\gamma$.
    For the Higgs-photon coupling, these imaginary parts are always smaller than $1\%$ of
    the real part of the amplitude and neglected here.  The numbers in
    parentheses ignore the modification of the Higgs-fermion couplings,
    allowing us to separately analyze how well the $H^\pm$ loop
    is captured by $\ope{\gamma}$.}
  \label{tab:2HDM_couplings_loop}
\end{table} 

The situation in Higgs pair production resembles the observations
in the singlet model.  The agreement can be worse already at threshold
if the inaccurate truncation leads to differences in the Higgs--top
couplings between the full and effective model.\medskip

\begin{figure}[tp]
  \centering
  \includegraphics[width=0.32\textwidth,clip=true,trim=1.27cm -0.4cm 0 0]{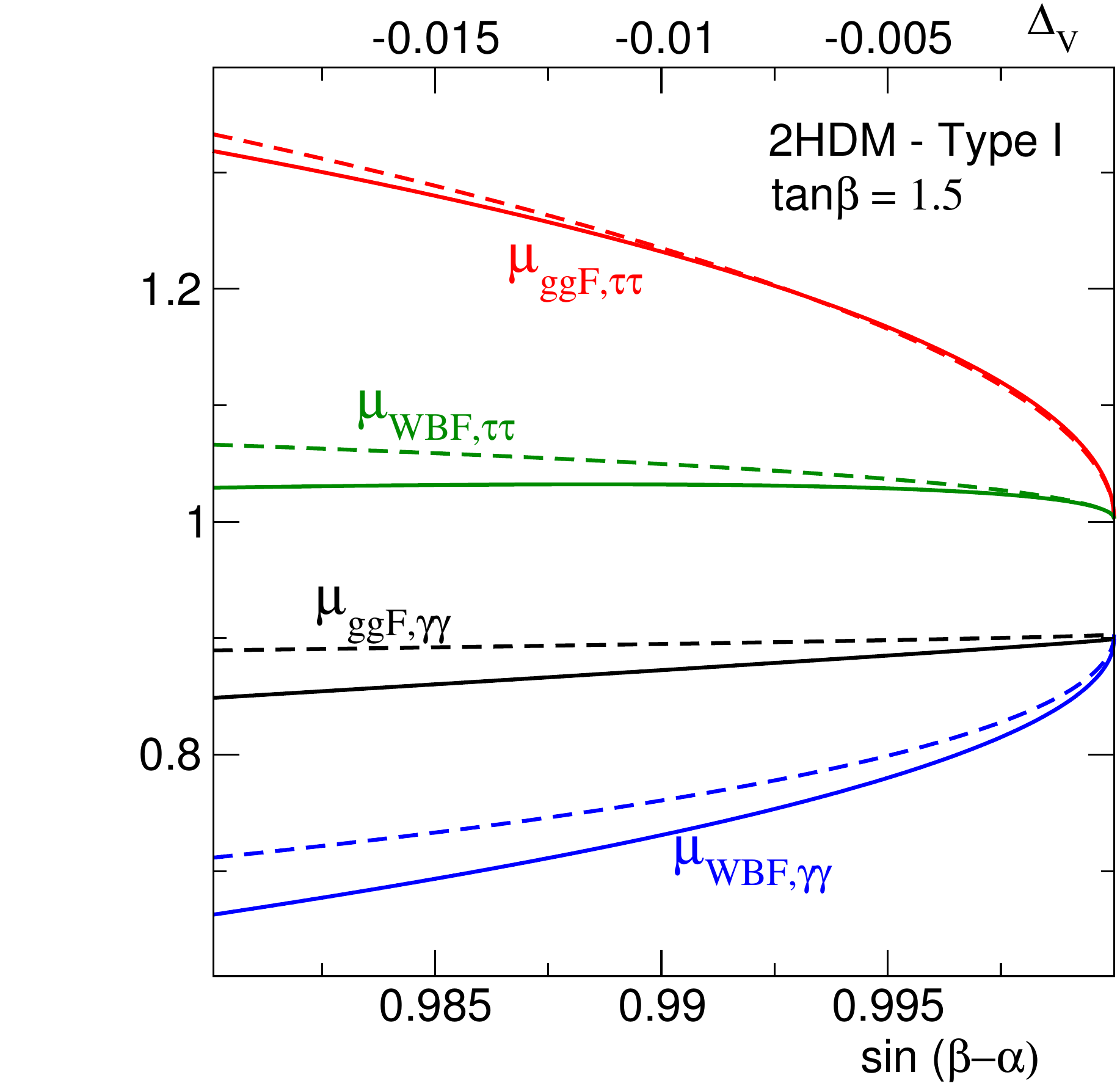}
  \includegraphics[width=0.32\textwidth,clip=true,trim=-0.8cm -0.4cm 0 0]{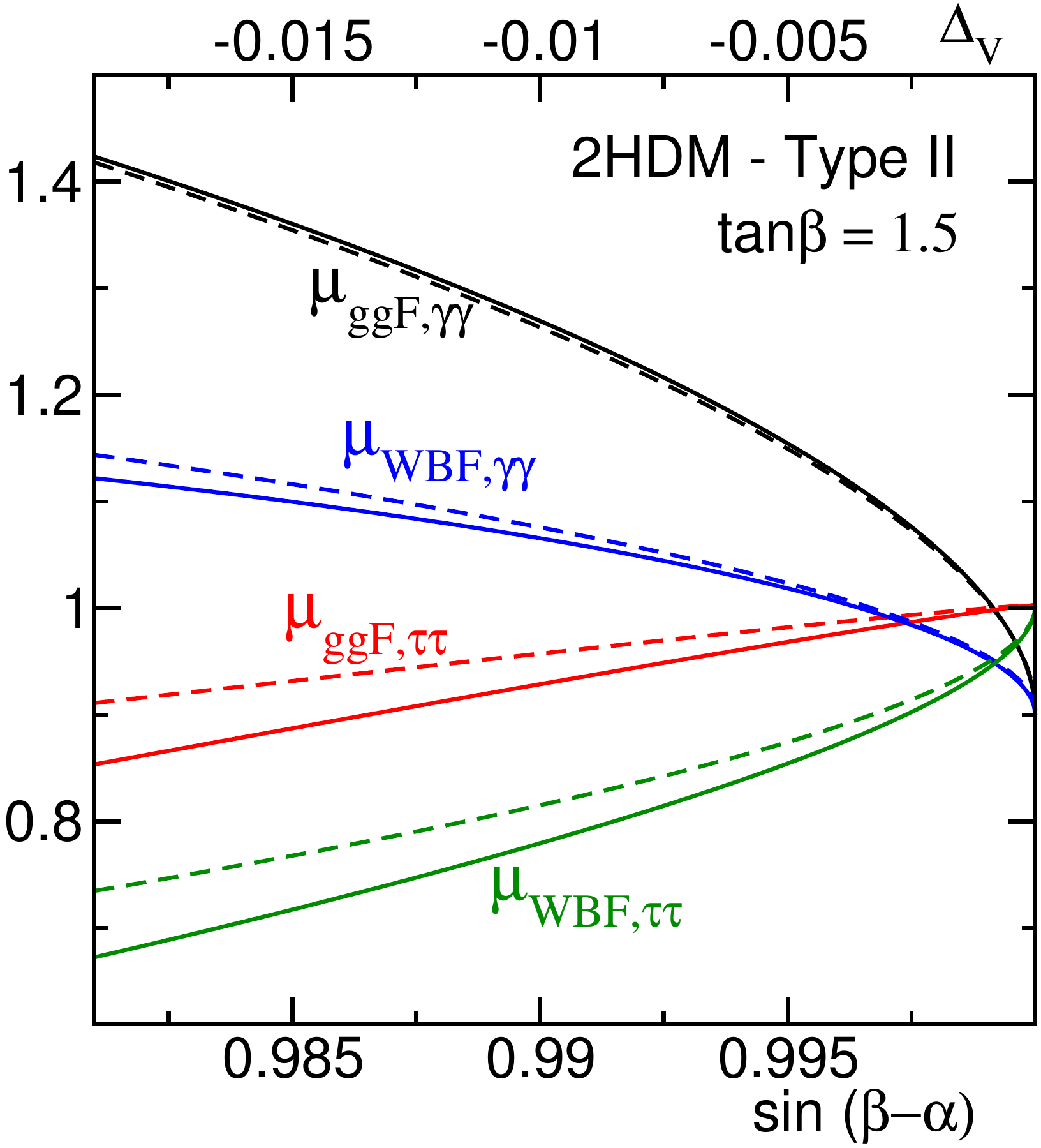}
  \includegraphics[width=0.32\textwidth,clip=true,trim=-0.8cm -0.4cm 0  0]{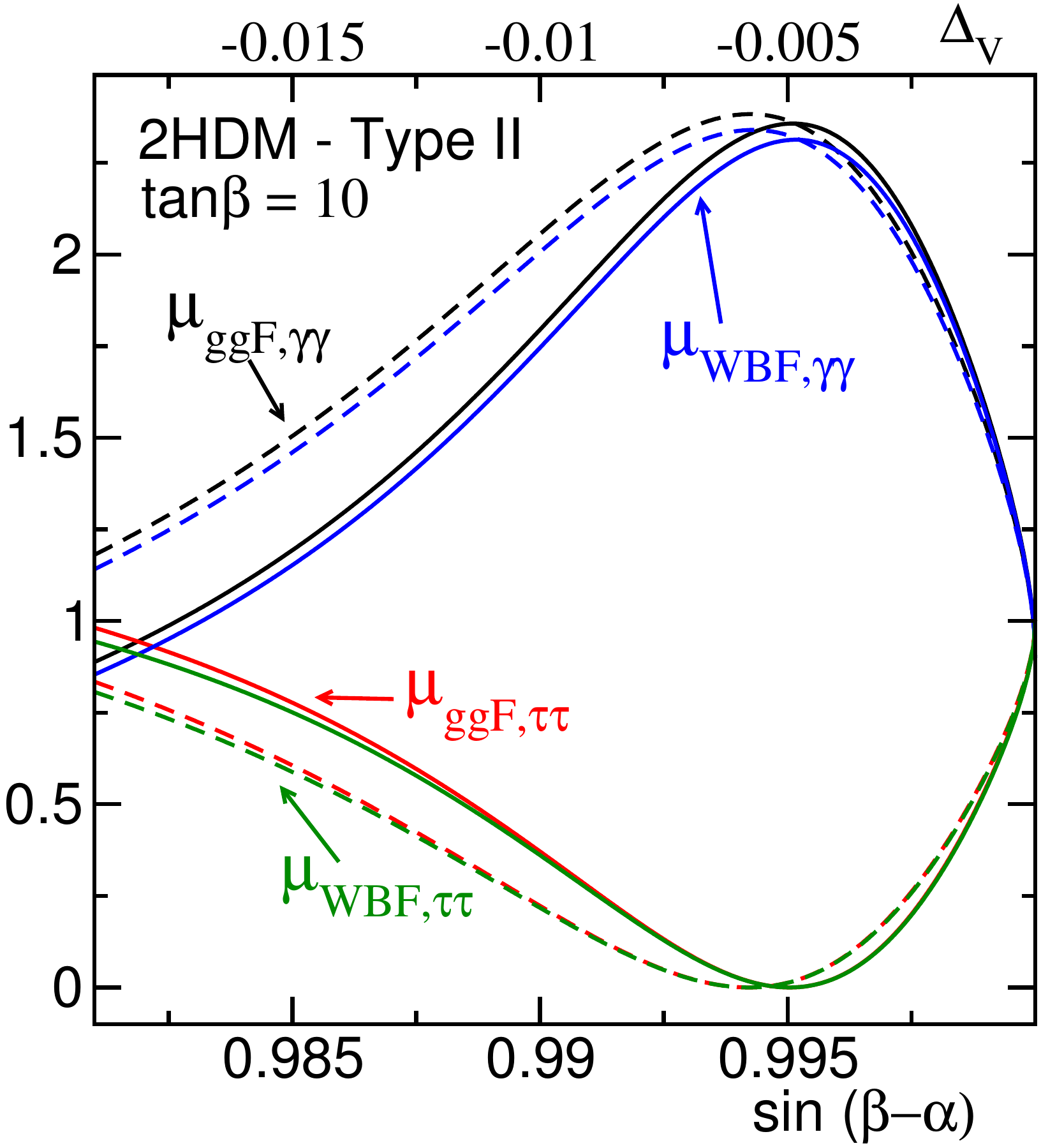}\\
  \includegraphics[width=0.32\textwidth,clip=true,trim=-0.2cm 0 0 0]{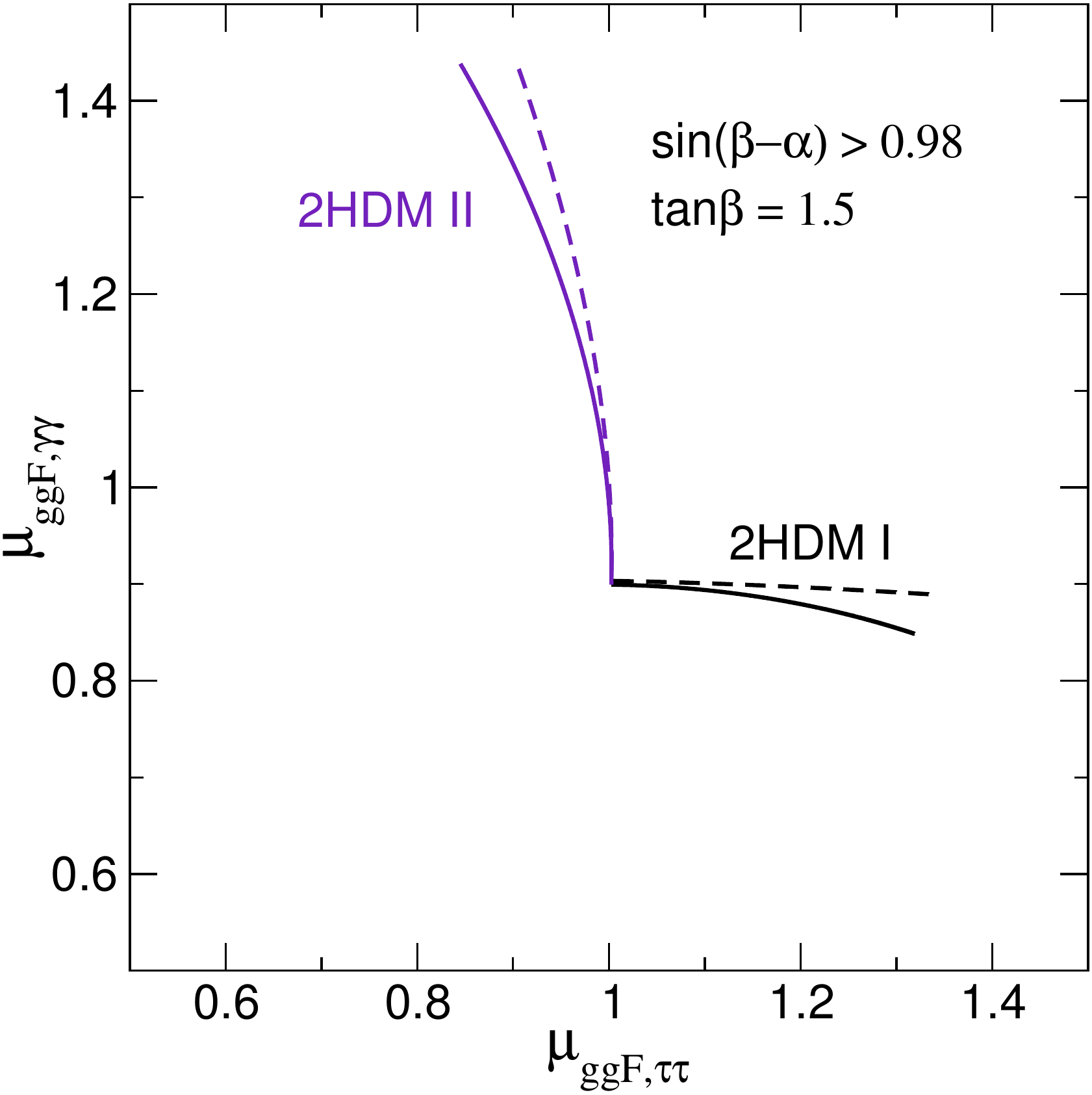}
  \includegraphics[width=0.32\textwidth,clip=true,trim=-0.55cm 0 0 0]{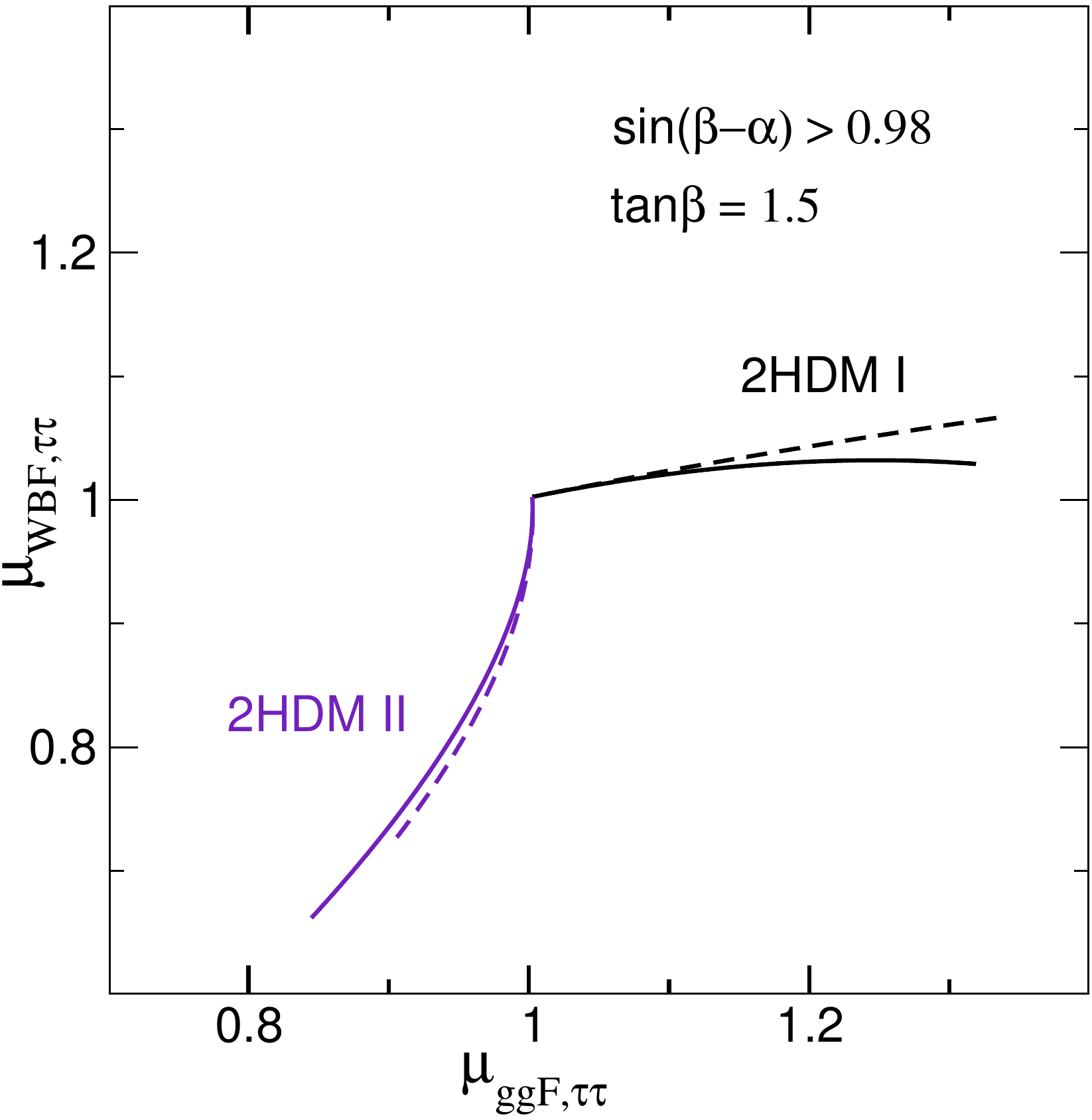}
  \includegraphics[width=0.32\textwidth,clip=true,trim=-0.35cm 0 0 0]{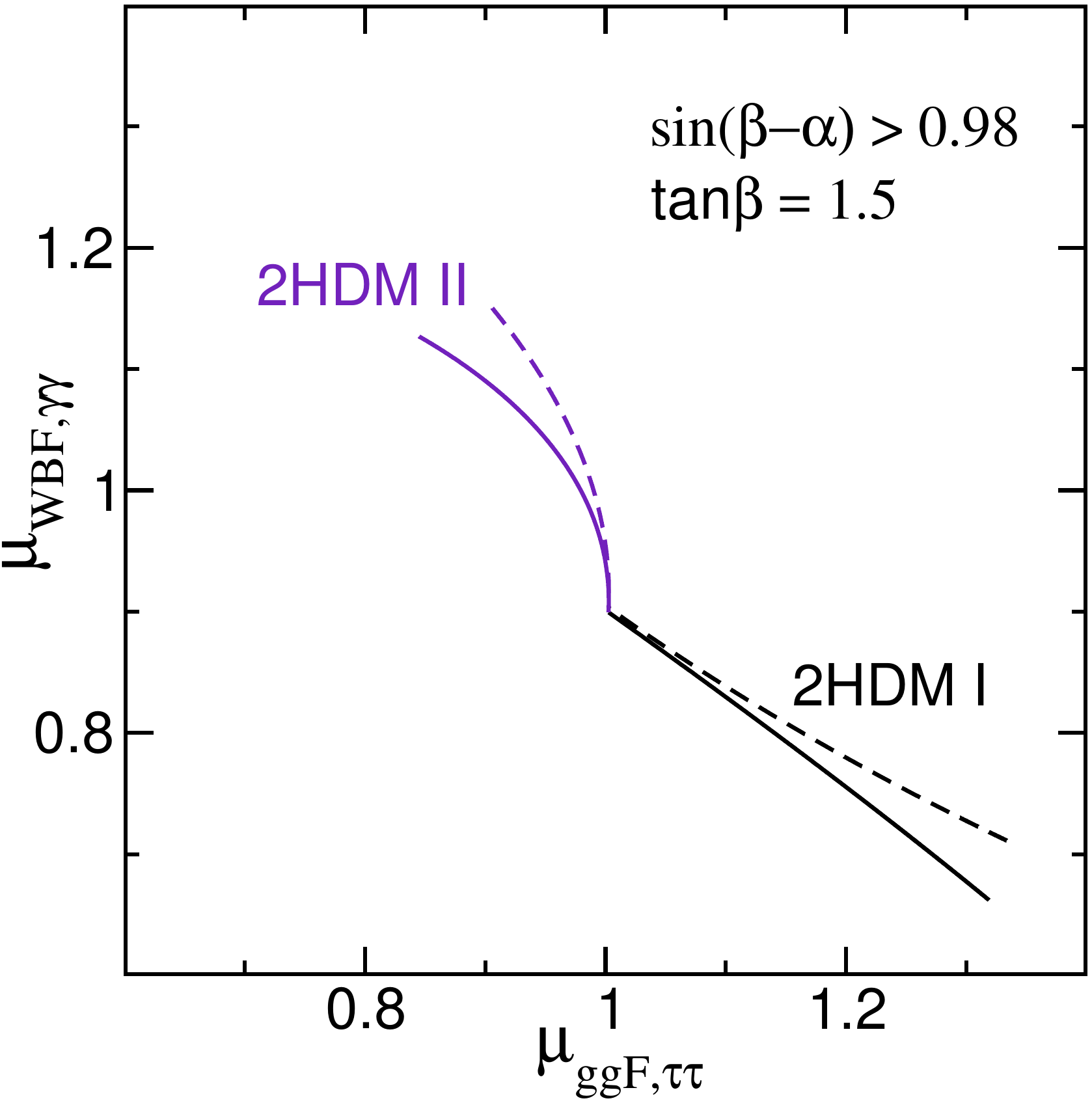}
  \caption{Signal strength modifications in the 2HDM.
    The solid lines show the full model, while the dashed lines give the EFT predictions.
    Top: signal strength $\mu_{p,d}$ for different Higgs production modes and decay channels
    in exemplary 2HDM setups, as a function of $\sin (\beta - \alpha)$.
    In the upper horizontal axis we track down the distance with respect
    to the SM-like limit through the coupling shift $\Delta_V$\,\eqref{eq:2hdm_decoup}.
    Bottom: signal strength correlations
    $\mu_{p_1,d_1}$ versus $\mu_{p_2,d_2}$ between different channels
    for variable  $\sin(\beta-\alpha)$. }
  \label{fig:2HDM_correlations}
\end{figure}

Leaving the discussion of individual benchmarks behind,
in Fig.~\ref{fig:2HDM_correlations} we demonstrate how deviations in
the signal rates $\mu_{p,d}$ can be correlated,
cf. Ref.~\cite{Cranmer:2013hia}.  The upper panels illustrate the
dependence on the decoupling parameter $\sin (\beta - \alpha)$. In all
cases we choose $\tan\beta = 1.5$, $m_{12} = 0$,
degenerate heavy Higgs masses $m_{H^{\pm},H^0,A^0}= 500$~GeV,
and restrict ourselves to $\sin(\beta-\alpha) \ge 0.98$. All
signal strength deviations are obtained by rescaling the SM production
cross section, branching ratio and total
width~\cite{Heinemeyer:2013tqa}.

In the limit $\sin (\beta - \alpha)\to 1$ or $\Delta_V \to 0$ we find
perfect agreement between the full model and the $v$-improved
dimension-6 model.  The
latter also captures the non-decoupling part of the Higgs-photon
coupling in the SM limit, $\mu_{\gamma \gamma} \ne 1$.  
Away from the
SM-like limit the dimension-6 model slightly overestimates the
signal strengths. This can for instance be attributed to $\Delta_V$;
it remains zero in the EFT while it decreases via
$\ord(v^4/\Lambda^4)$ corrections in the full model. Through the
$W$ loop this is also the main reason for the deviation in the
$\gamma\gamma$ final states. Truncated negative $\ord(v^4/\Lambda^4)$
corrections to $\Delta_\tau$ are also in part responsible for the
slight upward shift of $\mu_{\text{ggF},\tau\tau}$ in the dimension-6 model.  The
behavior of the down-type Yukawas in type-II models, which are
governed by $\Delta_{b,\tau} = -\cos(\beta - \alpha) \tan\beta +
\ord(v^4/\Lambda^4)$, leads to the strongly increased $\gamma\gamma$
rates at large $\tan \beta$, a feature which is well reproduced by the EFT.\medskip

Eventually, the 2HDM discussion leads us to the same conclusion as the
singlet model: as long as the mixing is small, the new resonances do
not contribute significantly, all the LHC probes in single Higgs
production is a set of three coupling modifications $\Delta_x$. New
Lorentz structures do not play any role for the models considered. Barring the special case of
Higgs pair production~\cite{Hespel:2014sla,higgspair2hdm} the EFT
captures most relevant aspects of Higgs phenomenology.  A naive construction
of the EFT by matching the effective dimension-6 Lagrangian to the 
2HDM in the gauge symmetric phase fails to correctly describe
the modified Higgs boson dynamics in typical 2HDM scenarios, since
formally suppressed terms in $v^2/\Lambda^2$ as well as delayed
decoupling or additional scales can become important for the phenomenologically
relevant scenarios to be tested at the LHC.

\subsection{Scalar top partners}
\label{sec:partners}

\begin{table}[t]
\renewcommand{\arraystretch}{1.2}
\centering
\begin{tabular}{c c rrrr c rrr}
  \toprule 
  \multirow{2}{*}{Benchmark}
  &\hspace*{1em}& \multicolumn{8}{c}{Scalar top-partner model}  \\
  \cmidrule{3-10} 
  && $M$ & $\kLL$ & $\kRR$ & $\kLR$
                  &\hspace*{1em}& $m_{\stone}$ & $m_{\sttwo}$ & $\theta_{\tilde{t}}$ \\
  \midrule
  P1 && 500 & $-1.16$ & 2.85 & 0.147 && 500 & 580 & $-0.15$ \\
  P2 && 350 & $-3.16$ & $-2.82$ & 0.017 && 173 & 200 & $-0.10$ \\   
  P3 && 500 & $-7.51$ & $-7.17$ & 0.012 && 173 & 200 & $-0.10$ \\   
  \bottomrule
 \end{tabular}
 \caption{Scalar top-partner Lagrangian parameters (left) and physical
   parameters (right) for representative model benchmarks. All masses are in GeV.}
  \label{tab:partner-benchmarks}
\end{table}

\begin{table}[t]
  \renewcommand{\arraystretch}{1.2}
  \setlength{\tabcolsep}{0.45em}
  \centering
  \begin{tabular}{c c rrrr c rrrr}
    \toprule 
    \multirow{2}{*}{Benchmark}
    &\hspace*{1em}& \multicolumn{4}{c}{EFT}
    &\hspace*{1em}& \multicolumn{4}{c}{EFT ($v$-improved)} \\
    \cmidrule{3-6} \cmidrule{8-11}
    && $\Lambda$ & $\bar{c}_H$ & $\bar{c}_W$ &  $\bar{c}_{HW}$
    && $\Lambda$ & $\bar{c}_H$ & $\bar{c}_W$ &  $\bar{c}_{HW}$ \\
    \midrule
    P1 && 500 & 0.0062 & $-3.11 \cdot 10^{-7}$  & $3.99 \cdot 10^{-7}$ && 500 & 0.0062 & $-3.11 \cdot 10^{-7}$  & $3.99 \cdot 10^{-7}$ \\
    P2 && 350 & 0.0043  & $-2.55 \cdot 10^{-4}$  & $2.55 \cdot 10^{-4}$ && 173 & 0.0176 & $-1.04 \cdot 10^{-3}$  & $1.04 \cdot 10^{-3}$\\   
    P3 && 500 & 0.0166 & $-2.97 \cdot 10^{-4}$  & $2.97 \cdot 10^{-4}$ && 173 & 0.1388 & $-2.48 \cdot 10^{-3}$  & $2.48 \cdot 10^{-3}$ \\   
    \bottomrule
  \end{tabular}
  \setlength{\tabcolsep}{0.5em}
  \caption{Matching scales (in GeV) and selected Wilson coefficient for the top partner benchmarks,
    both for default and $v$-improved matching.}
  \label{tab:partner-EFT}
\end{table}

New colored scalar particles are, strictly speaking, not an extension of the
SM Higgs sector, but they can lead to interesting modifications of the
LHC observables. We consider a scalar top-partner sector mimicking the
stop and sbottom sector of the MSSM. Its Lagrangian has the form
\begin{alignat}{5}
 \lag \supset& \,  (D_{\mu}\,\Qtilde)^\dagger\,(D^\mu\Qtilde) + (D_\mu\,\TR)^*\,(D^\mu\,\TR)
 - \tilde{Q}^\dagger\,M^2\,\tilde{Q}\, - M^2\,\TR^*\,\TR \notag \\
&-\kLL\,(\phi \cdot \Qtilde)^\dagger\,(\phi \cdot\Qtilde) -\kRR\,(\TR^*\TR)\,(\phi^\dagger\,\phi) 
 - \left[ \kLR \, M \, \TR^*\,(\phi \cdot \Qtilde) + \text{h.c.} \right]
 \label{eq:partner_lag}.
 \end{alignat}
Here, $\Qtilde$ and $\TR$ are the additional isospin doublet and
singlet in the fundamental representation of $SU(3)_C$. Their mass terms
can be different, but for the sake of simplicity we unify them to a
single heavy mass scale $M$. The singlet state $\sbR$ is assumed to be
heavier and integrated out. This leaves us with three physical degrees
of freedom, the scalars $\tilde{t}_1$, $\tilde{t}_2$ and
$\tilde{b}_2= \tilde{b}_L$. The eigenvalues of the stop mass matrix
\begin{alignat}{5}
 \begin{pmatrix}
 \kLL\dfrac{v^2}{2} + M^2 & \kLR\,\dfrac{vM}{\sqrt{2}} \\
 \kLR\,\dfrac{vM}{\sqrt{2}} & \kRR\,\dfrac{v^2}{2}\, + M^2 
 \end{pmatrix}
  \label{eq:partner_mass}
\end{alignat}
define two masses $m_\stone < m_\sttwo$ and a mixing angle
$\theta_{\tilde{t}}$. Again, we provide a detailed description of the
model setup in Appendix~\ref{sec:ap-partners}.\medskip

The main new physics effects in the Higgs sector are loop-induced
modifications of the Higgs interactions, most significantly to
$\Delta_g$, $\Delta_\gamma$, $\Delta_V$, possibly including new
Lorentz structures.  The Yukawa couplings do not change at one loop,
because we do not include gauge boson partners. As a side remark, 
the 2HDM described in Sec.~\ref{sec:2hdm} combined with the scalar top partners
given here corresponds to the effective description of the Minimal Supersymmetric Standard Model
in the limit of infinitely heavy gauginos, sleptons, and light-flavor squarks.
\medskip

Adding the top parters, the correction to the $hVV$ coupling
in the limit of small $\theta_{\tilde{t}}$ scales like
\begin{align}
\Delta_V \approx \frac{\kLL^2}{16 \pi^2} \, \left( \frac{v}{m_\stone} \right)^2 \,.
\label{eq:partner_decoup}
\end{align}
This shift can be sizeable for relatively low stop and sbottom masses, but also
for large couplings $\kappa_{ij}$ to the Higgs sector.

As already noted for the 2HDM, the decoupling of the heavy scalars becomes
non-trivial in the presence of a Higgs VEV. Following
Eq.\,\eqref{eq:partner_mass} the masses of the heavy scalar are not only
controlled by $M$ in the gauge symmetric phase, but they receive additional contributions of the
type $\kLR \, vM$, $\kLL v^2$, or $\kRR v^2$ after electroweak symmetry breaking. This leads to a mass
splitting of order $v$ between masses of order $M$. Large values of
$\kLR$ increase this splitting. This means that in the full model the
decoupling is best described in terms of $m_\stone < M$.\medskip

This motivates us to again define two different
matching schemes. First,  we stick to our default prescription and
carry out the matching of the linear EFT
Lagrangian to the full model in the unbroken
phase. The matching scale $\Lambda$ it then dictated
by the intrinsic heavy field mass scale $M$, and 
completely unrelated to $v$. The suppression scale of loop effects
in the complete model and this matching scale in the EFT only agree in
the limit $M - m_\stone \ll M$.

In this dimension-6 approach the stop loops generate a number of
operators,
\begin{align}
\overline{c}_{g} &=  
 \cfrac{\mw^2}{24\,(4\pi)^2\,M^2}\,\left[ \kLL + \kRR  - \kLR^2\right] 
&\overline{c}_{\gamma} &=  
 \cfrac{\mw^2}{9\,(4\pi)^2\,M^2}\,\left[ \kLL + \kRR  - \kLR^2\right] \notag \\
\overline{c}_{B} &=  
 -\cfrac{5\mw^2}{12\,(4\pi)^2\,M^2}\,\left[\kLL - \cfrac{31}{50} \kLR^2 \right] 
&\overline{c}_{W} &=  
 \cfrac{\mw^2}{4\,(4\pi)^2\,M^2}\,\left[\kLL - \cfrac{3}{10} \kLR^2 \right] \notag \\
\overline{c}_{HB} &=  
 \cfrac{5\mw^2}{12\,(4\pi)^2\,M^2}\,\left[\kLL - \cfrac{14}{25} \kLR^2 \right] 
&\overline{c}_{HW} &=  -
 \cfrac{\mw^2}{4\,(4\pi)^2\,M^2}\,\left[\kLL - \cfrac{2}{5} \kLR^2 \right] \notag \\
 \overline{c}_H &= 
 \cfrac{v^2}{4(4\pi)^2\,M^2}\,\Bigg[ 2\kRR^2-\kLL^2 - 
 \left( \kRR - \frac{1}{2}\kLL \right) \kLR^2 + \cfrac{\kLR^4}{10} \Bigg] \hspace*{-2cm}&& \notag \\
 \overline{c}_T &= \cfrac{v^2}{4(4\pi)^2\,M^2}\,\left[\kLL^2 - \cfrac{\kLL\,\kLR^2}{2} + \cfrac{\kLR^4}{10}\right] \,. 
 \label{eq:c-ew}
\end{align}

In addition, we define a $v$-improved matching at the scale $\luv = m_\stone$ in the
broken phase. The Wilson coefficients we obtain are the same as in
Eq.\,\eqref{eq:c-ew}, except that $M$ is replaced by $m_\stone$.
\medskip

Unlike in the previous two models, the top partner loops do not only
induce modifications to the SM Higgs couplings, but induce new Lorentz
structures.  In Tab.~\ref{tab:partner-benchmarks} we define a set of
parameter space configurations, all with light and almost degenerate
states and small mixing. The corresponding Wilson coefficients in our
two matching schemes are given in Tab.~\ref{tab:partner-EFT}.
Unrealistic parameter choices with strong
couplings are necessary to generate sizable loop corrections to the
$hVV$ couplings~\cite{Hollik:2008xn}. For fixed masses and mixing, the
Higgs couplings to the top partners depend on the interplay between
$M^2$ and the coupling constants $\kappa$. For small mixing and large $M^2$, light top
partner masses require large four-scalar couplings $\kappa_{ii}$.
Conversely, if $M^2$ is close to the physical masses, the Yukawa
couplings can be small.  This illustrates the balance between the
VEV-dependent (non-decoupling) and the explicit (decoupling) mass
contributions.

\begin{figure}[t]
  \centering
  \includegraphics[width=0.49\textwidth]{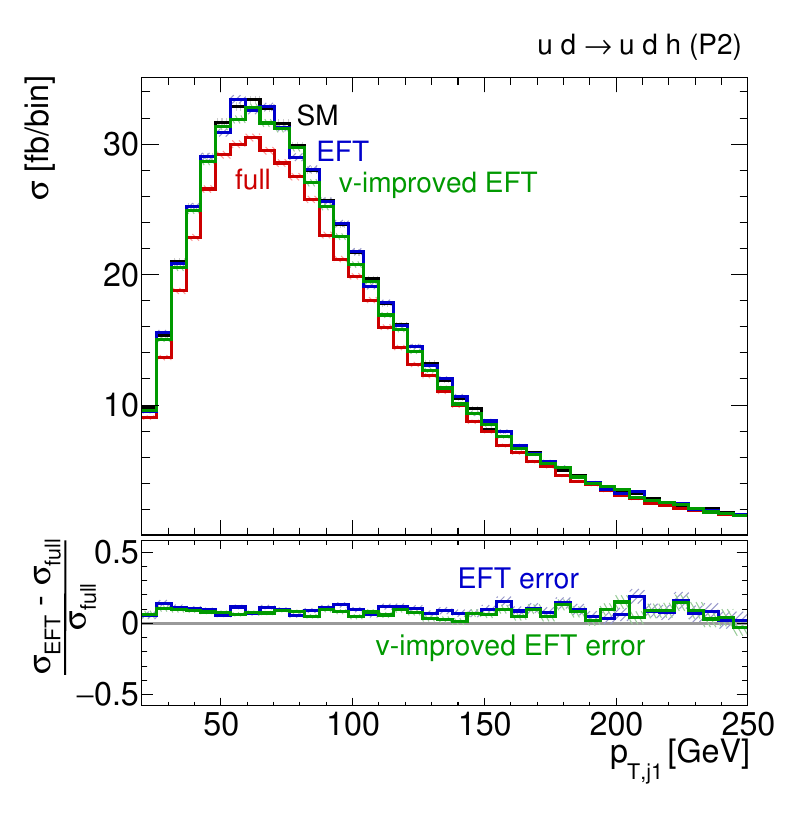}
  \includegraphics[width=0.49\textwidth]{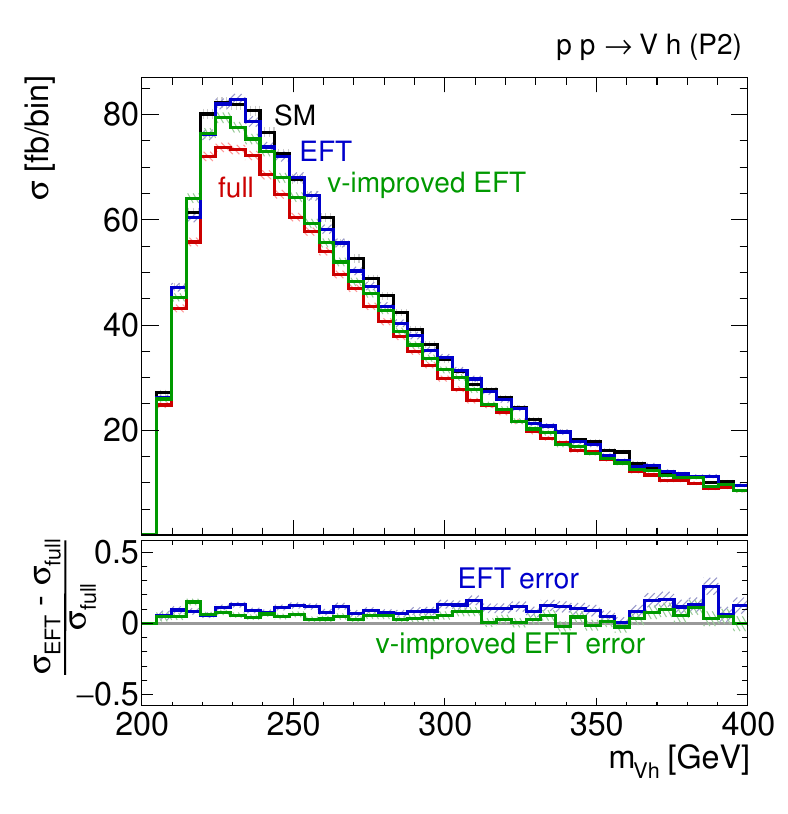}
  \caption{Kinematic distributions for the top partner model in benchmark P2.
    Left: tagging jet properties in WBF Higgs production.
    Right: $m_{Vh}$ distribution in Higgs-strahlung.}
  \label{fig:partners_distributions}
\end{figure}

\begin{table}[t]
  \renewcommand{\arraystretch}{1.2}
  \centering
    \begin{tabular}{c c rr c rr}
      \toprule
      \multirow{2}{*}{Benchmark} &\hspace*{1em}& \multicolumn{2}{c}{$\sigma_\text{EFT} / \sigma_\text{triplet}$}
      &\hspace*{1em}& \multicolumn{2}{c}{$\sigma_\text{$v$-improved EFT} / \sigma_\text{triplet}$} \\
      \cmidrule{3-4}\cmidrule{6-7}
      && WBF & $Vh$ && WBF & $Vh$ \\
      \midrule
      P1 && 1.000 & 0.999 && 1.000 & 0.999 \\
      P2 && 1.095 & 1.100 && 1.074 & 1.049 \\
      P3 && 2.081 & 1.904 && 1.749 & 1.363 \\
      \bottomrule
    \end{tabular}
  \caption{Cross section ratios of the matched dimension-6 EFT
    approximation to the full scalar top-partner model at the LHC.
    We give the results both for the default matching scheme
    with matching scale $\Lambda = M$ as well as for the
    $v$-improved matching at $\Lambda = m_\stone$. The
    statistical uncertainties on these ratios are below 0.4\%.}
  \label{tab:partners_rates}
\end{table}

Since the contributions from scalar top partners to the Higgs
production in gluon fusion are well known~\cite{hgg-toppartners}, we
focus on corrections to the $hVV$ coupling in WBF and Higgs-strahlung,
shown in Tab.~\ref{tab:partners_rates}.  In benchmark P1 the WBF cross section
is reduced by about $0.6 \%$ compared to the Standard Model, with good
agreement between effective and full description. Such a scenario is
not relevant for LHC measurements in the foreseeable future. In more
extreme corners of the parameter space the loop effects in the full model grow,
higher-dimensional terms in the EFT become larger, the validity of the latter
worsens, and discrepancies between both increase. 
In benchmarks P2 and P3 the WBF rate is reduced by $9.1\%$
and $43.5\%$ with respect to the Standard Model.  In the left panel of
Fig.~\ref{fig:partners_distributions} we show that this change in the
total rate does not have dramatic effects in the kinematic
distributions.
By construction, the EFT based on the default matching captures only the formally
leading term at $\ord(v^2/\Lambda^2)$, only giving a reduction of
$0.5\%$ and $2.0\%$. The corresponding difference is again independent
for example of the tagging jet's transverse momentum.
With the $v$-improved matching, the cross section is reduced by 
$2.4 \%$ and $17.7 \%$, still far from the result of the full model. 

The results for Higgs-strahlung look similar: in the moderate
benchmark P1 the predictions of the full model and the dimension-6 Lagrangian agree
within $0.1 \%$, but in this scenario the overall deviation from the
Standard Model is negligible. In scenarios with larger loop effects,
the dimension-6 predictions fails to capture most of the full top partner
loops.  We demonstrate this in the right panel of
Fig.~\ref{fig:partners_distributions}.  As for WBF, the agreement
between EFT and full model becomes even worse in benchmark P3, with
numerical results similar to those given for WBF Higgs
production. Again the $v$-improved matching
performs better than the default matching.\medskip

To summarize, the top partner model for the first time generates a
large set of dimension-6 operators through electroweak loops. However,
in realistic scenarios with a large scale separation the loop
corrections for example to the $hVV$ vertex are tiny.  Pushing for
loop effects that are large enough to leave a visible imprint in WBF
and Higgs-strahlung requires breaking the scale separation between the
observed Higgs scalar and the top partners. In that case the EFT
fails already for the total rates, kinematic distributions
hardly add to this discrepancy.

\subsection{Vector triplet}
\label{sec:triplet}

\begin{table}[t]
  \renewcommand{\arraystretch}{1.2}
  \setlength{\tabcolsep}{0.3em}
  \centering
    \begin{tabular}{c c rrrrrr}
      \toprule
      \multirow{2}{*}{Benchmark} &\hspace*{1em}& \multicolumn{6}{c}{Triplet model} \\
      \cmidrule{3-8}
      && $M_V$~[GeV] & $g_V$ & $c_H$ & ${c}_{F}$ & ${c}_{VVHH}$  & $m_\xi$~[GeV] \\
      \midrule
      T1 && 591 & 3.0 & $-0.47$ & $-5.0$ & 2.0 & 1200  \\
      T2 && 946 & 3.0 & $-0.47$ & $-5.0$ & 1.0 & 1200  \\
      T3 && 941 & 3.0 & $-0.28$ & $3.0$ & 1.0 & 1200  \\
      T4 && 1246 & 3.0 & $-0.50$ & $3.0$ & $-0.2$ & 1200 \\
      T5 && 846 & 1.0 & $-0.56$ & $-1.32$ & $0.08$ & 849 \\
      \bottomrule
    \end{tabular}
  \caption{Benchmark points for the vector triplet model.}
  \label{tab:triplet_benchmarks}
  \setlength{\tabcolsep}{0.5em}
\end{table}

\begin{table}[t]
  \renewcommand{\arraystretch}{1.2}
  \setlength{\tabcolsep}{0.3em}
  \centering
    \begin{tabular}{c c rrrrr c rrrrr}
      \toprule
      \multirow{2}{*}{Benchmark}
      &\hspace*{1em}& \multicolumn{5}{c}{EFT} &\hspace*{1em}& \multicolumn{5}{c}{EFT ($v$-improved)}\\
      \cmidrule{3-7} \cmidrule{9-13}
      && $\Lambda$~[GeV] & $\bar{c}_W$ & $\bar{c}_H$ & $\bar{c}_6$ & $\bar{c}_f$
      && $\Lambda$~[GeV] & $\bar{c}_W$ & $\bar{c}_H$ & $\bar{c}_6$ & $\bar{c}_f$ \\
      \midrule
      T1 && 591 & $-0.044$ & 0.000 & 0.000 & 0.000 && 1200 & $-0.011$ & 0.000 & 0.000 & 0.000 \\
      T2 && 946 & $-0.017$ & 0.000 & 0.000 & 0.000 &&  1200 & $-0.011$ & 0.000 & 0.000 & 0.000 \\
      T3 && 941 & 0.006 & 0.075 & 0.100 & 0.025 &&  1200 & 0.004 & 0.046 & 0.061 & 0.015 \\
      T4 && 1246 & 0.006 & 0.103 & 0.138 & 0.034 &&  1200 & 0.007 & 0.111 & 0.149 & 0.037\\
      T5 && 846 & $-0.007$ & $-0.020$ & $-0.027$ & $-0.007$ && 849 & $-0.007$ & $-0.020$ & $-0.027$ & $-0.007$ \\
      \bottomrule
    \end{tabular}
  \caption{Matching scales and Wilson coefficients for the effective theory
   matched to the vector triplet model. We give these results both for the EFT matching
   in the unbroken phase as well as for the $v$-improved matching with $\Lambda = m_{\xi^0}$.}
  \label{tab:triplet_eft}
  \setlength{\tabcolsep}{0.5em}
\end{table}

Heavy vector bosons appear in many new physics scenarios and possibly
also in data~\cite{ww_resonance}. Their properties can be tested in
Higgs measurements, provided they are connected to the gauge-Higgs
sector of the Standard
Model~\cite{Low:2009di,Biekoetter:2014jwa,Pappadopulo:2014qza}.  For
these analyses the key property of new vector resonances are their SM
charges.  We analyze a massive vector field $V^a_\mu$ which is a
triplet under $SU(2)$, couples to a scalar and fermion currents, and
kinetically mixes with the weak gauge bosons of the Standard
Model~\cite{Pappadopulo:2014qza,Biekoetter:2014jwa}. The Lagrangian
includes the terms
\begin{alignat}{5}
\lag \supset& \,
  - \dfrac{1}{4}\,V_{\mu\nu}^a\,V^{\mu\nu\, a}
  + \dfrac{M_V^2}{2}\,V_\mu^a\,V^{\mu\,a}
  + i\,\frac{\gV}{2} \,\ch\,\vmua\,\left[\phi^\dagger \sigma^a \,\overleftrightarrow{D}^\mu\,\phi\,\right]
  +\dfrac{\gw^2}{2 \gV}\,\vmua\,\sum_\text{fermions}\, \cF \overline{F}_L\,\gamma^\mu\, \sigma^a \,F_L
 \notag \\
 &+ \dfrac{\gV}{2}\,\cvvv\,\epsilon_{abc}\,V_\mu^a\,V_\nu^b\,D^{[\mu}V^{\nu]c}
  + \gV^2\,\cvvhh\,\vmua\,V^{\mu a}\,\phibarphi\,
  - \dfrac{\gw}{2}\,\cvvw\,\epsilon_{abc}\,W^{\mu\nu}\,V_\mu^b\,V_\nu^c \,.
 \label{eq:lag-vectortriplet}
\end{alignat}
The new field-strength tensor is $V_{\mu\nu}^a = D_\mu\vnua -
D_\nu\,\vmua$ and the covariant derivative acts on the triplet as
$D_\mu\,V_\nu^a = \partial_\mu\,V_\nu^a+\gV \epsilon^{abc}\,V^b_\mu
V_\nu^c$.  The coupling constant $\gV$ is the characteristic strength
of the heavy vector-mediated interactions, while $\gw$ denotes the
$SU(2)$ weak gauge coupling.  It will turn out that $c_{VVW}$ and
$c_{VVV}$ are irrelevant for Higgs phenomenology at the LHC.  We give
details of the model and the matching to the corresponding EFT in
Appendix~\ref{sec:ap-triplet}.\medskip

The feature setting the vector triplet apart from the singlet,
doublet, and top partner models is that it directly affects the weak
gauge bosons.  The mixing of the new states with the $W$ and $Z$
bosons has two consequences: $i)$ a modification of the Higgs
couplings to SM particles, and $ii)$ new heavy states $\xi^0$,
$\xi^\pm$.

The definition of mass eigenstates from the heavy vector and the
SM-like gauge fields links the observable weak coupling $g$ and the
Lagrangian parameter $g_w$. For the coupling modifications this shift
in the gauge coupling and the direct heavy vector coupling to the
Higgs doublet combine to
\begin{align}
\Delta_V 
&\approx \frac{g^2 c_F c_H}{4}  \, \left( \frac{v}{M_V} \right)^2 
- \frac{3 g_V^2 c_H^2}{8}  \, \left( \frac{v}{M_V} \right)^2 \notag \\
\Delta_f 
&\approx \frac{g^2 c_F c_H}{4}  \, \left( \frac{v}{M_V} \right)^2 
- \frac{g_V^2 c_H^2}{8}  \, \left( \frac{v}{M_V} \right)^2 \,.
\label{eq:vector_decoup}
\end{align}
The contribution from the shift in the weak coupling is identical for
both coupling modifications.  In addition, contributions from virtual
heavy states $\xi$ modify the phase-space behavior of Higgs signals in
many ways.\medskip

Just as for the 2HDM and the top partners, the mass matrix for the
massive vectors contains both the heavy scale $M_V$, which will
eventually become the matching scale, and terms proportional to some
power of $v$ multiplied by potentially large couplings. The new
vector states have roughly degenerate masses
\begin{align}
\frac{m_\xi^2}{M_V^2} 
\approx 1 + g_V^2 c_{VVHH} \, \left( \frac{v}{M_V} \right)^2 
          + \frac{g_V^2 c_H^2}{4} \, \left( \frac{v}{M_V} \right)^2  \,.
  \label{eq:triplet_mxi_M}
\end{align}
Even if there appears to be a clear scale separation $M_V \gg v$,
large values of $g_V$, $c_{VVHH}$, or $c_H$ can change $m_\xi$
significantly and effectively induce a second mass scale.  Just as
for the top partners, a problem for the dimension-6 approach arises from
virtual $\xi$ diagrams contributing for example to WBF Higgs
production. If $m_\xi < M_V \equiv \luv$ the lightest new particles
appearing in Higgs production processes have masses below the matching
scale of the linear representation.
The way out of a poor agreement between the full model and its dimension-6
description is again switching to a $v$-improved matching in the broken
phase with matching scale $\luv = m_\xi$.

Integrating out the heavy vector triplet at tree level
leaves us with dimension-6 Wilson coefficients
\begin{align}
\bar{c}_{H} &= \dfrac{3\,g^2\,v^2}{4\,M_V^2}\,\left[\ch^2\dfrac{g_V^2}{g^2}  - 2\,\cF\,\ch\right] 
&\bar{c}_{6} &= \dfrac{g^2\,v^2}{M_V^2}\,\left[\ch^2\dfrac{g_V^2}{g^2} - 2\,\cF\,\ch \right] \notag \\
\bar{c}_{f} &= \dfrac{g^2\,v^2}{4\,M_V^2}\,\left[\ch^2\dfrac{g_V^2}{g^2}  - 2\,\cF\,\ch \right] 
&\bar{c}_{W} &= - \dfrac{m_W^2}{M_V^2}\, \cF\ch \,,
 \label{eq:triplet_coefficients}
\end{align}
and four-fermion contributions that are irrelevant for Higgs
physics.
Additional loop-induced contributions will be further suppressed and
do not add qualitatively new features, so we neglect them.
As in the 2HDM, we compare this default matching to an alternative $v$-improved
matching with matching scale $\luv = m_{\xi^0}$. The coefficients in
Eq.\,\eqref{eq:triplet_coefficients} remain unchanged, except that
$M_V$ is replaced by $m_{\xi^0}$.

The main
phenomenological features of this model reside in the Higgs-gauge
interactions. In the dimension-6 description, these modifications are mapped
(amongst others) onto $\ope{W}$, which induces momentum-dependent
changes to the $hWW$ and $hZZ$ couplings. Therefore, our analysis
focuses on WBF Higgs production and Higgs-strahlung, where the
intermediate $t$-channel and $s$-channel gauge bosons can transfer
large momenta.\medskip

As for the other models we study a set of benchmark points, 
defined in Tab.~\ref{tab:triplet_benchmarks} and Tab.~\ref{tab:triplet_eft}.
Some of them are meant
to emphasize the phenomenological possibilities of the vector triplet
model. For those we ignore experimental constraints or parameter
correlations from an underlying UV completion:
\begin{itemize}
\item[T1-2] All dimension-6 EFT operators except for $\ope{W}$ vanish
  along the line $c_H/c_F = 2 g^2/g_V^2$.  We aim for a large effect
  only in the $hVV$ couplings.  The large couplings induce different
  scales $M_V$ and $m_\xi$.
\item[T3] The sign in front of $\ope{W}$ changes on another line in
  the $(c_H, c_F)$ space. The remaining operators do not vanish.
\item[T4] The vector triplet couplings and masses satisfy the leading
  constraints from direct collider searches. For weak couplings ($\gV
  \leq 1$) resonances typically decaying to di-lepton and neutrino
  final states have to stay above 3~TeV.  For the strongly interacting
  case ($\gV >1$) decays to di-bosons tend to exclude masses below
  $1-1.5$~TeV~\cite{Pappadopulo:2014qza,Kaminska:2015ora}.
 \item[T5] A weakly coupled UV completion can be based on the 
   gauge group $SU(3) \times SU(2) \times SU(2) \times
   U(1)$~\cite{Barger:1980ti}, arising for instance from deconstructed
   extra dimensions~\cite{ArkaniHamed:2001nc}. Its vector triplet
   phenomenology is effectively described by the parameter
   $\alpha = g_V / \sqrt{g_V^2 - g_w^2}$ together with the 
   symmetry breaking scale $f$~\cite{Pappadopulo:2014qza},
 \begin{alignat}{3}
   M_V^2 &= \alpha^2 \gV^2 f^2 \,, & \qqquad
   \ch &= - \alpha  \dfrac {\gw^2}{g_V^2} \,, & \qqquad
   \cvvhh &= \alpha^2 \left[ \dfrac {\gw^4} {4g_V^4}  \right] \,, \notag \\
   \cF &= - \alpha \,,  & \qquad
   \cvvw &= 1 \,,  & \qquad
   \cvvv &= - \dfrac{\alpha^3}{g_V} \left[1 - \dfrac{3\gw^2}{g_V^2}  + \dfrac {2\gw^2} {g_V^4}  \right] \,.
 \end{alignat}
\end{itemize}
\medskip

\begin{figure}[t]
  \centering
  \includegraphics[width=0.49\textwidth]{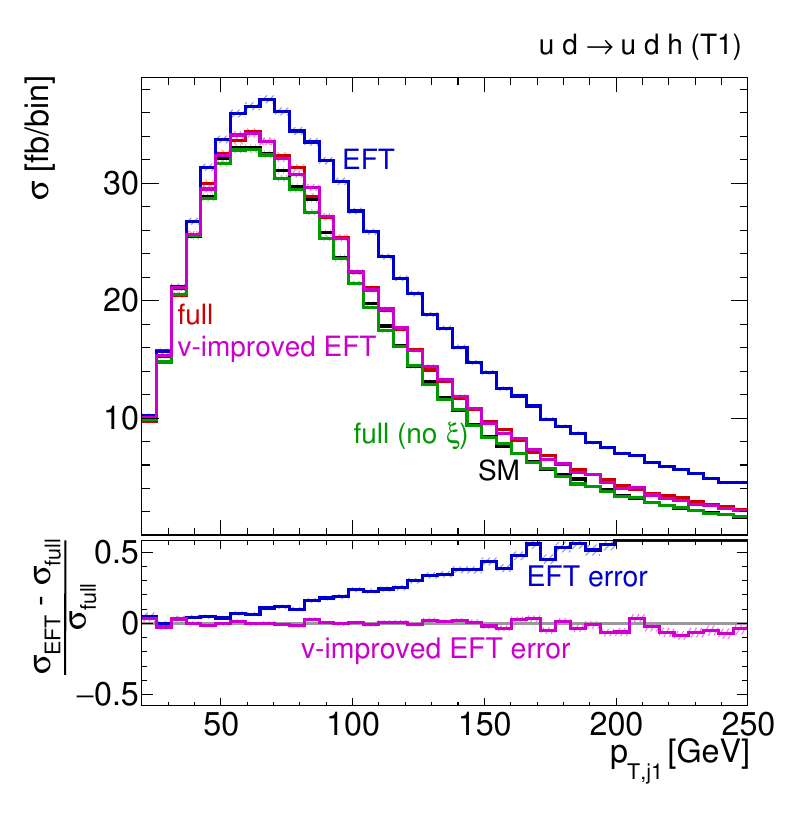}
  \includegraphics[width=0.49\textwidth]{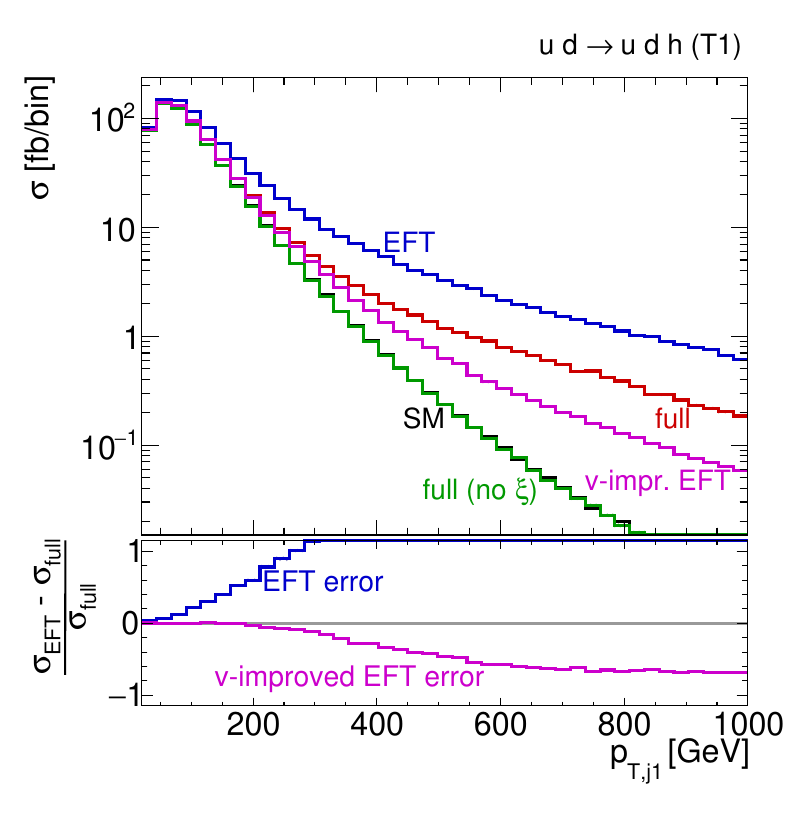} \\
  \includegraphics[width=0.49\textwidth]{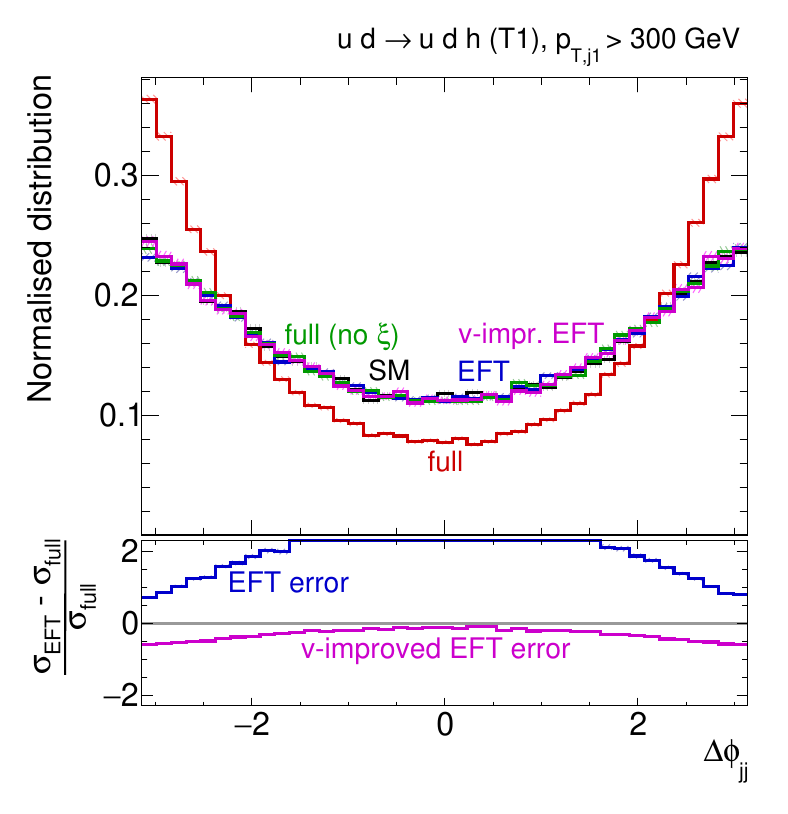}
  \includegraphics[width=0.49\textwidth]{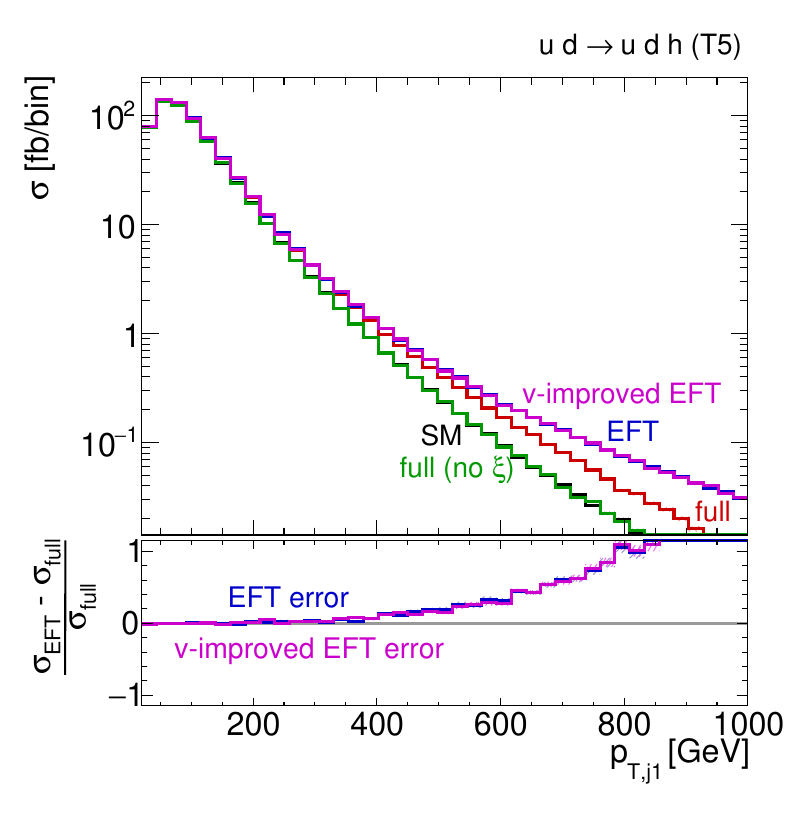}
  \caption{Tagging jet distributions in WBF Higgs production in the
    vector triplet model.  Top: $p_{T,j1}$ distribution in benchmark
    T1, focusing on the low (left) and high (right) transverse
    momentum regions.  Bottom left: $\Delta \phi_{jj}$ distribution
    above a certain $p_{T,j1}$ threshold for T1.  Bottom right:
    $p_{T,j1}$ distribution for scenario T5.}
  \label{fig:triplet_wbf}
\end{figure}

\begin{table}[t]
  \renewcommand{\arraystretch}{1.2}
  \centering
    \begin{tabular}{c c rr c rr}
      \toprule
      \multirow{2}{*}{Benchmark} &\hspace*{1em}& \multicolumn{2}{c}{$\sigma_\text{EFT} / \sigma_\text{triplet}$}
      &\hspace*{1em}& \multicolumn{2}{c}{$\sigma_\text{$v$-improved EFT} / \sigma_\text{triplet}$}\\
      \cmidrule{3-4} \cmidrule{6-7}
      && WBF & $Vh$ && WBF & $Vh$ \\
      \midrule
      T1 && 1.299 & 0.299 && 0.977 & 0.794 \\
      T2 && 1.045 & 0.737 && 0.992 & 0.907 \\
      T3 && 0.921 & 1.066 && 0.966 & 1.024 \\
      T4 && 1.026 & 0.970 && 1.012 & 0.978 \\
      T5 && 1.001 & 1.043 && 1.002 & 1.043 \\
      \bottomrule
    \end{tabular}
    \caption{Cross section ratios of the matched dimension-6 EFT
    approximation to the full vector triplet at the LHC.
    To avoid large contributions from the $\xi$ resonance in the $Vh$
    channel, we only take into account the region $m_{Vh} < 600$~GeV.
    The statistical uncertainties on these ratios are below 0.4\%.}
  \label{tab:triplet_rates}
\end{table}

In Fig.~\ref{fig:triplet_wbf} we show a set of kinematic distributions
in WBF Higgs production. In addition to the predictions of the full
vector triplet model and the matched EFT, we show distributions of the
vector triplet model without contributions from $\xi$ propagators.
The corresponding production cross section ratios between full vector
triplet model and EFT are given in Tab.~\ref{tab:triplet_rates}.
For the full model we observe a significant modification of the rate
relative to the Standard Model, especially towards large momentum
transfers. They can be traced to the $\xi$ fusion and mixed $W$-$\xi$
fusion diagrams, which increase strongly with energy. In comparison,
the modification of the $hWW$ coupling only leads to a relatively mild
rescaling.  These contributions from $\xi$ propagators can become
relevant already at energy scales well below $m_\xi$. The weak boson
virtualities inducing a momentum flow into the Higgs coupling are not
the only source of deviation from the Standard Model; the azimuthal
correlation between the tagging jets is well known to be sensitive to
the modified Lorentz structure of the $hWW$ vertex~\cite{phi_jj}.

Qualitatively, the dimension-6 approach captures the features of the full
model, driven by $\ope{W}$.  In T1 and T2 a negative Wilson
coefficient yields a non-linear increase of the cross section with
energy. Conversely, the positive coefficient in T3 reduces the rate
with energy, eventually driving the combined amplitude through zero.

Comparing full and effective model for the more realistic benchmark
points T4 and T5 we see good agreement in the bulk of the distribution.  The
deviations from the Standard Model are captured by the dimension-6
operators, including the momentum dependence coming from the $\xi$
diagrams. Only at very large momentum transfer the validity of the EFT
breaks down. For our realistic benchmark points the LHC is likely not
sensitive to these subtle effects.

In the more strongly coupled benchmark points T1\,--\,T3, the full model
predicts shifts in the jet distributions that are large enough to be relevant
for the upcoming LHC run. We find good agreement between the full model and the
default EFT only at low momentum transfer, where the effects of new physics are small.
This naive dimension-6 approach loses its validity already around
$p_{T,j} \gtrsim 80$~GeV, a phase space
region highly relevant for constraints on new
physics~\cite{sfitter_last}.\footnote{Note, however, that these scenarios
are already in tension with bounds from electroweak precision observables, but
we nevertheless show them to illustrate the qualitative aspects of EFT
breakdown.} This does not signal a breakdown of the $E / \Lambda$ expansion,
but a too large $c_i v^2 / \Lambda^2$. It is linked to the difference between
the scales $m_\xi$ and $M_V$ as given in Eq.\,\eqref{eq:triplet_mxi_M},
which the default matching procedure is blind to.
Indeed, with the $v$-improved matching the agreement is significantly
better, and the dimension-6 description departs from the full model
only at high energies, $p_{T,j1} \gtrsim 300$~GeV.\medskip

\begin{figure}
  \centering
  \includegraphics[width=0.49\textwidth]{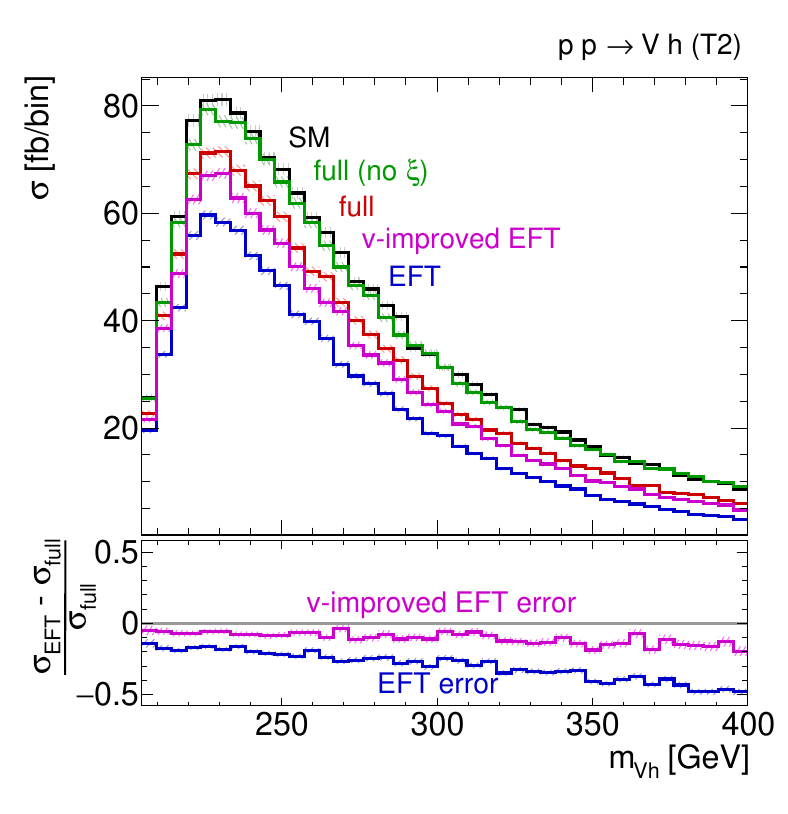}
  \includegraphics[width=0.49\textwidth]{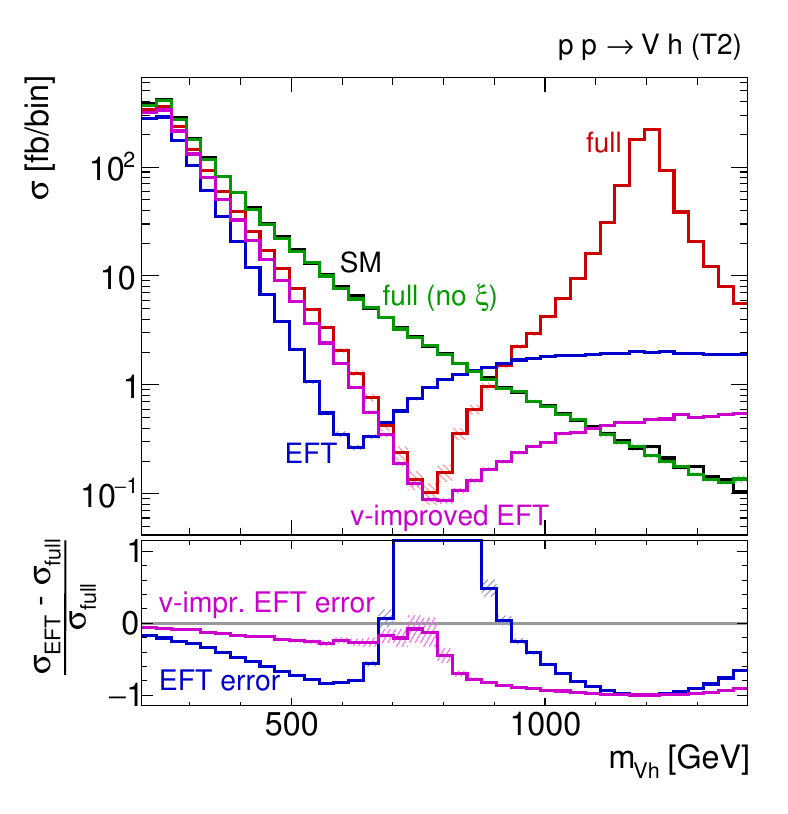} \\
  \includegraphics[width=0.49\textwidth]{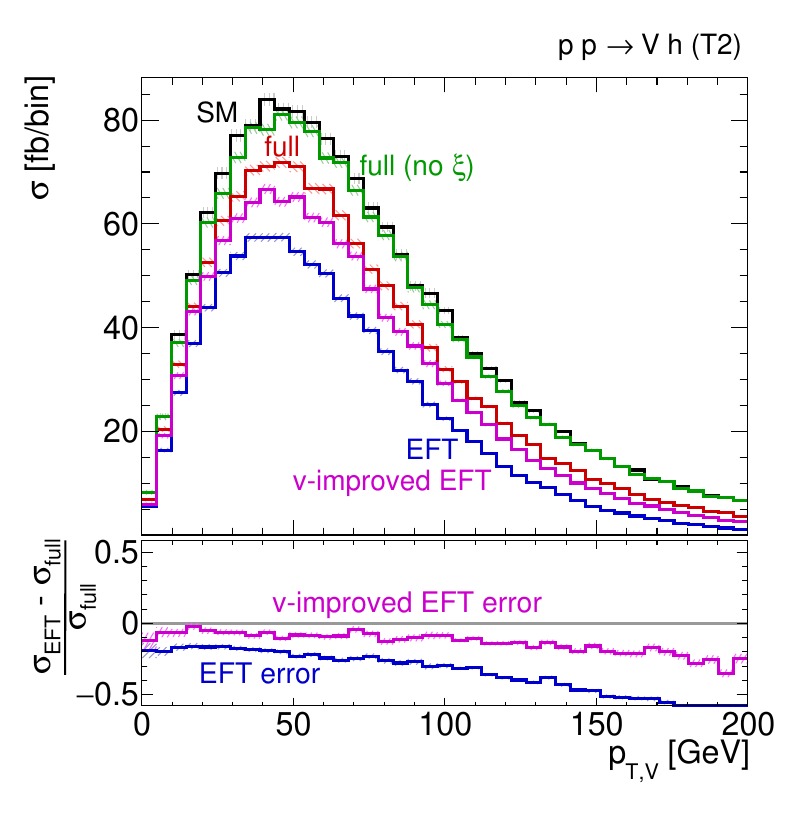}
  \includegraphics[width=0.49\textwidth]{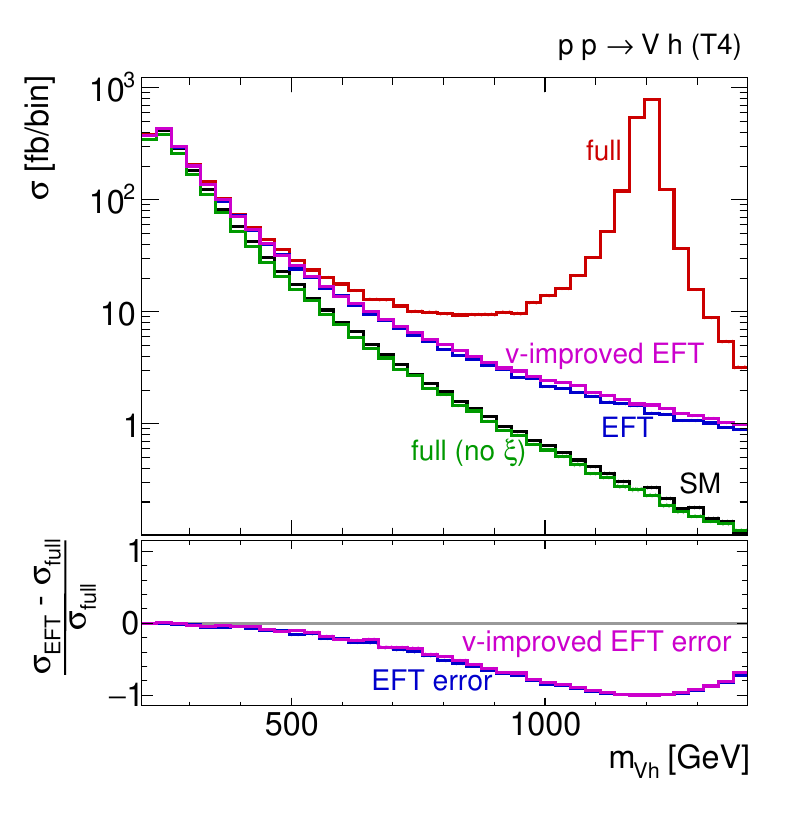}
  \caption{Higgs-strahlung distributions in the vector triplet model.
    Top: $m_{Vh}$ distribution for benchmark T2, focusing on the low (left) and high (right)
    invariant mass regions.
    Bottom left: $p_{T,V}$ distribution for the same benchmark. 
    Bottom right: $m_{Vh}$ distribution for T4.}
  \label{fig:triplet_vh}
\end{figure}

The situation is similar in Higgs-strahlung, shown in
Fig.~\ref{fig:triplet_vh}.  In the full model the $\xi$ propagators
again dominate over the the modified $hWW$ interaction.  In addition,
the interference with the $\xi$-mediated diagrams leads to a significant change
of the rate and introduces a momentum dependence already far below the
actual resonance.  The relative sign of the interference between $\xi$
amplitudes and SM-like diagrams is opposite to that in WBF.

In the EFT the operator $\ope{W}$ induces the corresponding strong
energy dependence.  A positive Wilson coefficient leads to a non-linear
increase of the cross section with the energy scale, probed by either
$m_{Vh}$ or the $p_{T,V}$. A negative coefficient leads to a
decreasing amplitude with energy, including a sign flip. Like for the full
model, these $\ope{W}$ terms have the opposite effect on the rate as
in WBF.

The full and effective models agree relatively well 
in the more weakly coupled benchmarks at low energies. In the
realistic scenarios T4 and T5, this agreement
extends over the most relevant part of the phase space, and the EFT
successfully describes how the $\xi$ propagators shift the
Higgs-strahlung kinematics. With increasing energy, momentum-dependent
effects in both the full model (due to the resonance) and the EFT (due
to $\ope{W}$) become more relevant. While the sign of the effect is
the same in full model and EFT, the size and energy dependence 
is different, and the EFT eventually fails to be a good
approximation. At even higher energies, the ``dips''
at different energies in the full model and EFT as well as the $\xi$
resonance in the full model mark the obvious failure of the effective
theory.

For benchmark T1 to T3, where the effects are numerically much more relevant
for the LHC, the range of validity of the default EFT is limited. The
couplings are so large that in spite of a resonance mass $m_\xi \sim
1$~TeV the dimension-6 description already fails at $m_{Vh} \gtrsim 220~\gev$.
Switching to the $v$-improved matching again
ameliorates the dimension-6 approximation.
Even then, this mismatch between full model and EFT is more pronounced in
Higgs-strahlung than in WBF, because $\xi$ contributions play a larger
role in these $s$-channel diagrams than in the $t$-channel WBF
amplitudes. \medskip

\begin{figure}
  \centering
  \includegraphics[width=0.49\textwidth,clip=true,trim=-1cm 0 -1cm 0]{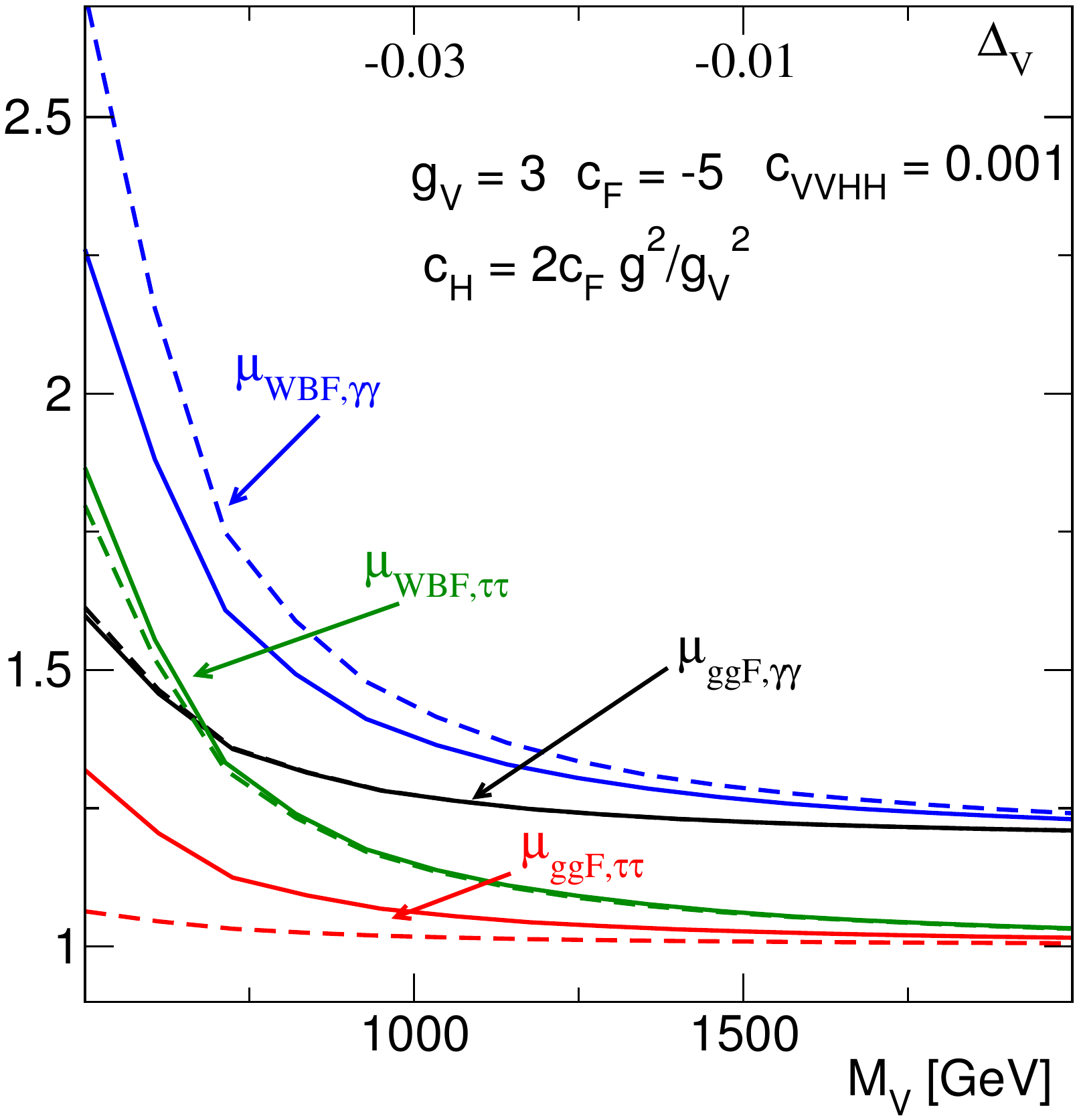} 
  \includegraphics[width=0.49\textwidth,clip=true,trim=-1cm -0.92cm -1cm 0]{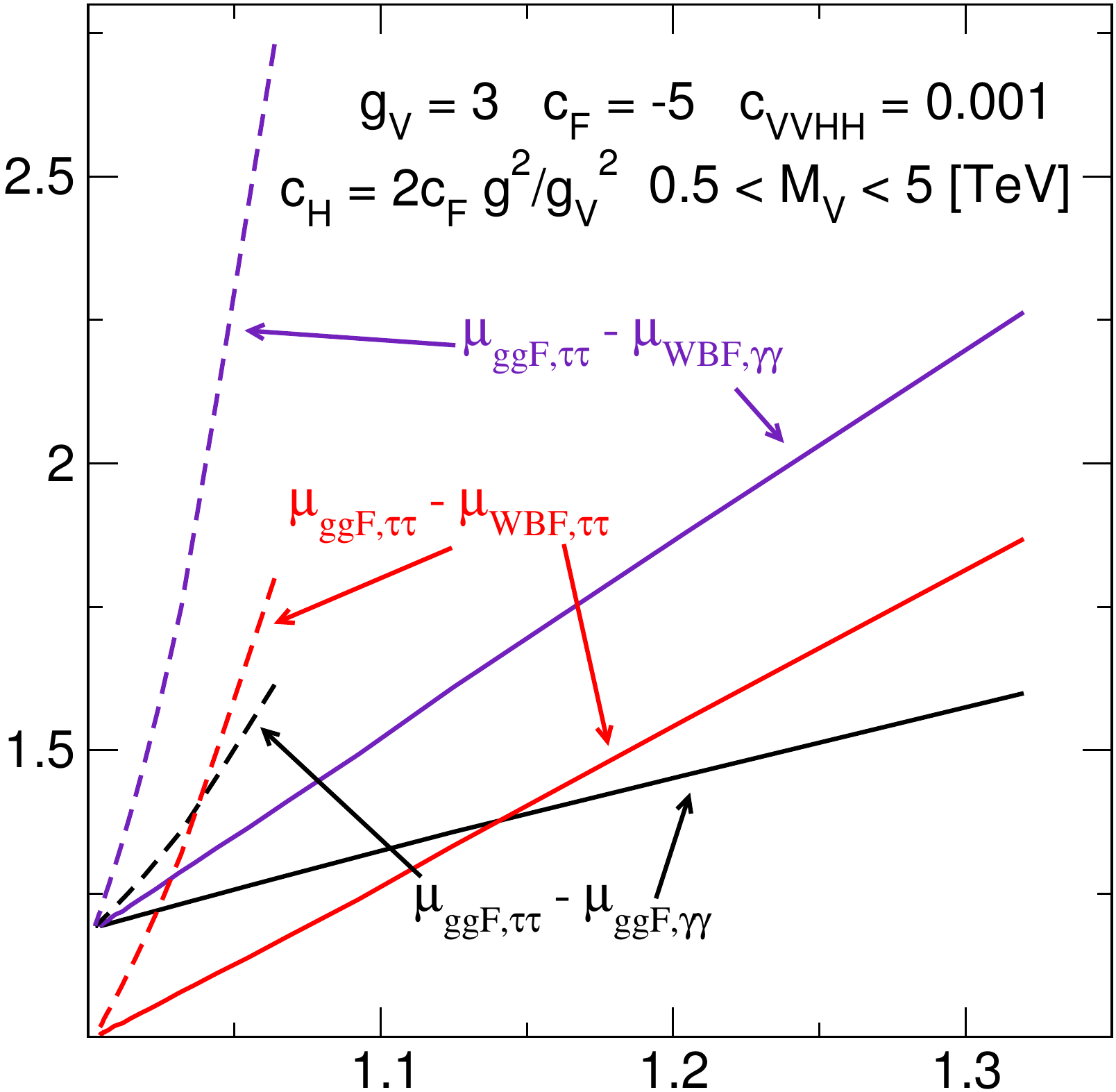} 
  \caption{Signal strength modifications in the vector triplet.  The
    solid lines show the full model, while the dashed lines give the
    dimension-6 predictions for the default matching.  Left: signal strength $\mu_{p,d}$ for different
    Higgs production modes and decay channels for an exemplary vector
    triplet setup as a function of $M_V$. In the upper horizontal axis
    we show the deviation from the SM-like limit through the coupling
    shift $\Delta_V$, Eq.\,\eqref{eq:vector_decoup}.  Right: signal
    strength correlations $\mu_{p_1,d_1}$ versus $\mu_{p_2,d_2}$
    between different channels for variable $M_V$.  }
  \label{fig:triplet_correlations}
\end{figure}

In Fig.~\ref{fig:triplet_correlations} we again go beyond individual benchmark
points, and examine the agreement
between full model and its dimension-6 description in terms of signal
strengths, correlated for different Higgs production modes and decay
channels. For definiteness, we assume vector triplet parameters in
line with the benchmarks T1 and T2, and
vary the heavy vector mass scale $M_V = 0.5 \dots 5$~TeV. The
dimension-6 coefficients are based on the default matching.

The huge deviations in the WBF signal strength are due to the sizable
momentum-dependent effects in the fusion process.  As discussed above,
this behavior is poorly captured by the EFT for large vector
couplings, and fails dramatically for light mass scales.  The same
differences are visible from the different trajectories in
the correlated signal strength plane, shown in the left panel.  The
mild offset from $\mu_{p,\gamma\gamma} = 1$ in the limit $M_V \gg v$
can be traced back to the non-decoupling $\xi^{\pm}$-mediated
contribution to the $h\gamma\gamma$ loop.  The
$\ord(\cF\ch\,v^4/m_V^4)$ contributions of dimension eight and higher
are responsible for the additional upward enhancement of the fermion
Yukawas in the full model, which is in particular visible for
$\mu_{gg,\tau\tau}$, where the full model predictions systematically
surpass the EFT. Finally, we find that an enhanced top-$W$
interference in $\Delta_\gamma$ pulls the full model $\gamma\gamma$
rates below the dimension-6-based predictions.  The accidental counterbalance
of the higher dimension effects missing in the EFT explains the
remarkable agreement with the full model for
$\mu_{\text{ggF},\gamma\gamma}$.\medskip

Like the additional scalar models discussed before, the vector
triplet model offers regions in parameter space where the EFT works up
to large momentum transfer for realistic scenarios. It successfully captures the virtual $\xi$
contributions in the momentum dependent contribution from $\ope{W}$,
but these numerical effects are small.  Relevant effects for the LHC
occur if the separation of scales is spoiled by large couplings or
light new particles. In this case we find substantial dimension-6 departures
from the full model predictions for example in the bulk of the
WBF distributions, which typically further increase with the energy scale.
A modified dimension-6 description incorporating $v$-dependent effects
improves the EFT accuracy such that large deviations only occur
in the high-energy tails of distributions.

\section{Summary}
\label{sec:summary}

An effective field theory for the Higgs sector offers a theoretically
well-defined, efficient, and largely model-independent language to
analyze extensions of the Standard Model in both rate measurements and
kinematic distributions. A fit of dimension-6 operators to LHC Higgs
measurements works fine~\cite{sfitter_last} and constitutes the
natural extension of the Higgs couplings analyses of Run~I.  Most of
the relevant higher-dimensional operators correspond to simple
coupling modifications, supplemented by four operators describing new
Lorentz structures in the Higgs coupling to weak
bosons~\cite{sfitter_last}.

In this paper we have studied the validity of this approach from the theoretical side.
We know that at the LHC a clear hierarchy of electroweak and
new physics scales cannot be guaranteed, the question is whether
dimension-6 operators nevertheless capture the phenomenology of specific
UV-complete theories with sufficient accuracy.  We have systematically compared a singlet Higgs
portal model, a two-Higgs doublet model, scalar top partners, and a
heavy vector triplet to their dimension-6 EFT descriptions, based on
the linear realization of electroweak symmetry breaking with a Higgs doublet. 
We have analyzed the main Higgs production and decay signatures, covering
rates as well as kinematic distributions.\medskip

We have found that the dimension-6 operators provide an adequate description in almost
all realistic weakly coupled scenarios. Shifts in the total rates are well described by
effective operators.  Kinematic distributions typically do not probe
weakly interacting new physics with sufficient precision in the high-energy
tails to challenge
the effective operator ansatz.  This is obvious for the extended scalar models, where
new Lorentz structures and momentum-dependent couplings with dramatic
effects in LHC distributions only appear at the loop level. 
A loop-suppressed effective scale suppression $E^2/(4
\pi \Lambda)^2$ has to be compared with on-shell couplings
modifications proportional to $v^2/\Lambda^2$.  Only phase space regions
probing energies around $4 \pi v \approx 3$~TeV significantly
constrain loop contributions in the Higgs sector and eventually lead to
breakdown of the effective field theory. In turn, a simple dimension-6
descriptions will capture all effects that are expected to be measurable with sufficient statistics at the LHC Run II.  
On the other hand, the vector triplet
model shows that modifications of the gauge sector can generate
effects in LHC kinematics at tree level. However, we again find that
for weakly interacting models and phenomenologically viable benchmark points they are described well by an appropriate
set of dimension-6 operators.\medskip

\begin{table}[t]
\renewcommand{\arraystretch}{1.2}
\centering
\begin{tabular}{ll c ccc}
  \toprule
  Model & Process &\hspace*{1em}& \multicolumn{3}{c}{EFT failure} \\ 
  \cmidrule{4-6}
      & && resonance & kinematics & matching \\
  \midrule
  singlet & on-shell $h \to 4 \ell$, WBF, $Vh$, \dots && & & \largex  \\
      & off-shell WBF, \dots && & \brlargex & \largex \\
      & $hh$ && \largex &  \largex & \largex \\
  2HDM    & on-shell  $h \to 4 \ell$, WBF, $Vh$, \dots && & & \largex  \\
      & off-shell $h \to \gamma \gamma$, \dots && & \brlargex & \largex \\
      & $hh$ && \largex &  \largex & \largex \\
  top partner & WBF, $Vh$ && & & \largex \\
  vector triplet & WBF  &&  & \brlargex & \largex \\ 
      & $Vh$ && \largex & \brlargex & \largex \\       
  \bottomrule
\end{tabular}
 \caption{Possible sources of failure of dimension-6 Lagrangian at the
   LHC.  We use parentheses where deviations in kinematic distributions
   appear, but are unlikely to be observed in realistic scenarios.}
 \label{tab:differences}
\end{table}

Three sources for a possible breakdown of the dimension-6 description
are illustrated in Tab.~\ref{tab:differences}\footnote{Forcing the EFT
  approach into a spectacular breakdown was the
  original aim of this paper, but to our surprise this did not happen.}:
First, the EFT cannot describe light new resonances. Such a signature at the LHC would
be an obvious signal to stop using the EFT and switch to appropriate simplified models.
Second, selected kinematic distributions fail to be described by the
dimension-6 Lagrangian, in particular for Higgs pair production.
Deviations in the high-energy tails of WBF and Higgs-strahlung
distributions on the other hand are too small to be relevant in realistic
weakly coupled scenarios. These two cases do not threaten LHC analyses in practice.

The third issue with the dimension-6 EFT description is linked to
matching in the absence of a well-defined scale hierarchy.  Even with only one
heavy mass scale in the Lagrangian, the electroweak VEV together with
large couplings can generate several new physics scales,
defined by the masses of the new particles.
A linear EFT description, which is justified by the SM-like properties of the newly
discovered Higgs boson, should in principle be matched in the phase where
the electroweak symmetry is unbroken. Such a procedure is
blind to additional scales induced by the electroweak VEV, potentially
leading to large errors in the dimension-6 approximation.
Including $v$-dependent terms in the Wilson coefficients,
which corresponds to matching
in the broken phase, can significantly improve the EFT
performance. We have explicitly demonstrated this for all the models considered
in this paper.

None of these complications with the dimension-6 description
presents a problem in using effective operators to fit LHC Higgs data.
They are purely theoretical issues that need to be considered
for the interpretation of the results.


\subsubsection*{Acknowledgments}

We would like to thank Juan Gonzalez-Fraile for fruitful discussions
over many cups of coffee.
TP would like to thank the HDays workshop in Santander for
great discussions contributing to the presentation of this paper. 
JB is grateful to the DFG for his funding through the
Graduiertenkolleg \emph{Particle physics beyond the Standard Model}
(GRK~1940).  DLV is funded by the F.R.S.-FNRS \emph{Fonds de la
  Recherche Scientifique} (Belgium). TP acknowledges support by the
DFG Forschergruppe \emph{New Physics at the LHC} (FOR~2239). The work of AF is funded in part by
the U.S.\ National Science Foundation under grants PHY-1212635 and PHY-1519175.


\clearpage
\input{appendix}


\end{document}

%% file: declare.tex
\newcommand{\toolfont}[1]{\texttt{#1}}
\newcommand{\Hpm}{{\ensuremath{H^\pm}}}
\newcommand{\Hp}{{\ensuremath{H^+}}}
\newcommand{\Hm}{{\ensuremath{H^-}}}
\newcommand{\ord}{\ensuremath{\mathcal{O}}}
\newcommand{\ope}[1]{\ensuremath{\mathcal{O}_{#1}}}
\newcommand{\largex}{\ensuremath{\large \boldsymbol{\times}}}
\newcommand{\brlargex}{\ensuremath{\large (\boldsymbol{\times})}}
\newcommand{\mheavy}{\ensuremath{M}}

\newcommand{\lag}{\ensuremath{\mathcal{L}}}
\newcommand{\delx}{\ensuremath{\Delta}}

\newcommand{\met}{\ensuremath{\slashed{E}_T}}

\newcommand{\sw}{\ensuremath{s_w}}
\newcommand{\swd}{\ensuremath{s^2_w}}
\newcommand{\cw}{\ensuremath{c_w}}
\newcommand{\cwd}{\ensuremath{c^2_w}}

\newcommand{\tanb}{\ensuremath{\tan\beta}}
\newcommand{\cotb}{\ensuremath{\cot\beta}}

\newcommand{\br}{\text{BR}}
\newcommand{\qqquad}{\qquad \qquad}
\newcommand{\qqqquad}{\qquad \qquad \qquad}

\newcommand{\st}[1]{\tilde{t}_{#1}}

\newcommand{\gev}{{\ensuremath\rm GeV}}
\newcommand{\tev}{{\ensuremath\rm TeV}}

\def\slashchar#1{\setbox0=\hbox{$#1$}           
   \dimen0=\wd0                                 
   \setbox1=\hbox{/} \dimen1=\wd1               
   \ifdim\dimen0>\dimen1                        
      \rlap{\hbox to \dimen0{\hfil/\hfil}}      
      #1                                        
   \else                                        
      \rlap{\hbox to \dimen1{\hfil$#1$\hfil}}   
      /                                         
   \fi}

\def\eg{{e.\,g.}\ }
\def\ie{{i.\,e.}\ }

\renewcommand{\vec}[1]{{\bf #1}}

\newcommand{\ohb}{\ensuremath{\hat{\mathcal{O}}_{HB}}}
\newcommand{\ohw}{\ensuremath{\hat{\mathcal{O}}_{HW}}}
\newcommand{\ohgam}{\ensuremath{\hat{\mathcal{O}}_{\gamma}}}
\newcommand{\ob}{\ensuremath{\hat{\mathcal{O}}_B}}
\newcommand{\ow}{\ensuremath{\hat{\mathcal{O}}_W}}
\newcommand{\oh}{\ensuremath{\hat{\mathcal{O}}_H}}
\newcommand{\ogam}{\ensuremath{\hat{\mathcal{O}}_{\gamma}}}
\newcommand{\oT}{\ensuremath{\hat{\mathcal{O}}_T}}
\newcommand{\og}{\ensuremath{\hat{\mathcal{O}}_g}}
\newcommand{\ou}{\ensuremath{\hat{\mathcal{O}}_u}}
\newcommand{\od}{\ensuremath{\hat{\mathcal{O}}_d}}
\renewcommand{\ol}{\ensuremath{\hat{\mathcal{O}}_\ell}}
\newcommand{\of}{\ensuremath{\hat{\mathcal{O}}_f}}
\newcommand{\osix}{\ensuremath{\hat{\mathcal{O}}_6}}
\newcommand{\pbp}{\ensuremath{\phi^\dagger\,\phi}}

\newcommand{\luv}{\ensuremath{\Lambda}}

\newcommand{\phibarphi}{\ensuremath{\phi^\dagger\,\phi}}
\newcommand{\ephys}{\ensuremath{E_{\text{phys}}}}

\newcommand{\lageff}{\ensuremath{\lag_{\text{eff}}}}
\renewcommand{\met}{\ensuremath{\slashed{E}_T}}

\newcommand{\vmua}{\ensuremath{V_{\mu}^a}}
\newcommand{\vnua}{\ensuremath{V_{\nu}^a}}
\newcommand{\ch}{\ensuremath{c_H}}
\newcommand{\cF}{\ensuremath{c_F}}
\newcommand{\gV}{\ensuremath{g_V}}
\newcommand{\gw}{\ensuremath{g_w}}
\newcommand{\cvvv}{\ensuremath{c_{VVV}}}
\newcommand{\cvvw}{\ensuremath{c_{VVW}}}
\newcommand{\cvvhh}{\ensuremath{c_{VVHH}}}
\newcommand{\tch}{\ensuremath{\tilde c_H}}
\newcommand{\tcF}{\ensuremath{\tilde c_F}}
\newcommand{\tcvvv}{\ensuremath{\tilde c_{VVV}}}
\newcommand{\tcvvw}{\ensuremath{\tilde c_{VVW}}}
\newcommand{\tcvvhh}{\ensuremath{\tilde c_{VVHH}}}

\newcommand{\mv}{\ensuremath{M_V}}
\newcommand{\tmv}{\ensuremath{\tilde{M}_V}}

\newcommand{\mw}{\ensuremath{m_W}}

\newcommand{\cvwtilde}{\ensuremath{\tilde{c}_{WV}}}

\newcommand{\lrd}{\ensuremath{\overleftrightarrow{D}}}

\newcommand{\jhdown}{\ensuremath{J_\mu^{H\,a}}}
\newcommand{\jfdown}{\ensuremath{J_\mu^{F\,a}}}
\newcommand{\jwdown}{\ensuremath{J_\mu^{W\,a}}}

\newcommand{\jhup}{\ensuremath{J_H^{\mu,a}}}
\newcommand{\jfup}{\ensuremath{J_F^{\mu,a}}}
\newcommand{\jwup}{\ensuremath{J_W^{\mu,a}}}

\newcommand{\ohfprime}{\ensuremath{\hat{\mathcal{O}}'_{HF}}}

\newcommand{\orop}{\ensuremath{\hat{\mathcal{O}}_{r}}}
\newcommand{\odop}{\ensuremath{\hat{\mathcal{O}}_{D}}}

\newcommand{\stone}{{\ensuremath{\tilde{t}_{1}}}}
\newcommand{\sttwo}{{\ensuremath{\tilde{t}_{2}}}}

\newcommand{\Qtilde}{\ensuremath{\tilde{Q}}}
\newcommand{\TR}{\ensuremath{\tilde{t}_R}}

\newcommand{\stL}{\ensuremath{\tilde{t}_{L}}}
\newcommand{\stR}{\ensuremath{\tilde{t}_{R}}}
\newcommand{\sbL}{\ensuremath{\tilde{b}_{L}}}
\newcommand{\sbR}{\ensuremath{\tilde{b}_{R}}}

\newcommand{\kLL}{\ensuremath{\kappa_{LL}}}
\newcommand{\kLR}{\ensuremath{\kappa_{LR}}}
\newcommand{\kRR}{\ensuremath{\kappa_{RR}}}

\newcommand{\MLL}{\ensuremath{M_{LL}}}
\newcommand{\MLR}{\ensuremath{M_{LR}}}
\newcommand{\MRR}{\ensuremath{M_{RR}}}
\newcommand{\MRL}{\ensuremath{M_{RL}}}

\newcommand{\ctd}{\ensuremath{\cos^2\theta_{\tilde{t}}}}
\newcommand{\std}{\ensuremath{\sin^2\theta_{\tilde{t}}}}

\newcommand{\ct}{\ensuremath{\cos\theta_{\tilde{t}}}}
\renewcommand{\st}{\ensuremath{\sin\theta_{\tilde{t}}}}

%% file: appendix.tex
\appendix

\section{Models and matching}

\subsection{Operator bases}
\label{sec:ap-eft}

As mentioned in Sec.~\ref{sec:theory_eff}, we here adopt the notation
and conventions of Ref.~\cite{Alloul:2013naa}, which is based on the
SILH framework with the decomposition and
normalization of the Wilson coefficients defined in
Ref.~\cite{silh}. For our purposes, it is enough to single
out the subset that encodes all possible new physics contributions to
the Higgs sector compatible with CP conservation and the flavor
structure of the SM. These are given in Tab.~\ref{tab:operators} and
correspond to the Lagrangian in Eq.~\eqref{eq:EFT}.
 
\begin{table}[b!] 
  \centering
    \renewcommand{\arraystretch}{1.3}
    \begin{tabular}[t]{r @{${}={}$}l} 
      \toprule
      \multicolumn{2}{c}{Higgs fields}  \\
      \midrule
      $\oh$ & $\partial^\mu(\phi^\dagger\,\phi)\,\partial_\mu\,(\phi^\dagger\,\phi)$ \\ 
      $\osix$ & $(\pbp)^3$  \\
      $\oT$ & $(\phi^\dagger\,\overleftrightarrow{D}^\mu\,\phi)\,(\phi^\dagger\,\overleftrightarrow{D}_\mu\,\phi)$ \\
      \bottomrule
      \toprule
      \multicolumn{2}{c}{Higgs and fermion fields} \\
      \midrule
      $\ou$ & $(\pbp)\,(\phi^\dagger\cdot\,\overline{Q}_L)\,u_R$ \\
      $\od$ & $(\pbp)\,(\phi\, \overline{Q}_L)\,d_R$ \\
      $\ol$ & $(\pbp)\,(\phi\, \overline{L}_L)\,l_R$ \\
      \bottomrule 
    \end{tabular}
    \hspace*{1cm}
    \begin{tabular}[t]{r @{${}={}$}l} 
      \toprule
      \multicolumn{2}{c}{Higgs and gauge boson fields} \\
      \midrule
      $\ohb$ & $(D^\mu\phi^\dagger)\,(D^\nu\phi)\,B_{\mu\nu}$ \\
      $\ohw$ & $(D^\mu\phi^\dagger)\,\sigma^k\,(D^\nu\,\phi)\,W^k_{\mu\nu}$ \\ 
      $\og$ & $(\pbp)\,G^A_{\mu\nu}\,G^{\mu\nu\, A}$ \\
      $\ogam$ & $(\pbp)\,B_{\mu\nu}\,B^{\mu\nu}$ \\
      $\ob$ & $(\phi^\dagger\,\overleftrightarrow{D}^\mu\,\phi)\,(\partial^\nu\,B_{\mu\nu})$ \\ 
      $\ow$ & $\left(\phi^\dagger\,\sigma^k\,\overleftrightarrow{D}^\mu\phi\right)\,(D^\nu\,W^k_{\mu\nu})$ \\
      \bottomrule
    \end{tabular}
  \caption{Dimension-6 operators considered in our analysis. These
    correspond to a subset of the most general effective operator basis
    ~\cite{silh} describing new physics effects to the SM
    Higgs sector with CP-invariance and SM-like fermion structures.}
  \label{tab:operators}
\end{table}

The conventions for how covariant derivatives act on the Higgs,
fermion and gauge vector fields are fixed as follows:
\begin{align}
  D_\mu \phi &= \partial_\mu \phi - \dfrac {i g'} 2 B_\mu \phi - i g \dfrac {\sigma^a} 2 \, W^a_\mu \, \phi \,,\notag \\
  D_\mu F_L &= \partial_\mu F_L - i g' \dfrac{Y_{F_L}} 2 B_\mu F_L - i g \dfrac {\sigma^a} 2 \, W^a_\mu \, F_L \,,\notag \\
  D_\mu V_\nu^a &= \partial_\mu V_\nu^a + g \, \varepsilon^{abc} \, W_\mu^b \, V_\nu^c \,,\notag \\
  D_\mu W^a_{\nu \rho} &= \partial_\mu W_{\nu \rho}^a + g \, \varepsilon^{abc} \,W_\mu^b \, W_{\nu\rho}^c \,. 
\label{eq:covariant}
\end{align}
While the effective Lagrangian in Eq.~\eqref{eq:EFT} is written in
terms of the fundamental SM gauge fields, the connection to physics
observables is more easily seen in the mass-eigenstate basis, which we
can write as
\begin{alignat}{5}
 \lag \supset & -\dfrac{m^2_{H}}{2v}\,g^{(1)}_{HHH}\,HHH  + \dfrac{1}{2}\,g_{HHH}^{(2)}\,H(\partial_\mu H)\,(\partial^\mu H) \notag \\
  & - \dfrac{1}{4}\,g_{ggH}\,G^{\mu\nu\,A}\,G_{\mu\nu}^A\,H - \dfrac{1}{4}\,g_{\gamma\gamma H}\,F^{\mu\nu}\,F_{\mu\nu}\,H \notag \\
  & -\dfrac{1}{4}\,g_{Z}^{(1)}\,Z_{\mu\nu}\,Z^{\mu\nu}\,H -g_{Z}^{(2)}\,Z_{\nu}\,\partial_\mu \,Z^{\mu\nu}\,H + \dfrac{1}{2}\,g^{(3)}_{Z}\,Z_\mu Z^{\mu}\,H \notag 
  \\
  & - \dfrac{1}{2}\,g_{W}^{(1)}\,W^{\mu\nu}\,W^\dagger_{\mu\nu}\,H - 
  \left[g_{W}^{(2)} W^\nu\,\partial^\mu W^\dagger_{\mu\nu} H + \text{h.c.} \right] + g_{W}^{(3)}\,m_W\,W^\dagger_\mu W^\mu\, H \notag \\ 
  & -\left[g_{u}\dfrac{1}{\sqrt{2}}\left(\bar{u}P_R u\right)H + g_{d}\dfrac{1}{\sqrt{2}}\left(\bar{d}P_R d\right)H + g_{\ell}\dfrac{1}{\sqrt{2}}\left(\bar{\ell}P_R \ell\right)H + \text{h.c.}
  \right]
 \label{eq:masslag},
\end{alignat}
with the different effective couplings $g_i$ quoted in
Tab.~\ref{tab:coefficients}.  More details on the notation and
conventions can be found in Ref.~\cite{Alloul:2013naa}.

\begin{table}[t] 
  \renewcommand{\arraystretch}{1.5}
  \centering
  \begin{tabular}{r l l}
    \toprule
    Coupling & Operators & Expression \\
    \midrule
    $g^{(1)}_{Z} $ & $\ohb, \ohw, \ohgam$ & $\frac{2g}{m_W\cwd}\left[\bar{c}_{HB}\swd-4\bar{c}_{\gamma}\sw^4+\cwd \,\bar{c}_{HW} \right] $ \\
    $g^{(2)}_{Z} $ & $\ohw, \ohb, \ow, \ob$ & $ \frac{g}{m_W\cwd}\left[(\bar{c}_{HW} + \bar{c}_W)\cwd + (\bar{c}_B + \bar{c}_{HB})\swd\right]$\\  
    $g^{(3)}_{Z} $ & $\oh, \oT, \ogam$ & $\frac{g m_W}{\cw^2}\left[1-\frac{1}{2}\bar{c}_H - 2\bar{c}_T + 8\bar{c}_\gamma\frac{\sw^4}{\cwd} \right]$ \\
    \midrule
    $g^{(1)}_{W} 
    $ & $\ohw$ & $\frac{2g}{m_W}\,\bar{c}_{HW} $ \\
    $g^{(2)}_{W}$ & $\ohw,\ow$ & $\frac{g}{m_W}\,\left[\bar{c}_W + \bar{c}_{HW}\right]$ \\  
    $g^{(3)}_{W}$ & $\oh$ & $g(1-\frac{1}{2}\bar{c}_H)$\\
    \midrule
    $g_{f}$ & $\oh, \of \quad (f = u,d,\ell)$ & $\frac{\sqrt{2}m_f}{v}\left[1-\frac{1}{2}\bar{c}_H + \bar{c}_f\right] $ \\
    \midrule
    $g_{g} $ & $\oh, \og$ & $g_H - \frac{4 \bar{c}_g g_s^2 v}{m_W^2}$ \\
    \midrule
    $g_{\gamma}$ & $\oh, \ogam$ & $a_H - \frac{8 g \bar{c}_\gamma \swd}{m_W}$\\
    \midrule
    $g^{(1)}_{HHH}$  & $\oh, \osix$ & $1+\frac{5}{2} \bar{c}_6 - \frac{1}{2}\bar{c}_{H}$ \\ 
    $g^{(2)}_{HHH} $  & $\oh$ & $\frac{g}{m_W}\,\bar{c}_H$ \\
    \bottomrule
  \end{tabular} 
  \caption{Subset of the dimension-6 operators which enter the different
    leading-order Higgs couplings which are relevant for LHC
    phenomenology, in the notation and conventions of
    Ref.~\cite{Alloul:2013naa} (see text).  The different superscripts
    denote the various terms in the Lagrangian in Eq.~\eqref{eq:masslag} and
    correspond to either a SM-like interaction with a rescaled coupling
    strength or to genuinely new Lorentz structures. The weak coupling
    constant is written as $g \equiv e/\sw$.  The SM contribution to the
    loop-induced Higgs coupling to the gluons (photons) is denoted by
    $g_H$ ($a_H$).}
  \label{tab:coefficients}
\end{table}

Note that the Higgs-fermion coupling shift is given by $g_f \propto
y_f (1 - \bar{c}_H/2  +  3 \bar{c}_f/2)$, but $\of$ also
shifts the fermion masses to $m_f = y_f v (1 + \bar{c}_f/2) /
\sqrt{2}$, yielding the result given above. Similarly, $\oh$ and
$\ogam$ generate additional contributions to the Higgs-boson and
gauge-boson kinetic terms, which are restored to their canonical form
by the field re-definitions
\begin{align}
H &\to H \, \left(1-\dfrac{1}{2}c_H \right), & 
Z_\mu &\to Z_\mu \, \left( 1+\dfrac{4\sw^4}{\cw^2}c_\gamma \right) \,,\notag \\
&& 
A_\mu &\to A_\mu \, \left( 1+4\sw^2 \, c_\gamma \right) -
  Z_\mu \left( \dfrac{8\sw^3}{\cw} c_\gamma \right).
\end{align}
None of the operators considered in this basis affects the relations
between $g$, $m_W$, $v$ and $G_F$, so the SM relations
\begin{align}
  m_W &= \dfrac {g\,v} 2 \,, &
  G_F &=\dfrac {\sqrt{2} \, g^2} {8\, m_W^2} = \dfrac 1 {\sqrt{2} v^2} \,,
\end{align}
can always be used to translate these coupling shifts from one scheme
of input parameters to another.

Dimension-6 operators result in a modified pattern of Higgs
interactions, leading to coupling shifts $g_{xxH} \equiv
g_{xxH}^\text{SM}(1+\Delta_x)$ and also genuinely novel Lorentz
structures. Interestingly, in general more than one of the
effective operators in Tab.~\ref{tab:operators} contributes to a
given Higgs interaction in the mass basis, implying that it is in
general not possible to establish a one-to-one mapping between Wilson
coefficients and distorted Higgs couplings.

Note that the Wilson coefficients of the operators $\oT$ and $\ob+\ow$
are strongly constrained by electroweak precision data~\cite{silh}. In
this work, we allow ourselves, on occasion, to ignore these bounds to
more distinctly illustrate the effects in the Higgs sector.\medskip

Translations between effective operator bases can be performed with
the help of equations of motion, field redefinitions, integration by
parts and Fierz identities. Here we quote a number of such relations
which turn out to be particularly useful for the practitioner. For
example, in addition to the effective operators in the SILH basis,
we often find the operators
\begin{align}
  \orop &= \pbp\, (D_\mu \phi)^2\,, &
  \ohfprime &= \left( \bar{f}_L \, \gamma^\mu \, \sigma^a \, f_L \right) \left( \phi^\dagger \, \sigma^a \, \overleftrightarrow{D}_\mu \, \phi \right)  \,,
  \notag \\
  \odop &= \left( D^2 \, \phi \right)^2\,, &
  \hat{\ord}'_{HH} &= \left( \phi^\dagger \, \sigma^a \, \overleftrightarrow{D}_\mu \, \phi \right)\left( \phi^\dagger \, \sigma^a \, \overleftrightarrow{D}_\mu \, \phi \right)\,.
  \label{eq:EFT_OHf}
\end{align}
$\hat{\ord}'_{HH}$ can be replaced by using the completeness relation of the Pauli matrices, which for arbitrary SU(2) doublets $\xi,\,\chi,\,\eta,\,\psi$ leads to
\begin{align}
(\xi^\dagger \sigma^a \chi)(\eta^\dagger \sigma^a \psi) &=
\sum_{ijkl} \xi^*_i \sigma^a_{ij} \chi_j \, \eta^*_k \sigma^a_{kl} \psi_l \notag \\
&= \sum_{ijkl} (2\delta_{il}\delta_{jk}-\delta_{ij}\delta_{kl})\xi^*_i \chi_j \, \eta^*_k \psi_l = 2(\xi^\dagger\psi)(\eta^\dagger\chi)-(\xi^\dagger \chi)(\eta^\dagger\psi)\,.
\end{align}
Thus we find
\begin{align}
\hat{\ord}'_{HH} &= (\phi^\dagger\,\sigma^a\,D^\mu \phi)^2
+((D^\mu\phi^\dagger)\,\sigma^a\,\phi)^2
-2((D^\mu\phi^\dagger)\,\sigma^a\,\phi)(\phi^\dagger\,\sigma^a\,D_\mu \phi) \notag \\
&= (\phi^\dagger\,D^\mu \phi)^2
+ ((D^\mu\phi^\dagger)\,\phi)^2
-2\left[ 2((D^\mu\phi^\dagger)\,D^\mu \phi)(\phi^\dagger\phi) - 
 ((D^\mu\phi^\dagger)\,\phi)(\phi^\dagger\,D^\mu \phi) \right]
\notag \\
&= \oh - 4 \orop \,.
  \label{eq:EFT_JHJH_replacement}
\end{align}
The equation of motion for the $W$ fields,
\begin{align}
D^\nu W^a_{\mu \nu} 
= - i g \, \phi^\dagger \dfrac {\sigma^a} {2} \overleftrightarrow{D}_\mu \phi 
  - g \sum_f \bar{f}_L \dfrac {\sigma^a} {2} \gamma_\mu f_L  \,,
 \label{eq:eom-w}
\end{align}
gives rise to the identity
\begin{align}
  \sum_f \ohfprime = \dfrac {2} {g} \, \ow - {i} \, \oh + 4 i \, \orop  \,.
  \label{eq:EFT_OHf_replacement}
\end{align}
A global redefinition $\phi \to \phi + \alpha \, (\pbp) \phi / v^2$ generates a shift in the Wilson coefficients
\begin{alignat}{5}
 c_H \to c_H + 2\alpha\,, \quad
 c_r \to c_r + 2\alpha\,, \quad c_6 \to c_6 {+} 4\alpha\,, \quad c_f \to c_f {+} \alpha\,,
 \label{eq:wilson-shift}
\end{alignat}
so that
with the choice $\alpha=-c_r/2$ one can eliminate the operator
$\ope{r}$ in favor of other operators:
\begin{align}
  \orop \leftrightarrow \biggl\{ - \dfrac 1 2 \, \oh + 2 \, \lambda \, \osix + \sum_f \left[ \dfrac 1 2 \, y_f \, \of + \text{h.c.}\right] \biggr\} \,.
  \label{eq:EFT_Or_replacement}
\end{align}
Finally, $\odop$ can be exchanged for others using the equation of motion for
$\phi$,
\begin{align}
  D^2 \phi &= - \mu^2  \, \phi - 2\, \lambda \, \pbp \, \phi - \sum_\text{gen.} \left[y_u \, \bar{Q}^T_L \, u_R + y_d \, \bar{d}_R \, Q_L + y_\ell \, \bar{\ell}_R \, L_L \right] \,. 
\end{align}
This leads to
\begin{align}
  \odop = \mu^4 \, \pbp 
 + 4 \, \lambda \, \mu^2 \, (\pbp)^2 
 + \mu^2 \, \sum_f y_f \, \bar{f}_L \phi f_R  \
 + 4 \, \lambda^2 \, (\pbp)^3 + 2 \, \lambda \, \sum_f y_f \, \pbp \, \left( \bar{f}_L \phi f_R \right) \,.
\end{align}
The first three terms lead to a renormalization of the SM parameters
$\mu$, $\lambda$, $y_f$, without any impact on physical
observables. The last two terms, however, means that $\ope{D}$ is
equivalent to the combination
\begin{align}
  \odop \leftrightarrow 4 \, \lambda^2 \, \osix + 2 \, \lambda \, \sum_f \left( y_f \, \of + \text{h.c.} \right)\,.
  \label{eq:EFT_OD_replacement}
\end{align}

\subsubsection*{HLM basis}

Aside from the relatively simple case of the multi-Higgs sector
extensions, we make use of the covariant derivative
expansion~\cite{Gaillard:1985uh,Cheyette:1987qz} to analytically carry out
the matching between the different UV completions to
their corresponding EFT description.  The method has been recently
reappraised in Ref.~\cite{Henning:2014wua} and employed in a number of
studies~\cite{heft_limitations,heft_limitations2,Chiang:2015ura,Huo:2015exa}.
By applying this method, the Wilson coefficients are readily obtained
in a different operator basis (henceforth dubbed HLM),
\begin{align}
  \lag_\text{HLM} = \sum_i \dfrac{k_i} {\Lambda^2} \ope{i}''.
\end{align}
The HLM operators involving Higgs fields and their interaction with
gauge bosons are listed in Tab.~\ref{tab:ops2}.  In addition, the HLM
basis contains a subset of operators with no direct correspondence to
the bosonic SILH operators, which must be rewritten with the help of
equations of motion and field redefinitions, as we discuss below.

\begin{table}[tb] 
  \renewcommand{\arraystretch}{1.3}
  \setlength{\tabcolsep}{1ex}
  \centering
    \begin{tabular}[t]{r @{${}={}$}l} 
    \toprule
    \multicolumn{2}{c}{HLM basis}  \\
    \midrule
    $\ope{H}''$ & $\frac{1}{2} \, \partial^\mu (\pbp) \,\partial_\mu (\pbp)$ \\ 
    $\ope{6}''$ & $(\pbp)^3$  \\
    $\ope{T}''$ & $\frac{1}{2} \left(\phi^\dagger \, \overleftrightarrow{D}^\mu \, \phi \right) \, \left(\phi^\dagger \, \overleftrightarrow{D}_\mu \, \phi \right)$ \\
    $\ope{B}''$ & $\frac{i g'}{2} \left( \phi^\dagger \, \overleftrightarrow{D}^\mu \, \phi \right) \, \partial^\nu \, B_{\mu\nu}$ \\
    $\ope{W}''$ & $\frac{i g}{2} \left(\phi^\dagger\,\sigma^k\,\overleftrightarrow{D}^\mu\phi\right)\,(D^\nu\,W^k_{\mu\nu})$ \\
    $\ope{GG}''$ & $g_s^2 (\pbp)\,G^A_{\mu\nu}\,G^{\mu\nu\, A}$ \\ 
    $\ope{BB}''$ & $g'^2(\pbp)\,B_{\mu\nu}\,B^{\mu\nu}$ \\
    $\ope{WW}''$ & $g^2 (\pbp)\,W^k_{\mu\nu}\,W^{\mu\nu\, k}$ \\
    $\ope{WB}''$ & $g g' \left(\phi^\dagger\sigma^k\phi \right)  \, B_{\mu\nu}\,W^{\mu\nu\, k}$ \\
    \bottomrule 
  \end{tabular}
  \hspace{1cm}
  \begin{tabular}[t]{r @{${}={}$}l} 
    \toprule
    \multicolumn{2}{c}{HISZ basis} \\
    \midrule
    $\ope{\phi1}'$  & $(D_\mu\phi)^\dagger\phi\,\phi^\dagger(D^\mu\phi)$ \\
    $\ope{\phi2}'$  & $\frac{1}{2}\partial^\mu(\phi^\dagger\phi)\,\partial_\mu(\phi^\dagger\phi)$ \\ 
    $\ope{\phi3}'$  &  $\frac{1}{3}(\pbp)^3$ \\
    $\ope{GG}'$  &  $(\pbp)\,G^A_{\mu\nu}\,G^{\mu\nu\, A}$ \\
    $\ope{BB}'$  &  $\phi^\dagger\,\hat{B}_{\mu\nu}\,\hat{B}^{\mu\nu}\,\phi  = -\frac{g'^2}{4}\pbp\,B_{\mu\nu}\,B^{\mu\nu}$ \\
    $\ope{WW}'$  &  $\phi^\dagger\,\hat{W}_{\mu\nu}\,\hat{W}^{\mu\nu}\,\phi  = -\frac{g^2}{4}\pbp\,W^k_{\mu\nu}\,W^{\mu\nu\, k}$ \\
    $\ope{BW}'$  &  $\phi^\dagger\,\hat{B}_{\mu\nu}\,\hat{W}^{\mu\nu}\,\phi  = -\frac{g\,g'}{4}(\phi^\dagger\sigma^k\phi)\,B_{\mu\nu}\,W^{\mu\nu\, k}$ \\
    $\ope{B}'$  &  $(D^\mu\phi)^\dagger \hat{B}_{\mu\nu} (D^\nu\phi)  = i \frac{g}{2}(D^\mu\phi^\dagger)(D^\nu\phi)\,B_{\mu\nu}$ \\
    $\ope{W}'$  &  $(D^\mu\phi)^\dagger \hat{W}_{\mu\nu} (D^\nu\phi)  = i \frac{g}{2}(D^\mu\phi^\dagger)\sigma^k( D^\nu\phi)\,W_{\mu\nu}^k$ \\
    \bottomrule
  \end{tabular}
  \caption{Bosonic CP-conserving Higgs operators in the HLM basis
    (left) and the HISZ basis (right). Here $\hat{B}_{\mu\nu}=i g'/2
    B_{\mu\nu}$ and $\hat{W}_{\mu\nu}=i g \sigma^k/2 W_{\mu\nu}^k$.}
  \label{tab:ops2}
\end{table}

The operators in Tab.~\ref{tab:ops2} translate to the SILH basis via
\begin{align}
\begin{aligned}
  \ope{H}'' &= \frac 1 2 \oh\,, \quad &
  \ope{6}'' &= \ope{6}\,, \quad &
  \ope{T}'' &= \frac 1 2 \oT\,, \quad &
  \ope{B}'' &= \frac {i g'} 2 \ob \,, \quad &
  \ope{W}'' &= \frac {i g} 2 \ow\,,  \\
  \ope{GG}'' &= g_s^2 \og\,, &
  \ope{BB}'' &= g'^2 \ogam\,, \quad &
  \ope{WB}'' &= 2i g' \ob - 4i g' \ohb - g'^2 \ogam\,, \hspace{-10em} \\
  \ope{WW}'' &= -2i g' \ob + 2i g \ow + 4i g' \ohb - 4 i g \ohw  + g'^2 \ogam  \, . \hspace{-20em}
\end{aligned}
\end{align}
In addition, the HLM basis contains extra operators with no SILH counterpart, 
\begin{align}
  \ope{R}'' = \pbp \left(D_\mu \, \phi\right)^\dagger \left(D^\mu \, \phi \right)\,, \qqquad
  \ope{D}'' = \left( D^2 \phi \right)^2 \, ,
  \label{eq:HLM_add_ops}
\end{align}
which can be eliminated using Eq.~\eqref{eq:EFT_Or_replacement}
and Eq.~\eqref{eq:EFT_OD_replacement}, respectively.  The Wilson
coefficients $k_i$ of the HLM basis translate to the SILH coefficients
$\bar{c}_i$ as follows:
\begin{align}
  \bar{c}_H &= \dfrac {v^2} {\Lambda^2} \, \left( k_H - k_R \right) \,, &
  \bar{c}_B &= \dfrac {v^2} {\Lambda^2} \, \dfrac {g^2} 4 \, \left(k_B + 4 \, k_{WB} - 4 \, k_{WW} \right) \notag \,, \\
  \bar{c}_T &= \dfrac {v^2} {\Lambda^2} \, k_T \,, & 
  \bar{c}_W &= \dfrac {v^2} {\Lambda^2} \, \dfrac {g^2} 4 \, \left(k_W + 4  \, k_{WW} \right) \notag \,, \\
  \bar{c}_6 &= - \dfrac {v^2} {\Lambda^2} \, \left( \dfrac {k_6} {\lambda} + 2\, k_R + 4 \, \lambda \, k_D \right) \,, & 
  \bar{c}_{HB} &= \dfrac {v^2} {\Lambda^2} \, g^2 \, \left(  k_{WW} - k_{WB} \right) \notag \,, \\
  \bar{c}_g &= \dfrac {v^2} {\Lambda^2} \, \dfrac {g^2} 4 \, k_{GG} \,,  &   
  \bar{c}_{HW} &= - \dfrac {v^2} {\Lambda^2} \, g^2 \, k_{WW} \notag \,, \\
  \bar{c}_\gamma &= \dfrac {v^2} {\Lambda^2} \, \dfrac {g^2} 4 \, \left( k_{BB} - k_{WB} + k_{WW} \right) \,, &
  \bar{c}_f &= - \dfrac {v^2} {\Lambda^2} \, \left( \dfrac 1 2 \, k_R + 2 \, \lambda \, k_D \right) \,,
\end{align}
where for the sake of completeness we have included the coefficients
of the redundant operators given in Eq.~\eqref{eq:HLM_add_ops}.

\subsubsection*{HISZ basis}

We also give the conversion to the popular HISZ
basis~\cite{Hagiwara:1993ck} (see also
Refs.~\cite{Corbett:2012ja,sfitter_last} for recent studies in this
framework)
\begin{align}
  \lag_\text{HISZ} = \sum_i \dfrac{f_i} {\Lambda^2} \ope{i}' \, ,
\end{align}
with Higgs-gauge operators given in Tab.~\ref{tab:ops2}.  We use the
same conventions for the covariant derivative as above (note that this
is not the case in some of the cited literature). The operators can
then be translated via the relations
\begin{align}
\oh  &= 2\ope{\phi2}'\,, \qquad &
\ow  &= \dfrac{2 i}{g}  \left( \ope{WW}' +  \ope{BW}' - 2\ope{W}'   \right)\,,  \quad & 
\ohw &= -\dfrac{2i }{g} \ope{W}'\,,  \notag \\ 
\oT  &= 2\ope{\phi2}'  -  4\ope{\phi1}'\,,  &
\ob  &= \dfrac{2 i}{g'}  \left( \ope{BB}' +  \ope{BW}' - 2\ope{B}' \right)\,, & 
\og  &= \ope{GG}' \,,  \notag \\
\osix &= 3 \ope{\phi3}' \,, & 
\ohb  &= -\dfrac{2 i }{g'}\ope{B}'\,, & 
\ogam &= -\dfrac{4}{g'^2}  \ope{BB}'  \,.
\end{align}
The HISZ basis also includes the redundant operator $\ope{\phi4}' =
(D_\mu\phi)^\dagger(D^\mu\phi)\,\phi^\dagger\phi$, which can be
removed using Eq.~\eqref{eq:EFT_Or_replacement}.  For the coefficients,
we find
\begin{align}
  \bar{c}_H &= \dfrac{v^2}{\Lambda^2}\,\left(\dfrac{1}{2}f_{\phi1}+f_{\phi2}\right)\,, &
  \bar{c}_W &= -\dfrac{v^2}{\Lambda^2}\,\dfrac{g^2}{4}f_{WW} \notag\,, \\
  \bar{c}_T &= -\dfrac{v^2}{\Lambda^2}\,\dfrac{1}{2}f_{\phi1}\,, & 
  \bar{c}_B &= \dfrac{v^2}{\Lambda^2}\,\dfrac{g^2}{4} (f_{WW}-f_{BW}) \notag\,, \\
  \bar{c}_6 &= -\dfrac{v^2}{\Lambda^2}\,\dfrac{1}{3\lambda}f_{\phi3}\,, &
  \bar{c}_{HW} &= \dfrac{v^2}{\Lambda^2}\,\dfrac{g^2}{8} (f_{W}+2f_{WW}) \notag\,, \\
  \bar{c}_g &= \dfrac{v^2}{\Lambda^2}\,\dfrac{g^2}{4g_s^2}f_{GG}\,, &
  \bar{c}_{HB} &= \dfrac{v^2}{\Lambda^2}\,\dfrac{g^2}{8} (f_{B}+2f_{BW}-2f_{WW}) \notag\,, \\
  \bar{c}_\gamma &= \dfrac{v^2}{\Lambda^2}\,\dfrac{g^2}{16} (f_{BW}-f_{BB}-f_{WW}) \, .
\end{align}

\subsection{Singlet extension}
\label{sec:ap-singlet}

For the sake of simplicity we consider a minimal version of the
singlet model, in which a discrete $\mathbb{Z}_2$ parity precludes
additional (\eg cubic) terms in the potential. The SM is then extended
by including a real scalar singlet with the Lagrangian
\begin{align}
\lag &= (D_\mu\phi)^\dagger\,(D^\mu\,\phi) + (\partial_\mu\,S)^2 - V(\phi,S) \,,\notag \\
V(\phi,S) &= 
  \mu^2_1\,(\phi^\dagger\,\phi) 
+ \lambda_1\,|\phi^{\dagger}\phi|^2 
+ \mu^2_2\,S^2
+ \lambda_2\,S^4 
+ \lambda_3\,|\phi^{\dagger}\,\phi|S^2 \, .
\label{eq:singlet-potential-app}
\end{align}
The scalar doublet and singlets fields are expanded into components as
\begin{align}
\phi = \begin{pmatrix} G^+ \\[1mm] \dfrac{1}{\sqrt{2}} (v + l^0 + i G^0)
         \end{pmatrix} 
\qquad \text{and} \qquad S = \cfrac{1}{\sqrt{2}} (v_s + s^0) \, ,
\label{eq:fields-def}
\end{align}
where $v \equiv \sqrt{2}\langle \phi \rangle = 246$~GeV and
$v_s \equiv \sqrt{2}\langle S \rangle$ denote their respective
VEVs. The minimization condition for the potential of
Eq.~\eqref{eq:singlet-potential-app} can be used to eliminate the
parameters $\mu_{1,2}$ in favor of $v$ and $v_s$. The CP-even
components $l^0$ and $s^0$ mix to form a light ($h$) and a heavy
($H$) mass eigenstate,
\begin{align}
h &=l^0  \cos\alpha  -s^0 \sin\alpha\,, \nonumber \\[-2ex]
 H &= l^0 \sin\alpha  +s^0 \cos\alpha\,, \qquad \text{where} \quad 
 \tan(2\alpha) = \cfrac{\lambda_3vv_s}{\lambda_2 v_s^2 - \lambda_1v^2}\,.
\label{eq:masseigen-repeat}
\end{align}
Their masses are
\begin{alignat}{5}
 m^2_{h,H} &= \lambda_1\,v^2 + 
 \lambda_2\,v_s^2 \mp |\lambda_1\,v^2 - 
 \lambda_2\,v_s^2|\,\sqrt{1+\tan^2(2\alpha)}
\label{eq:masseigen}
\end{alignat}
with $m_{H}^2 \approx 2\lambda_2 v_s^2 \gg m_{h}^2$ in the
limit $v^2 \ll v_s^2$.\medskip

To perform the matching to the EFT,
we identify the UV scale $\Lambda \equiv \sqrt{2 \lambda_2} v_s \approx m_H$ for $v_s \gg v$. From the
singlet-doublet mixing one then finds a universal coupling shift of
the SM-like light Higgs to all other SM particles in
Eq.~\eqref{eq:shift2}, given by
\begin{alignat}{5}
 \Delta \approx -\frac{\sin^2 \alpha}{2} \approx
  -\dfrac{g_\text{eff}^2}{2} \,\biggl(\cfrac{v}{\Lambda} \biggr)^2 \,, \qquad
  g_\text{eff} = \dfrac{\lambda_3}{\sqrt{2\lambda_2}} \, .
 \label{eq:singlet-delta4}
\end{alignat}
Integrating out the heavy Higgs boson we find
\begin{align}
  \lageff \supset \dfrac {\sin^2 \alpha} {2v^2} \; 
   \partial^\mu (\phi^\dagger \phi) \partial_\mu (\phi^\dagger \phi) +
   \ord(\Lambda^{-4}) \,.
\label{eq:singlet-matching}
\end{align}
We thus see that, up to dimension-6 operators the
heavy-singlet--induced BSM effects in Higgs production and decay are
completely captured by the operator $\oh$
(cf.\ Tab.~\ref{tab:operators}) with coefficient
\begin{align}
  \bar{c}_H = \dfrac{\lambda_3^2}{2\lambda_2} \, \left(\dfrac {v} {\Lambda}\right)^2 + \ord\biggl(\frac{v^4}{\Lambda^4}\biggr) \,.
\end{align}
\medskip

The light Higgs couplings to fermions and gauge bosons in the singlet model 
are universally suppressed relative to the SM. In the full model and the
EFT, respectively, they are given by
\begin{align}
  1 + \Delta_x =  \cos \alpha \,,
\qqquad
  1 + \Delta_x^\text{EFT} = 1-\dfrac{1}{2}\bar{c}_H \,.
\end{align}

A more complex pattern emerges for the
self-interactions involving at least one heavy Higgs field.  We find
\begin{alignat}{5}
g_{hhH}
&= - \cfrac{g_\text{eff} \,(2m^2_{h} + m^2_{H})}{v_s}\,
      \left[ 1+ g_\text{eff}\,\cfrac{v^2}{v_s^2} + \ord\biggl(
  \frac{v^3}{v_s^3}\biggr) \right] 
\sim \lambda_3 v_s + \ord(v) \,,
\notag \\ 
g_{hHH} 
&=  \cfrac{g_\text{eff} v \,(m^2_{h} + 2 m^2_{H})}{v_s^2}\, 
   \left[ 1- g_\text{eff} + \ord\biggl( \frac{v}{v_s}\biggr) \right] 
\sim 2\lambda_3v \left( 1- \dfrac{\lambda_3}{2\lambda_2} \right) +
\ord\left(\dfrac{v^2}{v_s}\right) \, ,
 \label{eq:singlet-tripleheavy}
\end{alignat}
in which we observe a characteristic non-decoupling behavior which
manifests itself as a linear growth of $g_{hhH}$ with the heavy
Higgs mass.  In the EFT, the leading self-interaction contribution enters
via a dimension-8 operator, which is
neglected in our dimension-6
analysis. Therefore, the sole Wilson coefficient $\bar{c}_H = \sin^2 \alpha$ defines
the singlet model EFT up to dimension 6.\medskip

On the other
hand, let us emphasize a key structural difference between the
$\oh$-induced and the UV-complete singlet model contributions to the
Higgs self-coupling $hhh$. At variance with the latter, the effective
operators also induces a new momentum structure into the self
coupling, namely adding derivatives in the Lagrangian or energy
dependent terms in the Feynman rules
\begin{align}
 \lag \supset
&- \dfrac{m_{h}^2}{2v}\left[
   \left(1-\dfrac{c_H v^2}{2\Lambda^2} \right) h^3
   -\dfrac{2c_H v^2}{\Lambda^2 {m_{h}^2}} 
    h \, \partial_\mu h \, \partial^\mu h \right] \notag \\
&= - \dfrac{m_{h}^2} {2v} 
    \left( 1 - \dfrac{1}{2} \bar{c}_H \right) 
    h^3 
   + \dfrac {g}{2 m_W} \bar{c}_H \; h \partial_\mu h \partial^\mu h 
\label{eq:singlet-self},
\end{align}
which means that the SM-like $h^3$ term is not only rescaled but
also endowed with new Lorentz structures involving derivatives.  This
kind of momentum dependence is encoded in the split into
$g^{(1)}_{HHH}$ and $g^{(2)}_{HHH}$ in Eq.~\eqref{eq:masslag}.  This
effect does not correspond to the Higgs singlet mixing, where such a
momentum dependence can only be generated via loop-induced heavy
particle exchange with momentum-dependent couplings like a heavy
fermion triangle.

\subsection{Two-Higgs-doublet model}
\label{sec:ap-2hdm}
 
The most general gauge invariant, CP-conserving potential with two
scalar fields reads
\begin{alignat}{5}
 V(\phi_1,\phi_2) 
&= m^2_{11}\,\phi_1^\dagger\phi_1
 + m^2_{22}\,\phi_2^\dagger\phi_2
 - \left[ m^2_{12}\,\phi_1^\dagger\phi_2 + \text{h.c.} \right] \notag \\
&+ \dfrac{\lambda_1}{2} \, (\phi_1^\dagger\phi_1)^2
 + \dfrac{\lambda_2}{2} \, (\phi_2^\dagger\phi_2)^2
 + \lambda_3 \, (\phi_1^\dagger\phi_1)\,(\phi_2^\dagger\phi_2) 
 + \lambda_4 \, |\phi_1^\dagger\,\phi_2|^2 \notag \\
&+ \left[ \dfrac{\lambda_5}{2} \, (\phi_1^\dagger\phi_2)^2 
        + \lambda_6 \, (\phi_1^\dagger\phi_1) \, (\phi_1^\dagger\phi_2)
        + \lambda_7 \, (\phi_2^\dagger\phi_2)\,(\phi_1^\dagger\phi_2) + \text{h.c.} 
   \right] \, ,
\label{eq:2hdmpotential-app}
\end{alignat}
where the mass terms $m^2_{ij}$ and the dimensionless self-couplings
$\lambda_i$ are real parameters and $v_j = \sqrt{2}
\langle \phi_j^0 \rangle$.  The ratio of VEVs is denoted as $\tan\beta =
v_2/v_1$, whereas $v_1^2 + v_2^2 = v^2 = (246~\gev)^2$ to reproduce
the known gauge boson masses.  For the Yukawa couplings, there are
four possible scenarios that satisfy the SM flavor symmetry and
preclude tree-level flavor-changing neutral
currents~\cite{Glashow:1976nt}:
\begin{itemize}
\item type-I, where all fermions couple to just one Higgs doublet
  $\phi_2$;
\item type-II, where up-type (down-type) fermions couple
  exclusively to $\phi_2$ ($\phi_1$);
\item lepton-specific, with a type-I quark sector and a type-II
  lepton sector; and
\item flipped, with a type-II quark sector and a type-I lepton
  sector.
\end{itemize}
In all four cases, the absence of tree-level FCNCs is protected by a
global $\mathbb{Z}_2$ discrete symmetry $\phi_i \to (-1)^{i}\,\phi_i$ (for $i=1,2$). The
symmetry demands that $\lambda_{6,7}=0$ in
Eq.~\eqref{eq:2hdmpotential-app}, but it can be softly broken by
dimension-two terms in the Lagrangian, viz.\ $\lag_\text{soft}
\supset m^2_{12}\,\phi_1^\dagger\,\phi_2 + \text{h.c.}$\medskip

The Higgs mass-eigenstates follow from the set of rotations
\begin{align}
\begin{pmatrix} H^0 \\ h^0 \end{pmatrix} = R(\alpha)\,\begin{pmatrix} h^0_1 \\ h^0_2 \end{pmatrix},
\qquad
\begin{pmatrix} G^0 \\ A^0 \end{pmatrix} = R(\beta)\,\begin{pmatrix} a^0_1 \\ a^0_2 \end{pmatrix},
\qquad
\begin{pmatrix} G^\pm \\ H^\pm \end{pmatrix} = R(\beta)\,\begin{pmatrix} h^\pm_1 \\ h^\pm_2 \end{pmatrix} ,
\end{align}
where
\begin{align}
\phi_k &= \begin{pmatrix} h_k^+ \\[1mm] \dfrac{1}{\sqrt{2}} (v_k + h_k^0 + i a_k)
         \end{pmatrix}, &
R(\theta) = \begin{pmatrix} \cos\theta & \sin\theta \\
			-\sin\theta & \cos\theta \end{pmatrix}.
\end{align}
Since the two doublets contribute to giving masses to the weak gauge
bosons, custodial symmetry will impose tight constraints on the viable
mass spectrum of the model~\cite{Veltman:1976rt,oblique-2hdm}.
Analytic relations linking the different Higgs masses and mixing
angles with the Lagrangian parameters in Eq.~\eqref{eq:2hdmpotential-app}
can be found \eg in Appendix A of~\cite{Lopez-Val:2013yba}.
The conventions $0 <\beta < \pi/2 $ 
and $0 \leq \beta-\alpha
< \pi$ guarantee that the Higgs coupling to
vector bosons has the same sign in the 2HDM and in the SM.
As we will next show, 
the decoupling limit implies that the light Higgs
interactions approach the alignment limit, where $\cos\beta \sim
|\sin\alpha|$ and the couplings become SM-like~\cite{Gunion:2002zf}.
\medskip

A 2HDM with large mass hierarchy between the light Higgs $m_{h^0} =
\ord(v)$ and its heavier companions $m_{H^0,H^{\pm},A^0} \gg
m_{h^0}$ can be readily mapped onto an
EFT~\cite{Gunion:2002zf,eft2hdm,heft_limitations2}. In the unbroken phase, we
match by first rotating $\phi_1$ and $\phi_2$ into
the so-called Higgs basis, where only one Higgs doublet obtains a vacuum
expectation value, $\langle \phi_l \rangle = v/\sqrt{2}$, $\langle \phi_h
\rangle = 0$~\cite{heft_limitations2,Davidson:2005cw}. This doublet $\phi_l$ is then identified
with the SM-like Higgs doublet, while the other doublet $\phi_h$ is integrated out.
Its decoupling is described by the mass scale
\begin{align}
\Lambda^2 = \mheavy^2 = m^2_{11}\sin^2\beta + m^2_{22}\cos^2\beta + m^2_{12}
\sin (2\beta) \label{eq:2hdm-lambda}
\end{align}
and the expansion parameter
\begin{alignat}{5}
x \equiv \dfrac{v^2\sin 2\beta}{2\mheavy^2}
\left[\frac {\lambda_1} 2-\frac {\lambda_2} 2+\left(\frac {\lambda_1} 2+\frac
{\lambda_2} 2-\lambda_3-\lambda_4-\lambda_5 \right)\cos 2\beta \right] + {\cal
O}\left( \dfrac{v^4}{\mheavy^4} \right) \ll 1 \; 
\label{eq:2hdm-xi-unbroken}
\end{alignat}
where we assume perturbative couplings, $\lambda_i \lesssim
\ord(1)$.

As discussed in Section~\ref{sec:2hdm}, the dimension-6 EFT
defined this way does not provide a good approximation for scenarios where the
LHC will have sensitivity to discover new physics. A more appropriate
effective theory is obtained by matching at a physical mass instead
of $\mheavy$. Specifically, this $v$-improved EFT is given by replacing
$\mheavy \to m_{A^0}$ in Eqs.~\eqref{eq:2hdm-lambda} and \eqref{eq:2hdm-xi-unbroken}.

Similar to the singlet extension, see Eq.~\eqref{eq:singlet-matching},
mixing between the two CP-even Higgs boson at tree-level causes the
$h^0$ kinetic term to be rescaled, leading to
\begin{align}
\bar{c}_H = x^2 = {\cal O}(\Lambda^{-4}).
\end{align}
This corresponds to a dimension-8 term, which we neglect here.
However, there exists a dimension-6 contribution to the triple light Higgs scalar
interaction,
\begin{alignat}{5}
g^{(1)}_{h^0h^0h^0} &= 1 + x^2\left[\dfrac{3}{2} - 
\frac{4m_{12}^2}{m^2_{h^0}\sin 2\beta} \right] + \ord(x^3)
=  1 - x^2 \frac{\mheavy^2}{\lambda v^2} + \ord(\mheavy^{-3})
\, .
 \label{eq:2hdm-hhh}
\end{alignat}
Non-trivial contributions to dimension-6 operators also arise in the
Yukawa sector.  For definiteness, we concentrate on 2HDM type I and
II.  At tree-level and up to $\ord (\Lambda^{-2})$, we find for the
Wilson coefficients
\begin{align}
\text{type I\phantom{I}:} && \bar{c}_u &= x\cot\beta\,, &
                  \bar{c}_d &= x\cot\beta\,, &
                  \bar{c}_\ell &= x\cot\beta\,, \\
\text{type II:}&& \bar{c}_u &= x\cot\beta\,, &
                  \bar{c}_d &= -x\tan\beta\,, &
                  \bar{c}_\ell &= -x\tan\beta\,. 
\end{align}
The above expressions hold both in the standard EFT and the
$v$-improved EFT, with the obvious replacement $\mheavy \to m_{A^0}$ for the
latter.
The operators $\ohb$, $\ohw$, $\ow$, $\ob$, $\oT$ and $\ogam$ receive
contributions only at loop-level, while $\og=0$ since there are no new
colored particles in the 2HDM. 
The operator $\ogam$ receives a
correction from the charged Higgs loop.  Expanding this contribution,
and using $m_{h^0}^2 / m_{H^\pm}^2 = \ord(x)$, we find
\begin{align}
  \Delta_\gamma 
 = \dfrac 1 {g_{H \gamma \gamma}^\text{SM}} \;  \dfrac {e^2} {720 \, \pi^2 \, v } 
 \Bigg[& 30 \left(1 - [\cot \beta + \tan \beta] \dfrac {m_{12}^2} {m_{H^\pm}^2}  \right) 
        + \left( 19 - 4  [\cot \beta + \tan \beta] \dfrac {m_{12}^2} {m_{H^\pm}^2} \right) \dfrac {m_{h^0}^2} {m_{H^\pm}^2} \notag \\
&       - 30 \cot (2 \beta)   [\cot \beta + \tan \beta]  \dfrac {m_{12}^2} {m_{H^\pm}^2} \; x 
 \Bigg] + \ord (x^2)\,,
\end{align}
where in the first row we identify characteristic non-decoupling terms
contributing to $\ord (x^0)$.  On the other hand, the operator
\begin{align}
  \lageff \supset \dfrac{g'^2 \bar{c}_\gamma } {m_W^2}  (\phi^\dagger \phi )  B_{\mu \nu} B^{\mu\nu}
\end{align}
leads to
\begin{align}
  \Delta^\text{EFT}_\gamma = \dfrac 1 {g_{H \gamma \gamma}^\text{SM}} \; \dfrac{16 \; \sw^2 \; \bar{c}_\gamma}{v} \,. 
\end{align}
Identifying these expressions, we find within the $v$-improved EFT framework
\begin{align}
  \bar{c}_\gamma = \frac {g^2} {11\,520 \, \pi^2  } \Bigg[ 
& 30 \left(1 - [\cot \beta + \tan \beta] \frac {m_{12}^2} {m_\Hpm^2}  \right) \notag \\
 &\!+  \left(19 - 4  [\cot \beta + \tan \beta] \frac {m_{12}^2} {m_\Hpm^2} \right) \frac {m_{h^0}^2} {m_\Hpm^2} - 30 \cot (2 \beta)   [\cot \beta + \tan \beta]  \frac {m_{12}^2} {m_\Hpm^2} \; x 
   \Bigg] \,.
   \label{eq:THDM_cgamma}
\end{align}
\medskip

In the full type-I 2HDM, the  tree-level couplings shifts $g^\text{2HDM}_{h^0 xx}/g^\text{SM}_{hxx} = 1+\Delta_x$  of the light Higgs are given by
\begin{align}
  1+\Delta_V &= \sin(\beta - \alpha)\,, \quad
  1+\Delta_t = \dfrac {\cos \alpha} {\sin \beta}\,, \quad
  1+\Delta_b = \dfrac {\cos \alpha} {\sin \beta}\,, \quad
  1+\Delta_\tau = \dfrac {\cos \alpha} {\sin \beta}\,,
\end{align}
while in the type-II 2HDM they read
\begin{align}
  1+\Delta_V &= \sin(\beta - \alpha)\,, \quad
  1+\Delta_t = \dfrac {\cos \alpha} {\sin \beta}\,, \quad
  1+\Delta_b = - \dfrac {\sin \alpha} {\cos \beta}\,, \quad
  1+\Delta_\tau = - \dfrac {\sin \alpha} {\cos \beta}\,, 
\end{align}
The light Higgs coupling to a charged Higgs pair is given in all cases
by 
\begin{align}
   \frac {g_{h^0\Hp\Hm}}{g^\text{SM}_{hhh}} &= \dfrac{1}{3 m_{h^0}^2} \left[ 
    \sin (\beta - \alpha) \left( 2 m_{H^\pm}^2 - m_{h^0}^2 \right) + \dfrac
    {\cos (\alpha + \beta)} {\sin (2\beta)} \left(2m_{h^0}^2 - \dfrac {2
    m_{12}^2}{\sin \beta \cos \beta} \right) \right] \, ,
\end{align}
with $g^\text{SM}_{hhh} = -3 m_h^2/v$.
Note that at tree level custodial symmetry ensures that both couplings
to the weak gauge bosons $V=W,Z$ scale with the same factor $\sin
(\beta - \alpha)$, a degeneracy that can be mildly broken by quantum
effects~\cite{Lopez-Val:2013yba}.  \medskip

In the effective model, we have\footnote{Note that the operator $\ogam$ introduces a new Lorentz
  structure for the $h^0 VV$ interaction, representing a charged Higgs
  loop.  The results in Section~\ref{sec:2hdm} reveal how large this
  effect turns out to be in practice.}
\begin{align}
  \Delta^\text{EFT}_V &= 0 \,, &
  \Delta^\text{EFT}_t &= \bar{c}_u \,, &
  \Delta^\text{EFT}_b &= \bar{c}_d \,, &
  \Delta^\text{EFT}_\tau &= \bar{c}_\ell \,.
\end{align}
The loop-induced couplings are more involved, giving
\begin{align}
1+\Delta_g &= \dfrac{1}{A_{gg}^\text{SM}} 
  \Bigg[\sum_{f = t,b}\,(1+\Delta_f)\,A_f(\tau_f)
  \Bigg]\,, \\
1+\Delta_\gamma &= \dfrac{1}{A_{\gamma \gamma}^\text{SM}} 
  \Bigg[ \sum_{f = t,b}\, N_C\,Q_f^2\,(1+\Delta_f)\,A_f(\tau_f) + Q^2_{\tau}\,(1+\Delta_\tau)\,A_{f}(\tau_\tau) + (1+\Delta_W)\, A_v(\tau_W)   \notag \\
 &\quad \quad \quad \quad \qquad \qquad - {g}_{h^0 \Hp \Hm} \; \dfrac {m_W \sw} {e m_{H^\pm}^2} \; A_s(\tau_{H^{\pm}}) \Bigg] \,,
\end{align}
where $A_{xx}^\text{SM}$ are the corresponding contributions in the SM.
The conventional loop form factors read 
\begin{alignat}{5}
 A_s(\tau) & = -\frac{\tau}{2}\,\left[1-\tau f(\tau) \right] = 1/6 + \mathcal{O}(\tau^{-1}), \notag \\
 A_{f}(\tau) &= \tau\left[1+(1-\tau)\,f(\tau) \right] = 2/3 + \mathcal{O}(\tau^{-1}), \notag \\
 A_v(\tau) &= -\frac{1}{2}\,\left[2+3\tau+3(2\tau-\tau^2)\,f(\tau) \right] = -7/2 + \mathcal{O}(\tau^{-1}) 
 \label{eq:loopfunctions},
\end{alignat}
\begin{align}
 f(\tau) =
\begin{cases}
  - \dfrac 1 4 \left[ \log \dfrac{1 + \sqrt{1- \tau}} {1 - \sqrt{1 - \tau}} - i \pi \right]^2 & \text{for } \tau < 1 \\
  \left[ \arcsin \dfrac 1 {\sqrt{\tau}} \right]^2 & \text{for } \tau \geq 1 \,,
\end{cases}
\label{eq:ftau}
\end{align}
and $\tau_x = 4 m_x^2 / m_{h^0}^2$. In the effective model, we find
\begin{align}
  1+\Delta_g^{\text{EFT}} &= \dfrac 1 {A_{gg}^\text{SM}} \Bigg[  \sum_{f = t,b}\,(1+\bar{c}_f)\,A_f(\tau_f)\Bigg],  \\
 1+\Delta_\gamma^{\text{EFT}} &= \dfrac 1 {A_{\gamma \gamma}^\text{SM}} \Bigg[  \sum_{f = t,b}\, N_C\,Q_f^2(1+\bar{c}_f)\, A_f(\tau_f) + Q^2_{\tau}\,(1+\bar{c}_\ell)\,A_{f}(\tau_\tau) + A_v(\tau_W) + \dfrac { 64 \, \pi^2 \bar{c}_\gamma} {g^2}\Bigg] \,.
\end{align}
The comparison of couplings in the full 2HDM and the EFT is summarized in Tab.~\ref{tab:2hdm-matching}.

\begin{table}[tb!]
  \renewcommand{\arraystretch}{1.3}
  \centering
  \begin{tabular}{c c lll} 
    \toprule
    Coupling &\hspace*{1em}& & 2HDM & EFT \\
    \midrule
    \multirow{2}{*}{$1+\Delta_t$} && type I: & $x\,\cotb - \frac{x^2}{2} + \ord(x^3)$ &   $\bar{c}_u = x\cotb$ \\
    && type II: &  $x\,\cotb - \frac{x^2}{2} + \ord(x^3)$ &   $\bar{c}_u = x\cotb$ \\
    \midrule
    \multirow{2}{*}{$1+\Delta_b$} && type I: & $x\,\cotb + \ord(x^3)$ &   $\bar{c}_d = x\cotb$ \\
    && type II: &  $-x\,\tanb + \ord(x^3) $ &   $\bar{c}_d = -x\tanb$ \\
    \midrule
    \multirow{2}{*}{$1+\Delta_\tau$} && type I: & $x\,\cotb + \ord(x^3)$ &   $\bar{c}_\ell = x\cotb$ \\
    && type II: &  $-x\,\tanb + \ord(x^3)$ &   $\bar{c}_\ell = -x\tanb$ \\
    \midrule
    $1 + \Delta_V$ && & $1-\frac{x^2}{2} + \ord(x^3)$ & $\ope{}^{d8}$ \\
    \midrule
    $1 + \Delta_{h^0}$ && &  $1- x^2  \Bigl(\frac{3}{2}-\frac{4m_{12}^2}{m^2_{h^0}\sin 2\beta}\Bigr) + \ord(x^3)$ & $\bar{c}_6 = -x^2\frac{\mheavy^2}{\lambda v^2}$ \\
    \bottomrule
  \end{tabular}
  \caption{Tree--level Higgs coupling shifts $\Delta_x$ as a function of the 2HDM parameters. In the last column,
    the Wilson coefficients for the relevant dimension-6 operators in Tab.~\ref{tab:operators} are matched to the
    2HDM in the limit of decoupling heavy scalars $x \simeq v^2/M^2 \ll 1$ (cf. Eq.~\eqref{eq:2hdm-xi-unbroken}).}
  \label{tab:2hdm-matching}
\end{table}

\subsection{Scalar top partners}
\label{sec:ap-partners}

The simplified scalar top-partner generation sector is described by the Lagrangian
\begin{alignat}{5}
 \lag &\supset  (D_{\mu}\,\Qtilde)^\dagger\,(D^\mu\Qtilde) + (D_\mu\,\TR)^*\,(D^\mu\,\TR)
 - \underbrace{\tilde{Q}^\dagger\,M^2\,\tilde{Q}\,
 - M^2\,\TR^*\,\TR}_{\lag_\text{mass}}
 \notag \\
& \qquad\quad  -\underbrace{\kLL\,(\phi\cdot\Qtilde)^\dagger(\phi\cdot\Qtilde)
 -\kRR\,(\TR^*\TR)\,(\phi^\dagger\,\phi) }_{\lag_\text{Higgs}} 
 -\underbrace{\left[ \kLR \, M \, \TR^*\,(\phi \cdot \Qtilde) + \text{h.c.} \right]}_{\lag_\text{mixing}} \,.
\label{eq:lag-partners}
 \end{alignat}
We use the customary notation for the $SU(2)_L$
invariant product $\phi^a\cdot \Qtilde^b \equiv
\epsilon_{ab}\,\phi^a\,\Qtilde^b$, with the help of the antisymmetric
pseudo-tensor $\epsilon^{ab} \equiv (i\sigma^2)^{ab}$, so that
$\epsilon^{12} = -\epsilon^{21} = 1$. 

Notice that the term $\lag_\text{Higgs}$ gives rise to scalar
partner masses proportional to the Higgs VEV, mirroring the
supersymmetric F-term contribution to the squark masses.  By a
similar token, the explicit mass terms $\lag_\text{mass}$ are
analogous to the squark soft-SUSY breaking mass terms; while
$\lag_\text{mixing}$ is responsible for the mixing between the gauge
eigenstates, as a counterpart of the MSSM $A$-terms.
In the absence of an underlying
supersymmetry, the Lagrangian in Eq.~\eqref{eq:lag-partners} features
no equivalent of the D-term contributions.\medskip

Collecting all bilinear terms from Eq.~\eqref{eq:lag-partners} we get
\begin{alignat}{5}
 \lag &\supset (\stL^*\; \stR^*)
 \,\begin{pmatrix} \MLL^2 & \MLR^2 \\ \MRL^2  & \MRR^2 \end{pmatrix}\begin{pmatrix} \stL \\ \stR \end{pmatrix}
  \label{eq:masses1}
\end{alignat}
where
\begin{alignat}{5}
 \MLL^2 &= \kLL\cfrac{v^2}{2} + M^2 \,, \qqquad 
 \MLR^2=\MRL^2 = \kLR \, M\cfrac{v}{\sqrt{2}} \,, \qqquad 
 \MRR^2 = \kRR\,\dfrac{v^2}{2}\, + M^2 \,.
 \label{eq:masses2}
\end{alignat}
Assuming all parameters in Eq.~\eqref{eq:lag-partners} to be real,
the above mass matrix can be diagonalized through the usual
orthogonal transformation $R(\theta_{\tilde{t}})$ which rotates the gauge
eigenstates $(\stL, \stR)$ onto the mass basis $(\stone,\sttwo)$,
\begin{alignat}{5}
R(\theta_{\tilde{t}})\,\mathcal{M}_{\tilde{t}}^2\,R^\dagger(\theta_{\tilde{t}}) = \text{diag}(m^2_{\stone}, m^2_{\sttwo})\,,
\qquad \begin{pmatrix} \stone \\ \sttwo \end{pmatrix} = R(\theta_{\tilde{t}}) \begin{pmatrix} \stL \\ \stR \end{pmatrix} =
\begin{pmatrix} \cos\theta_{\tilde{t}} & \sin\theta_{\tilde{t}} \\ -\sin\theta_{\tilde{t}}  & \cos\theta_{\tilde{t}}\end{pmatrix}
 \begin{pmatrix} \stL \\ \stR \end{pmatrix}.
\label{eq:rotation}
\end{alignat}
The physical scalar partner masses and the mixing angle are then given by
\begin{align}
m_{\stone}^2 &= \MLL^2\ctd + \MRR^2\,\std + 2\MLR^2\,\st\ct \,,\notag \\
m_{\sttwo}^2 &= \MLL^2\,\std + \MRR^2\,\ctd - 2\MLR^2\,\st\,\ct \, ,
\label{eq:masses3} \\
\tan(2\theta_{\tilde{t}}) &= \frac{2\MLR^2}{\MLL^2-\MRR^2} \, .
\label{eq:mixing}
\end{align}
As we assume the right-handed partner $\sbR$ to be heavy and thus decoupled, the
sbottom-like scalar eigenstate $\sbL$ undergoes no mixing and can be
readily identified with the physical eigenstate.\medskip

To derive the effective theory, we compute the effective action at one
loop with the help of the covariant derivative
expansion~\cite{Gaillard:1985uh,Cheyette:1987qz,Henning:2014wua},
which is fully consistent with our mass degeneracy setup.  Notice
that, since the Lagrangian Eq.~\eqref{eq:lag-partners} lacks any linear
terms in the heavy scalar fields $\Psi \equiv (\Qtilde, \TR^*)$, the
tree-level exchange of such heavy partners cannot generate any
effective interaction at dimension 6. 

Following our default matching prescription, we set the matching scale as
$\Lambda = M$. The relevant Wilson coefficients in the SILH basis
then read:
\begin{alignat}{5}
& \bar{c}_{g} =  
 \cfrac{\mw^2}{24\,(4\pi)^2\,M^2}\,\left[(\kLL + \kRR) - {\kLR^2}\right] \notag \\
& \bar{c}_{\gamma} =  
 \cfrac{\mw^2}{9\,(4\pi)^2\,M^2}\,\left[(\kLL + \kRR) - {\kLR^2}\right] \notag \\
& \bar{c}_{B} =  
 -\cfrac{5\mw^2}{12\,(4\pi)^2\,M^2}\,\left[\kLL - \cfrac{31}{50}{\kLR^2}\right] \notag \\
& \bar{c}_{W} =  
 \cfrac{\mw^2}{4\,(4\pi)^2\,M^2}\,\left[\kLL - \cfrac{3}{10}{\kLR^2}\right] \notag \\
& \bar{c}_{HB} =  
 \cfrac{5\mw^2}{12\,(4\pi)^2\,M^2}\,\left[\kLL - \cfrac{14}{25}{\kLR^2}\right] \notag \\
& \bar{c}_{HW} =  -
 \cfrac{\mw^2}{4\,(4\pi)^2\,M^2}\,\left[\kLL - \cfrac{2}{5}{\kLR^2}\right] \notag \\
 &  \bar{c}_{{H}} = 
 \cfrac{v^2}{4(4\pi)^2\,M^2}\,\left[(2\kRR^2-\kLL^2) - 
 {\Bigl( \kRR - \frac{1}{2}\kLL \Bigr) \kLR^2 + \cfrac{\kLR^4}{10}}\right]
 \notag \\
 & \bar{c}_{{T}}  = \cfrac{v^2}{4(4\pi)^2\,M^2}\,\left[\kLL^2 - \cfrac{\kLL\,\kLR^2}{{2}} + \cfrac{\kLR^4}{{10}}\right].
\label{eq:triplet_coefficients_ap}
\end{alignat}

We also consider a $v$-improved matching. The only difference
to the default matching is the choice of the matching scale $\Lambda = m_\stone$,
which manifests itself as a rescaling of the Wilson coefficients in~Eq.\,\eqref{eq:triplet_coefficients_ap}
by a factor of $M^2 / m_\stone^2$.
\medskip

The scalar partner couplings to the Higgs boson can be written as
\begin{alignat}{5}
 g_{h\stone\stone}/v &= \kLL\,\ctd + \kRR\,\std + \sin(2\theta_{\tilde{t}})\,\kLR \,, \notag \\
 g_{h\sttwo\sttwo}/v &= \kLL\,\std + \kRR\,\ctd - \sin(2\theta_{\tilde{t}})\,\kLR \,, \notag \\
 g_{h\sbL\sbL}/v & = \kLL \,.
 \label{eq:higgscouplings}
\end{alignat}

\subsection{Vector triplet}
\label{sec:ap-triplet}

We consider a real vector triplet field $V_\mu^{a=1,2,3}$ transforming
under the SM gauge group as $(r_c,r_L,r_Y) =
(\textbf{1},\textbf{3},0)$.  Its dynamics can be effectively described
by means of the Lagrangian~\cite{Pappadopulo:2014qza}
\begin{alignat}{5}
 \lag \supset& -\cfrac{1}{4}\,V_{\mu\nu}^a\,V^{\mu\nu\, a}\, + \cfrac{\mv^2}{2}\,V_{\mu}^a\,V^{\mu\,a}
 + i\,\gV\,\ch\,\vmua\,\left[\phi^\dagger\tau^a\,\overleftrightarrow{D}^\mu\,\phi\,\right]
  +\cfrac{\gw^2}{\gV}\,\vmua\,\cF\sum_F\, \overline{F}_L\,\gamma^\mu\,\tau^a\,F_L
 \notag \\
 &+
 \cfrac{\gV}{2}\,\cvvv\,\epsilon_{abc}\,V_{\mu}^a\,V_{\nu}^b\,D^{[\mu}V^{\nu]c}\, + \gV^2\,\cvvhh\,\vmua\,V^{\mu a}\,\phibarphi\,
 - \cfrac{\gw}{2}\,\cvvw\,\epsilon_{abc}\,W^{\mu\nu}\,V_\mu^b\,V_\nu^c \,,
 \label{eq:lag-vectortriplet-app}
\end{alignat}
where the vector triplet field-strength tensor is $V_{\mu\nu}^a \equiv
D_{\mu}\vnua - D_{\nu}\,\vmua$ and $\tau^a\equiv \sigma^a/2$ are the
$SU(2)_L$ generators in the fundamental representation.  The covariant
derivative acts on the vector triplet field as $D_\mu\,V_\nu^a =
\partial_\mu\,V_\nu^a+g\epsilon^{abc}\,V^b_{\mu}V_{\nu}^c$.

The coupling constant $\gV$ stands for the characteristic strength of
the heavy vector-mediated interactions, while $\gw$ denotes the
$SU(2)_L$ weak gauge coupling (which differs from the coupling
strength $g$ of the observable $W$ boson due to $W$-$V$ mixing, see
below).  The different
dimensionless coefficients $c_i$ quantify the relative strengths of
the individual couplings.  This parametrization weights the extra $V$
and $\phi$ field insertions by one factor of $\gV$ each, while gauge
boson insertions are weighted by one power of the weak coupling.  An
exception is made for the couplings to fermions, where an extra
weighting factor $\gw^2/\gV^2$ is introduced for a convenient power
counting in certain UV embeddings~\cite{Pappadopulo:2014qza}.  For
simplicity, it is assumed that the fermion current in
Eq.~\eqref{eq:lag-vectortriplet-app} is universal.

Equation~\eqref{eq:lag-vectortriplet-app} is the most general Lagrangian
compatible with the SM gauge group and CP invariance, provided that
$V_\mu^a$ transforms as $V_\mu^a(\vec{x},t)
\stackrel{\text{CP}}{\longrightarrow}
-(-1)^{\delta_{a2}}\,V_\mu^a(-\vec{x},t)$ as the SM vectors. 
Moreover, the Lagrangian obeys a
global $SO(4) = SU(2)_L \times SU(2)_R$ symmetry, which is typical of
strongly interacting dynamics.

Since $V_\mu^a$ is not manifestly gauged, this simplified vector
triplet model in itself is not renormalizable.  However, it can be
easily linked to a gauge-invariant theory \eg via the Higgs or the
St\"uckelberg mechanisms~\cite{Pappadopulo:2014qza}.

An alternative model setup, which is particularly useful to construct
the effective theory,
introduces an explicit kinetic $V$-$W$ mixing via the Lagrangian
\begin{alignat}{5}
 \lag \supset &-\cfrac{1}{4}\,V_{\mu\nu}^a\,V^{\mu\nu\, a}\, + \cfrac{\tmv^2}{2}\,V_{\mu}^a\,V^{\mu\,a}
 + \,\gV\,\tch\,\vmua\,\jhup
  +\cfrac{\gw^2}{2\gV}\,\vmua\,\tcF\sum_F\,\jfup + \cvwtilde\,\cfrac{\gw}{2\gV}\,D_{[\mu}\,V_{\nu]}^a\,W^{\mu\nu\, a}
 \notag \\
 &+
 \cfrac{\gV}{2}\,\tcvvv\,\epsilon_{abc}\,V_{\mu}^a\,V_{\nu}^b\,D^{[\mu}V^{\nu]c}\, 
 + \gV^2\,\tcvvhh\,\vmua\,V^{\mu a}\,\phibarphi\,
 - \cfrac{\gw}{2}\,\tcvvw\,\epsilon_{abc}\,W^{\mu\nu}\,V_\mu^b\,V_\nu^c \,
 \label{eq:lag-vectortriplet-tilded},
\end{alignat}
where for convenience we have introduced the Higgs, fermion and vector
current bilinears
\begin{align}
J_\mu^{H, a} &= \cfrac{i}{2}\,\left[\phi^\dagger\,\sigma^a\,\lrd_\mu
\,\phi \right],
 & J_\mu^{F, a} &= \overline{F}_L\gamma_\mu\,\sigma^a\,F_L\,,
 & J_\mu^{W,a} &= D^\nu\,W_{\mu\nu}^a\, \label{eq:currents}. 
\end{align}
An appropriate field redefinition absorbs the kinetic mixing term
$V^{\mu a} \left(D^\nu W_{\mu \nu} \right)^a$ ~\cite{delAguila:2010mx}
and connects the parameters in the \emph{tilded} basis of
Eq.~\eqref{eq:lag-vectortriplet-tilded} and \emph{untilded} basis of
Eq.~\eqref{eq:lag-vectortriplet-app} through the relations
\begin{align}
  \mv^2 &= \dfrac {g_V^2} {g_V^2 - \cvwtilde^2 \gw^2} \tmv^2 \,,\notag \\
  \ch &= \dfrac {g_V} {\sqrt{g_V^2 - \cvwtilde^2 \gw^2}} \left[ \tch + \dfrac {\gw^2}{g_V^2} \cvwtilde \right] ,\notag \\
  \cF &= \dfrac{g_V} {\sqrt{g_V^2 - \cvwtilde^2 \gw^2}} \left[ \tcF + \cvwtilde \right] ,\notag \\
  \cvvhh &= \dfrac {g_V^2} {g_V^2 - \cvwtilde^2 \gw^2} \left[ \tcvvhh + \dfrac {\gw^2} {2 g_V^2} \cvwtilde \tch + \dfrac {\gw^4} {4g_V^4} \cvwtilde^2 \right] ,\notag \\
  \cvvw &= \dfrac {g_V^2} {g_V^2 - \cvwtilde^2 \gw^2} \left[ \tcvvw - \dfrac{\gw^2}{g_V^2} \cvwtilde^2 \right] ,\notag \\
  \cvvv &= \dfrac {g_V^2} {\left(g_V^2 - \cvwtilde^2 \gw^2\right)^{3/2}} \left[
    \tcvvv - \dfrac{\gw^2}{g_V^2} \cvwtilde (\tcvvw + 2) + 2 \dfrac {\gw^2} {g_V^4} \cvwtilde^3 \right]  \,.
  \label{eq:VectorTriplet_tilded_to_untilded}
\end{align}

\subsubsection*{Spectrum}

The heavy vector sector in the gauge basis contains one neutral state
$V_\mu^{0}\equiv V_\mu^3$ and two charged states $V_\mu^{\pm} \equiv
(V_\mu^1\mp V_\mu^2)/\sqrt{2}$.  Upon EWSB only one vector state
remains massless, which we readily identify with the standard photon
field $A_\mu = \cw\,B_\mu + \sw\,W_\mu^3$.  Here, the Weinberg angle
is linked as usual to the electroweak gauge couplings $e = \gw\,\sw =
g'\,\cw$, although at this stage we cannot yet relate it to
electroweak observables before the mixing with the heavy vectors is
included.  The latter involves, for the neutral fields, the heavy
vector component $V^0$ and the linear combination of $B,W^3$
orthogonal to the photon field. A similar mixing pattern appears in
the charged sector, involving the field components $V^{1,2}_\mu,
W^{1,2}_\mu$.  The physical mass eigenstates can be written as
\begin{align}
  Z_\mu &= \cos \theta_N \left(-\sw B_\mu + \cw W_\mu^3 \right) + \sin \theta_N \, V^3_\mu \,,\notag \\
  \xi^0_\mu &= - \sin \theta_N \left(-\sw B_\mu + \cw W_\mu^3 \right) + \cos \theta_N  \, V^3_\mu \,,\notag \\
  W^\pm_\mu &= \cos \theta_C \, \dfrac{W^1_\mu \mp W^2_\mu} {\sqrt{2}} + \sin \theta_C \, \dfrac{V^1_\mu \mp V^2_\mu} {\sqrt{2}} \,,\notag \\
  \xi^\pm_\mu &= - \sin \theta_C \, \dfrac{W^1_\mu \mp W^2_\mu} {\sqrt{2}} + \cos \theta_C \, \dfrac{V^1_\mu \mp V^2_\mu} {\sqrt{2}} \,.
\end{align}  
The mass eigenvalues are given by
\begin{align}
  m_{Z/\xi^0}^2 &= \dfrac 1 2 \left[ \hat{m}_V^2 + \hat{m}_Z^2 \mp \sqrt{ \left( \hat{m}_Z^2 - \hat{m}_V^2 \right)^2 + \ch^2 \, g_V^2 \, \hat{m}_Z^2 \, \hat{v}^2} \, \right] \notag \\
    &=
  \begin{cases}
    \hat{m}_Z^2 \left(1 - \dfrac {\ch^2 g_V^2} {4} \, \dfrac {\hat{v}^2} {\hat{m}_V^2} + \ord(\hat{v}^4/ \hat{m}_V^4)  \right)  \\
    \hat{m}_V^2 \left(1 + \dfrac {\ch^2 g_V^2} {4} \, \dfrac {\hat{v}^2} {\hat{m}_V^2} + \ord(\hat{v}^4/ \hat{m}_V^4)  \right) \,,
  \end{cases}
  \label{eq:VectorTriplet_mZxi}
\end{align}
\begin{align}
  m_{W^\pm/\xi^\pm}^2 &= \dfrac 1 2 \left[ \hat{m}_V^2 + \hat{m}_W^2 \mp \sqrt{ \left( \hat{m}_W^2 - \hat{m}_V^2 \right)^2 + \ch^2 \, g_V^2 \, \hat{m}_W^2 \, \hat{v}^2} \, \right] \notag \\
    &=
  \begin{cases}
    \hat{m}_W^2 \left(1 - \dfrac {\ch^2 g_V^2} {4} \, \dfrac {\hat{v}^2} {\hat{m}_V^2} + \ord(\hat{v}^4/ \hat{m}_V^4)  \right) \\
    \hat{m}_V^2 \left(1 + \dfrac {\ch^2 g_V^2} {4} \, \dfrac {\hat{v}^2} {\hat{m}_V^2} + \ord(\hat{v}^4/ \hat{m}_V^4)  \right) \,.
  \end{cases} 
  \label{eq:VectorTriplet_mWxi}
\end{align}
For the mixing angles, we find
\begin{align}
  \tan (2 \theta_N) &= \dfrac {\ch \, g_V \, \hat{v} \, \hat{m}_Z } {\hat{m}_V^2 - \hat{m}_Z^2}  
  = \dfrac{\ch \, g \, g_V} {2 \, \cw} \, \dfrac {\hat{v}^2} {\hat{m}_V^2}  +  \ord(\hat{v}^4/ \hat{m}_V^4) \,,\notag \\
  \tan (2 \theta_C) &= \dfrac {\ch \, g_V \, \hat{v} \, \hat{m}_W } {\hat{m}_V^2 - \hat{m}_W^2} 
  = \dfrac{\ch \, g \, g_V} 2 \, \dfrac {\hat{v}^2} {\hat{m}_V^2}  + \ord(\hat{v}^4/ \hat{m}_V^4) \,,
  \label{eq:VectorTriplet_mixingangles}
\end{align}
or
\begin{align}
  \sin \theta_C = \dfrac {\ch \, g \, g_V} {4} \, \dfrac {v^2} {\mv^2}  + \ord(\hat{v}^4/ \hat{m}_V^4) \,.
\end{align}
Here we define
\begin{align}
  \hat{m}_Z  = \dfrac{\gw \, \hat{v} } {2 \, \cw} \qquad
  \hat{m}_W  = \dfrac{\gw \, \hat{v} } {2} \qquad 
  \hat{m}_V^2  = \mv^2 + g_V^2 \, \cvvhh \, \hat{v}^2
  \label{eq:VectorTriplet_mVhat}
\end{align}
where $\hat{v}$ is the actual vev of $\phi$, which does not necessarily have the SM value of $v = 2m_W/g \approx 246 \ \gev$. 

Notice that the $V$-$W$ mixing also affects the weak current interactions, which are no longer governed by $\gw$. Instead, the physical $Wff'$ coupling reads
\begin{align}
  g  = \cos \theta_C \, \gw - \sin \theta_C \, \cF \, \dfrac {\gw^2} {g_V} 
     = \gw \, \left(1 - \dfrac{\cF \, \ch \, \gw^2} 4  \, \dfrac {v^2}{\mv^2} \right)+ \ord(v^4 / \mv^4 ) \,.
  \label{eq:VectorTriplet_gratio}
\end{align}
The relation between $\hat{v}$ and $v$ can be read off from Eq.~\eqref{eq:VectorTriplet_mWxi}, giving approximately
\begin{align}
  \dfrac{\hat{v}} v = 1 + \dfrac {\ch^2 \, g_V^2} 8 \, \dfrac {v^2} {\mv^2}  -
  \dfrac {\cF \, \ch \, \gw^2} 4 \, \dfrac {v^2} {\mv^2}  + \ord(v^4 / \mv^4 ) \,.
  \label{eq:VectorTriplet_vevratio}
\end{align}

The global $SU(2)_V$ custodial symmetry connects the charged and
neutral current strengths through $m_W^2\,m^2_{\xi^{\pm}} =
\cwd\,m_Z^2\,m^2_{\xi^0}$, which generalizes the SM relation $m_W^2 =
\cwd\,m_Z^2$.  Compatibility with EWPO enforces nearly mass-degenerate
states $m_{\xi^0} \simeq m_{\xi^{\pm}}$ for phenomenologically viable
scenarios.  In practice, we set up our model in the $m_W$-$g$ scheme,
\ie taking as input parameters $g$, $m_W$, $\alpha$, $m_{h^0}$,
$\alpha_s$; the model-specific parameters $c_i$; as well as the
physical masses $m_{\xi^\pm}$.  The mass spectrum and mixing angles we
obtain by solving Eq.~\eqref{eq:VectorTriplet_mZxi} and
Eq.~\eqref{eq:VectorTriplet_mWxi} iteratively.

\subsubsection*{Effective theory}

To construct the vector triplet EFT following the default matching, we identify 
the new physics scale $\Lambda = \mv$.  Starting from the heavy
triplet Lagrangian defined by Eq.~\eqref{eq:lag-vectortriplet-tilded},
we first integrate by parts the kinetic mixing term,
\begin{alignat}{5}
\cvwtilde{\dfrac{\gw}{2\gV}}\,D_{[\mu}\,V_{\nu]}^a\,W^{\mu\nu\, a} =
\cvwtilde\,\cfrac{\gw}{\gV}\,V^{\mu, a}\,(D^\nu\,W^a_{\mu\nu}) = 
 \cvwtilde\,\cfrac{\gw}{\gV}\,V^{\mu, a}\,\jwdown \,  
\label{eq:kinmix}, 
\end{alignat}
such that we can rewrite it in terms of the gauge current from
Eq.~\eqref{eq:currents}.  Integrating out the heavy vector field
$V^a_\mu$ one obtains the effective Lagrangian
\begin{alignat}{5}
 \lageff &\supset \cfrac{\tmv^2}{2}V^{\mu,a}\,V_\mu^a + \vmua\left[\gV\,\tch\,\jhup
  +\cfrac{\gw^2}{2\gV}\,\tcF\,\sum_F\,\jfup +
  \cvwtilde\,\cfrac{\gw}{\gV}\,\jwdown \right] +  \ord(V^3)\,,
 \label{eq:lageff1}
\end{alignat}
where we neglect those contributions involving higher powers in the
heavy field, as they play no role in our analysis.

The Euler-Lagrange equation for \vmua,
\begin{alignat}{5}
& [\partial^\mu\partial^\nu - g^{\mu\nu}\,\partial^2 - \tmv^2] \,V^{a}_{\nu}
= \gV\,\tch\,\jhup+\cfrac{\gw^2}{2\gV}\,\tcF\sum_F\jfup\,+\cvwtilde\,\cfrac{\gw}{\gV}\,\jwup + 
 \,\text{h.o.\ terms in}\; \vmua\,, \notag
\end{alignat}
leads to
\begin{align}
V^{\mu,a} = 
-\cfrac{1}{\tmv^2}\,\left[\cvwtilde\,\cfrac{\gw}{\gV}\,\jwup +\gV\,\tch\,\jhup+\cfrac{\gw^2}{2\gV}\,\tcF\,\sum_F\,\jfup\,\right] {+ {\cal O}(p_V^2/\tmv^4) + {\cal O}(V^2)}  
 \label{eq:eom-v}.
\end{align}

Plugging Eq.~\eqref{eq:eom-v} into Eq.~\eqref{eq:lageff1}, $\lageff$ can
be expressed in terms of current products as
\begin{alignat}{5}
 \lageff &\supset {-}\cfrac{\gw^{{4}}\,\tcF^2}{8\gV^2\,\tmv^2}\,\jfup\,\jfdown - 
 \cfrac{\gV^2\,\tch^2}{2\,\tmv^2}\,\jhup\,\jhdown
 - \cfrac{\gw^2\,\tcF\,\tch}{2\,\tmv^2}\,\jhup\,\jfdown {-}  \cfrac{\gw\,\tch\,\cvwtilde}{\tmv^2}\,\jhup\,\jwdown
 \notag \\
& \qquad - \cfrac{\gw^2\,\cvwtilde^2}{{2}\,\gV^2\,\tmv^2}\,\jwup\,\jwdown {-} \cfrac{\gw^3\,\tcF\,\cvwtilde}{{ 2}\,\gV^2\,\tmv^2}\,\jwup\,\jfdown
 \label{eq:lageff2}. 
\end{alignat}
In the following, we disregard 4-fermion operators since they are
irrelevant for our analysis. The remaining five current products in
Eq.~\eqref{eq:lageff2} can be expressed in terms of two independent
ones by using Eq.~\eqref{eq:eom-w} (with the replacement $g \to \gw$),
which corresponds to $\jwup = \gw\jhup+ \gw\jfup/2$:
\begin{alignat}{5}
 \lageff &\supset - 
 \cfrac{(\gV^2\tch+\gw^2\cvwtilde)^2}{2\,\gV^2\,\tmv^2}\,\jhup\,\jhdown
 - \cfrac{\gw^2\,(\tcF+\cvwtilde)\,(\gV^2\tch+\gw^2\cvwtilde)}{2\,\gV^2\,\tmv^2}\,\jhup\,\jfdown + \text{4-fermion}
 \label{eq:lageffeom}.
\end{alignat}
Using Eq.(A.4) in~\cite{Pappadopulo:2014qza}, it can be checked that
this equation is invariant when changing between the tilded and
untilded bases.  With the help of Eqs.~\eqref{eq:EFT_JHJH_replacement},
\eqref{eq:EFT_OHf_replacement}, and \eqref{eq:EFT_Or_replacement}
(and again relabeling $g \to \gw$ in these relations) the two
independent current products can be expressed in terms of dimension-6
operators as follows:
\begin{alignat}{5}
\jhup\,\jhdown &= -\cfrac{1}{4}\,(\oh - 4\,\orop) = 
-\cfrac{1}{4}\,\left[ 3\oh -8\lambda\osix - 2\sum_f\left[y_f\of + \text{h.c.} \right] \right]
\notag \\
  \jfup\,\jhdown &= 
  \cfrac{i}{2}\,\ohfprime 
  = \cfrac{i\ow}{\gw} + \cfrac{1}{2}\,\left[ 3\oh - 8\lambda\osix - 2\sum_f\left[y_f\of + \text{h.c.}\right]\right]
\label{eq:current-summary1} 
\end{alignat}
where $y_f$ denotes the bare Yukawa coupling $y_f \equiv
\sqrt{2}m_f/v$.  Plugging the above into Eq.~\eqref{eq:lageffeom}, one
can easily read off the relevant Wilson coefficients of the EFT:
\begin{alignat}{5}
 \bar{c}_H &= \cfrac{3\,\gw^2\,v^2}{4\,\tmv^2}\,\left[\tch^2\cfrac{\gV^2}{\gw^2} -2 \tcF\,\cvwtilde\,\cfrac{g^2}{\gV^2} - 2\,\tcF\,\tch { -\cvwtilde^2\,\cfrac{g^2}{\gV^2}}\right] ,\notag \\
 \bar{c}_6 &= \cfrac{\gw^2\,v^2}{\tmv^2}\,\left[\tch^2\cfrac{\gV^2}{\gw^2} -2 \tcF\,\cvwtilde\,\cfrac{g^2}{\gV^2} - 2\,\tcF\,\tch {-\cvwtilde^2\,\cfrac{g^2}{\gV^2}}\right] ,\notag \\
 \bar{c}_f &= \cfrac{\gw^2\,v^2}{4\,\tmv^2}\,\left[\tch^2\cfrac{\gV^2}{\gw^2} -2 \tcF\,\cvwtilde\,\cfrac{g^2}{\gV^2} - 2\,\tcF\,\tch { -\cvwtilde^2\,\cfrac{g^2}{\gV^2}}\right] ,\notag \\
 \bar{c}_W &= \cfrac{m_W^2}{\tmv^2}\,\left[-\tcF\tch -\tch\,\cvwtilde - \tcF\,\cvwtilde\,\cfrac{\gw^2}{\gV^2}  -\cvwtilde^2\,\cfrac{g^2}{\gV^2}\right] \, .
 \label{eq:wilsonmatched}
\end{alignat}
In the untilded basis, these correspond to
\begin{align}
 \bar{c}_{H} &= \dfrac{3\,\gw^2\,v^2}{4\,M_V^2}\,\left[\ch^2\dfrac{g_V^2}{\gw^2}  - 2\,\cF\,\ch\right] , \notag \\
 \bar{c}_{6} &= \dfrac{\gw^2\,v^2}{M_V^2}\,\left[\ch^2\dfrac{g_V^2}{\gw^2} - 2\,\cF\,\ch \right] , \notag \\
 \bar{c}_{f} &= \dfrac{\gw^2\,v^2}{4\,M_V^2}\,\left[\ch^2\dfrac{g_V^2}{\gw^2}  - 2\,\cF\,\ch \right] , \notag \\
 \bar{c}_{W} &= - \dfrac{m_W^2}{M_V^2}\, \cF\ch \,.
  \label{eq:VectorTriplet_EFT2}
\end{align}
with $f = u, d, \ell$. Other than that, only four-fermion interactions are generated at tree level and at $\ord(v^2/M_V^2)$;
these are not relevant for our analysis and are not considered here.

As in the 2HDM and scalar partner models, we define an additional $v$-improved EFT by
$\Lambda = m_{\xi^0}$, leading the same Wilson coefficients as above
except that $M_V$ is replaced by $m_{\xi^0}$.

\subsubsection*{Higgs couplings}

On the EFT side, it is illustrative to discuss the origin of the Higgs
coupling shifts within two different approaches. First we consider the
EFT that keeps the fermionic operator \ohfprime\ (\ie instead of using
the conventional replacement in Eq.~\eqref{eq:EFT_OHf_replacement} that
maximizes the use of bosonic operators). In this case, similar to
Eq.~\eqref{eq:VectorTriplet_gratio}, a renormalization effect of the
weak coupling occurs from $V$-$W$ mixing,
\begin{alignat}{5}
g &= \gw(1-i\bar{c}'_{HF})
\end{alignat}
where $g$ is the observable coupling between the $W$ boson and SM
fermions. In this EFT and using the untilded basis, the relevant
Wilson coefficients are
\begin{alignat}{5}
\bar{c}_H &= \ch^2 \dfrac{3\gV^2v^2}{4M_V^2}, \qquad
\bar{c}_f &= \dfrac{1}{3}\bar{c}_{H}, \qquad
\bar{c}'_{HF}= -i\cF\ch\dfrac{\gw^2v^2}{4M_V^2}\, .
\end{alignat}
Instead, if we now consider the EFT with the bosonic operator \ow, \ie
after applying the replacement in Eq.~\eqref{eq:EFT_OHf_replacement},
there is no additional renormalization of the weak coupling, so that
$g=\gw$.  The relevant Wilson coefficient are given in Eq.~\eqref{eq:VectorTriplet_EFT2}.

Now we are in a position to determine the Higgs coupling shifts in
the three models. For the Yukawa couplings we find
\begin{align}
&\text{Full model:} & \Delta_f^\text{full} &= \frac{\gw}{g} \,\frac{v}{\hat{v}} -1 =
                           \frac{1}{c_{\theta_C}^{}-\cF\frac{\gw}{\gV}s_{\theta_C}^{}}\, \frac{v}{\hat{v}} -1 \notag \\
&&&=  \ch^2\frac{\gV^2 v^2}{8M_V^2} + \cF\ch\frac{g^2 v^2}{4M_V^2} + {\cal O}(M_V^{-4}) \notag \\[2ex]
&\text{EFT with \ohfprime:} &
\Delta_f^{\ohfprime} &= \frac{\bar{c}_{H}}{2}+ \bar{c}_{f} =  \frac{\bar{c}_{H}}{2}+ \bar{c}_{f} + i\bar{c}'_{HF}  \notag \\
&&&= \ch^2\frac{\gV^2 v^2}{8M_V^2} + \cF\ch\frac{g^2 v^2}{4M_V^2} \notag \\[2ex]
&\text{EFT with \ow:} &
\Delta_f^{\ow} &= \frac{\bar{c}_{H}}{2}+ \bar{c}_{f}  \notag \\
  &&&=  \ch^2\frac{\gV^2 v^2}{8M_V^2} + \cF\ch\frac{g^2 v^2}{4M_V^2}
\end{align}

Similarly for the Higgs coupling to on-shell $W$ bosons we get
\begin{align}
&\text{Full model:} &
\Delta_W^\text{full} &= \frac 1 {g m_W} \left( \frac{c_{\theta_C}^2 g^2 \hat{v}}{2(c_{\theta_C}^{}-\cF\frac{\gw}{\gV}s_{\theta_C}^{})^2}
-\ch\frac{s_{\theta_C}^{}c_{\theta_C}^{} g \gV \hat{v}}{c_{\theta_C}^{}-\cF\frac{\gw}{\gV}s_{\theta_C}^{}}+
2c_{ VVHH} s_{\theta_C}^2 \gV^2 \hat{v} \right) - 1 \notag \\
&&&= \ch^2\frac{3\gV^2 v^2}{8M_V^2} + \cF\ch\frac{g^2 v^2}{4M_V^2} + {\cal O}(M_V^{-4}) \notag \,, \\[2ex]
&\text{EFT with \ohfprime:} &
\Delta_W^{\ohfprime} &= \frac {\gw} {g}  \left( 1-\frac{\bar{c}_{H}}{2} \right) - 1 = \frac{\bar{c}_{H}}{2}+ i\bar{c}'_{HF}  \notag \\
&&&= \ch^2\frac{3\gV^2 v^2}{8M_V^2} + \cF\ch\frac{g^2 v^2}{4M_V^2} \,, \notag \\[2ex]
&\text{EFT with \ow:} &
\Delta_W^{\ow} &=  \frac{\bar{c}_{H}}{2}+ 2\bar{c}_{W} \notag \\
  &&&= \ch^2\frac{3\gV^2 v^2}{8M_V^2} + \cF\ch\frac{g^2 v^2}{4M_V^2}  \,.
\end{align}